\documentclass[aps,prd,fleqn,notitlepage,nofootinbib,preprintnumbers,nobalancelastpage,superscriptaddress,floatfix]{revtex4}

\usepackage{footmisc,multirow}
\usepackage{amsmath,slashed}
\usepackage{amssymb}
\usepackage{enumerate}

\usepackage[pdftex]{graphicx}
\usepackage{epstopdf}
\graphicspath{./}

\usepackage[pdfborder={0 0 0}]{hyperref}
\usepackage{xspace}
\usepackage{relsize}

\DeclareRobustCommand{\plusplus}{\raisebox{0.2ex}{\smaller++}}

\newcommand{\Alpgen}{A\protect\scalebox{0.8}{LPGEN}\xspace}
\newcommand{\Madgraph}{M\protect\scalebox{0.8}{AD}G\protect\scalebox{0.8}{RAPH}\xspace}

\newcommand{\Whizard}{WH\protect\scalebox{0.8}{I}Z\protect\scalebox{0.8}{ARD}\xspace}
\newcommand{\Pythia}{P\protect\scalebox{0.8}{YTHIA}\xspace}
\newcommand{\PythiaEight}{\Pythia~8\xspace}

\newcommand{\Sherpa}{S\protect\scalebox{0.8}{HERPA}\xspace}
\newcommand{\Comix}{C\protect\scalebox{0.8}{OMIX}\xspace}

\newcommand{\Amegic}{A\protect\scalebox{0.8}{MEGIC\plusplus}\xspace}
\newcommand{\Csshower}{CSS\protect\scalebox{0.8}{HOWER}\xspace}

\newcommand{\Root}{R\protect\scalebox{0.8}{OOT}\xspace}
\newcommand{\TLimit}{TL\protect\scalebox{0.8}{IMIT}\xspace}
\newcommand{\Rivet}{R\protect\scalebox{0.8}{IVET}\xspace}
\newcommand{\Fastjet}{F\protect\scalebox{0.8}{AST}J\protect\scalebox{0.8}{ET}\xspace}

\newcommand{\LHC}{LHC\xspace}
\newcommand{\ATLAS}{ATLAS\xspace}
\newcommand{\CMS}{CMS\xspace}

\newcommand{\be}{\begin{eqnarray*}}
\newcommand{\ee}{\end{eqnarray*}}

\newcommand{\bee}{\begin{eqnarray}}
\newcommand{\eee}{\end{eqnarray}}
\newcommand{\beeq}{\begin{equation}}
\newcommand{\eeeq}{\end{equation}}

\newcommand{\mr}{\mathrm}

\newcommand{\done}{{\rm d}}
\newcommand{\seff}{s_\text{eff}}
\newcommand{\hPO}{\hphantom{1}}

\hypersetup{
  pdfauthor={Frank Krauss, Petar Petrov, Marek Schoenherr, Michael Spannowsky},
  pdftitle={Measuring collinear W emissions inside jets}
}

\begin{document}
\title{Measuring collinear $W$ emissions inside jets}

\begin{abstract}
  Single and multiple emission of electroweak gauge bosons and in
  particular of $W$ bosons is discussed in the parton shower language.
  Algorithms and observables for the reconstruction of both leptonically
  and hadronically decaying $W$ bosons inside light quark jets are
  compared, and they are applied to a study of how emission rates of
  $W$ bosons in light-jet events at the LHC could be measured.
\end{abstract}

\author{Frank Krauss} \email{frank.krauss@durham.ac.uk}
\affiliation{Institute for Particle Physics Phenomenology, Department
  of Physics,\\Durham University, DH1 3LE, United Kingdom}
\author{Petar Petrov} \email{petar.petrov@durham.ac.uk}
\affiliation{Institute for Particle Physics Phenomenology, Department
  of Physics,\\Durham University, DH1 3LE, United Kingdom}
\author{Marek Sch\"onherr} \email{marek.schoenherr@durham.ac.uk}
\affiliation{Institute for Particle Physics Phenomenology, Department
  of Physics,\\Durham University, DH1 3LE, United Kingdom}
\author{Michael Spannowsky} \email{michael.spannowsky@durham.ac.uk}
\affiliation{Institute for Particle Physics Phenomenology, Department
  of Physics,\\Durham University, DH1 3LE, United Kingdom}

\preprint{IPPP/14/04}
\preprint{DCPT/14/08}
\preprint{MCNET-14-06}
\preprint{LPN14-054}

\maketitle

\section{Introduction}
\label{sec:intro}
After the recent discovery of a Standard Model-like Higgs boson by \ATLAS and
\CMS~\cite{Aad:2012tfa, Chatrchyan:2012ufa}, the main focus of the LHC's 
upcoming $14$ TeV run will be on on further precision studies of this newly 
found particle to fully establish all of its predicted properties.  In 
addition, of course, the hunt for new physics phenomena and, possibly, heavy
new particles will continue.  Many models for such new physics predict
the existence of short-lived TeV-scale resonances, which predominantly decay 
to those particles in the Standard Model that are thought to be most sensitive
to electroweak-scale particles, {\it e.g.}\ the top quark, and the $W$, $Z$, 
or Higgs bosons.  The anticipated large masses of the primary resonances imply 
that their decay products are highly boosted; that is, they tend to have 
relatively large momenta and therefore quite often momenta transverse to the 
beam axes which are significantly larger than their mass.  As a consequence,
many search strategies for such new particles by now concentrate on signatures
including a number of highly boosted weak gauge bosons accompanied by jets. 
However, the findings up to now, or, better put, their absence has placed a 
number of stringent lower mass bounds on such new heavy particles, in general 
of $\mathcal{O}(\mathrm{TeV})$.  This in turn pushes their anticipated masses 
even higher and it also translates into even larger boosts.  This poses an
interesting challenge, since obviously increasing the boost of a heavy decaying
particle directly translates into decreasing distances of its decay products,
and traditional strategies for their isolation start to become obsolete \cite{Butterworth:2008iy,Krohn:2009th, Ellis:2009me, Kim:2010uj, Thaler:2010tr, Soper:2011cr, Ellis:2012sn, Soper:2012pb}.  

In this situation,  precise background predictions are of utmost importance to 
maximize the sensitivity in searches with leptons, missing transverse energy 
and jets at large center-of-mass energies.  Especially when electroweak--scale 
resonances are highly boosted theory uncertainties become large: firstly, 
higher--order QCD corrections may open up hitherto unconsidered channels or 
allow for more extreme kinematics~\cite{Ferroglia:2013zwa, Rubin:2010xp};
in addition, virtual electroweak corrections will become increasingly important
with increasing scales tested in a process.  This can be seen from a rough 
estimate, approximating such corrections through their leading logarithmic
behavior as $\alpha_W \ln^2(\hat{s}/m_W^2)$, where $\hat{s}$ is the typical
scale of the process~\cite{Ciafaloni:1998xg,Kuhn:1999nn,Fadin:1999bq,
  Ciafaloni:2000df,Hori:2000tm,Denner:2000jv,Accomando:2001fn,
  Beenakker:2001kf,Denner:2003wi,Denner:2004iz,Denner:2008yn}.
The large logarithms typically discussed in this first round of publications
emerge due to a non-cancellation of real and virtual contributions.  Neglecting,
physically justified, the quite distinct emission of real heavy gauge bosons  
leads to large Sudakov--type logarithms of the form above, which can
be resummed through exponentiation, similar to the QCD case.  And, in quite
good analogy with the better known case of QCD, real emissions tend to cancel
these logarithms.  In contrast to QCD, however, the large masses of the heavy
gauge bosons prevent a complete cancellation due to the phase space constraints
imposed by them on the real radiation pattern.  This has been studied in some
detail in~\cite{Baur:2006sn}, showing that more careful consideration has to
be placed on how inclusive processes are being studied.   This first study
has been supplemented by a similar, but more precise consideration of such
partial cancellation effects, including terms of higher logarithmic accuracy 
in~\cite{Bell:2010gi}.  Of course, with the advent of improved calculational
technologies, especially for the virtual (loop) contributions, such effects
can be studied at complete next-to leading order accuracy in a very coherent
fashion, for some recent work, {\it cf.}~\cite{Dittmaier:2012kx,
  Kuhn:2013zoa,Bierweiler:2013dja}. 

However, in searches for new physics, the emission of {\em multiple} gauge 
bosons and the interplay of real and virtual corrections to such processes 
may become an important aspect, which clearly stretches beyond the next-to
leading order accuracy, including on one gauge boson only, marking the 
current level of typical precision calculations in the electroweak sector.  
And while fully automated tree-level matrix element generators such as 
\Alpgen~\cite{Mangano:2002ea}, \Madgraph~\cite{Alwall:2011uj}, 
\Amegic~\cite{Krauss:2001iv}, \Comix~\cite{Gleisberg:2008fv}, or 
\Whizard~\cite{Kilian:2007gr}
are well capable of forcefully evaluating cross sections for the production of,
say, up to 6 gauge bosons including their decays at leading order, only 
recently algorithms have been worked out, which use the $SU(2)$-algebra in a 
way as elegant as the one available for the gluons in QCD~\cite{Dai:2012jh}. 

Another, somewhat orthogonal, idea that could be borrowed from QCD is the 
parton shower paradigm which proved fairly successful in generating multiple 
particle emissions through a probabilistic picture and thereby effectively 
taking into account the leading higher-order corrections.  Only recently, 
first serious studies have been performed~\cite{Christiansen:2014kba}, which 
aim at validating and improving the treatment of weak gauge bosons in the 
parton shower picture, implemented in the \PythiaEight 
framework~\cite{Sjostrand:2007gs}.  

In this publication the radiation of weak bosons off other primary particles
is studied, with a special emphasis on their emergence as jets or parts of 
jets in a boosted environment.  In such a set-up large scales can be related 
to the process at the parton level, leading to the occurrence of large 
Sudakov logarithms of the type discussed above.  It will be interesting to
see, if the parton shower picture is able to reproduce this kind of effect
qualitatively and quantitatively and to investigate avenues of how such weak
Sudakov effects can be studied experimentally.

Therefore, this article is structured in the following way:  In 
Sec.~\ref{sec:ewshower} the implementation of the radiation of electroweak 
gauge bosons in the parton shower formalism as included in \Sherpa is 
discussed.  Some critical observables are defined which allow the 
reconstruction of the emitted gauge bosons in Sec.~\ref{sec:analysis}, where we 
focus both on leptonically or hadronically decaying $W$ bosons.  In 
Sec.~\ref{sec:results} we estimate how well LHC's multi-purpose experiments 
can measure emissions of collinear electroweak bosons in run 2.
We summarize and add a concluding discussion in Sec.~\ref{sec:conclusion}.

\section{Electroweak Shower and Event Generation}
\label{sec:ewshower}

To evaluate how well resummed electroweak corrections can be measured at the 
LHC we study the production of $W$ bosons in the parton shower on top of QCD 
dijet events, $pp \to jj$, at $\sqrt{s} = 14$ TeV. We use the 
\Sherpa~\cite{Gleisberg:2008ta} event generator with the modifications detailed
below implemented on top of version 2.1.0.  The hard scattering matrix elements
are computed by \Comix~\cite{Gleisberg:2008fv} and showered using the 
modified \Csshower~\cite{Nagy:2005aa,Schumann:2007mg}.  
Hadronization~\cite{Winter:2003tt} and the underlying 
event~\cite{Alekhin:2005dx} are taken into account, including hadron decays. 
Higher order QED corrections are accounted for both in hadron and leptonic 
gauge boson decays~\cite{Schonherr:2008av}.  The CT10 parton distribution 
functions~\cite{Lai:2010vv} have been used.

The following describes the implementation of the radiation of electroweak 
gauge bosons in the parton shower formalism. It is implemented as an extension 
of the \Csshower. In addition to the standard QCD and QED splitting functions, 
already incorporated in the parton shower algorithms \cite{Seymour:1991xa,
  Sjostrand:2006za,Sjostrand:2007gs,Schumann:2007mg,Bahr:2008pv,Hoeche:2009xc}, 
splitting functions for the 
radiation of electroweak gauge bosons off fermions~\cite{Denner:2006xxx} are 
implemented. They have already been used in determining the cluster history 
in \Sherpa's implementation of the multijet merging 
algorithm~\cite{Catani:2001cc,Hoeche:2009rj}\footnote{
  Further implementations of this or similar algorithms for QCD multi--particle
  final state have been described in \cite{Lonnblad:2001iq,Mangano:2006rw,
    Alwall:2007fs,Hamilton:2009ne,Lonnblad:2011xx,Lonnblad:2012ng,Platzer:2012bs}. 
}.
In this study they are also used to describe parton evolution.

Following the notation of~\cite{Schumann:2007mg,Hoeche:2009xc,Carli:2010cg} 
the parton shower approximation to the cross section of the emission of 
an electroweak gauge boson off an $n$ particle configuration can be 
written as
\begin{equation}\label{eq:def_ewps}
  \begin{split}
    \done\sigma_{n+V} \,=\;
    &\done\sigma_n \sum_f\sum_s^{n_\text{spec}}
      \frac{\done t}{t}\,\done z\,\frac{\done\phi}{2\pi}\;
      \frac{1}{n_\text{spec}}\;J(t,z)\;
      \mathcal{K}_{f(s)\to f^{(\prime)}V(s)}(t,z)\;,
  \end{split}
\end{equation}
wherein the labels $f$ and $s$ run over all fermions of the $n$-particle 
configuration and signify the emitter and spectator fermions, with 
$n_\text{spec}$ the number of spectators.  In the collinear limit any number 
of spectators, and in particular the choice of one spectator only would be 
a valid choice.  However, in order to maintain a dipole--\-like formulation
all particles with electroweak charges present acceptable choices as 
spectators.  The emission phase space is parametrized in terms of the 
evolution variable $t$, the splitting variable $z$ and the azimuthal 
angle $\phi$. Their precise definition, depending on whether $f$ and/or 
$s$ are in the initial or final state, as well as the values of the 
Jacobean factors $J(t,z)$ can be found in~\cite{Schumann:2007mg}.  The 
latter also contain a ratio of parton distribution functions to account 
for the possible change in initial state parton flavors and/or momenta. 

The splitting functions $\mathcal{K}$ are an adaptation of the expressions 
calculated in~\cite{Denner:2006xxx}, cast in a form suitable
for \Sherpa's \Csshower. For the collinear emission of an electroweak 
gauge boson, $W^\pm$ or $Z$, off a fermion or anti-fermion $f$ in the 
presence of a spectator $s$ in the high energy limit $E\gg m_V$ they 
read
\begin{equation}\label{eq:def_K}
  \begin{split}
    \mathcal{K}_{f(s)\to f'W(s)}(t,z)
    \,=\;&
    \frac{\alpha}{2\pi}
    \left[
      f_W\,c_\perp^W\, \tilde{\rm V}_{f(s)\to f'b(s)}^\text{CDST}(t,z)
      + f_h\,c_L^W\, \tfrac{1}{2}\,(1-z)
    \right]\\
    \mathcal{K}_{f(s)\to fZ(s)}(t,z)
    \,=\;&
    \frac{\alpha}{2\pi}
    \left[
      f_Z\,c_\perp^Z\, \tilde{\rm V}_{f(s)\to fb(s)}^\text{CDST}(t,z)
      + f_h\,c_L^Z\, \tfrac{1}{2}\,(1-z)
    \right]\;.
  \end{split}
\end{equation}
Therein, $\tilde{\rm V}_{f(s)\to f^{(\prime)}b(s)}^\text{CDST}$ are the spin--\-averaged 
Catani--\-Dittmaier--\-Seymour--\-Trocsanyi (massive) dipole splitting kernels, 
neglecting their color factors~\cite{Catani:1996vz,Catani:2002hc} and 
$\alpha$ is the QED coupling constant.  $c_\perp$ and $c_L$ are the coupling 
factors of the transverse and longitudinal gauge boson polarizations. 
They are given by 
\begin{equation}\label{eq:def_c}
  \begin{split}
    \begin{array}{ll}
      c_\perp^W\,=\;\seff\,\frac{1}{2s_W^2} \left|V_{ff'}\right|^2 \;,\qquad\qquad &
      c_\perp^Z\,=\;\seff\,\frac{s_W^2}{c_W^2}\,Q_f^2
                    +\left(1-\seff\right)\frac{(I_f^3-s_W^2Q_f)^2}{s_W^2c_W^2}\;, \\
      c_L^W\,=\;\frac{1}{2s_W^2} \left|V_{ff'}\right|^2 
                \left[\seff\,\frac{m_{f'}^2}{m_W^2}+\left(1-\seff\right)\frac{m_f^2}{m_W^2}\right]\;, &
      c_L^Z\,=\;\frac{I_f^3}{s_W^2}\,\frac{m_f^2}{m_W^2}\;,
    \end{array}
  \end{split}
\end{equation}
with $Q_f$, $I_f^3$ and $m_f$ the charge, the three--\-component of the weak 
isospin and the mass of the respective fermion. $s_W^2$ and $c_W^2$ are the 
squared sine and cosine of the Weinberg angle, and $m_W$ is the mass of the 
$W$ boson.  $V_{ff'}$ is the CKM matrix element in case of emissions off quarks 
and the unit matrix in case of emissions off leptons. $\seff$ is the fraction 
of left-handed fermions in the spin--\-averaged fermion line. As the coupling 
to $W^\pm$ and $Z$ bosons differs for left-- and right--\-handed fermions but 
the shower operates in a spin-averaged approximation, this factor $\seff$ is 
essential for a correct description of the splitting probabilities. It is 
important to note that a global definition of $\seff$ is only sensible in a 
limited set of processes: 
i) pure QCD/QED reactions which are left-right symmetric 
($\seff=\tfrac{1}{2}$), and ii) electroweak processes where all 
fermion lines are connected to the same type of electroweak gauge bosons, 
e.g.\ $q\bar q'\to e \nu_e$ ($\seff=1$). 
In EW--\-QCD mixed processes, e.g.\ $q q'\to e \nu_e q q'$, a global $\seff$ 
cannot be defined. This, of course, directly implies that in the present case 
of dijet production, after the first EW gauge boson is radiated the global 
$\seff$ is only correct for all fermion lines not connected to 
the one that radiated the first EW boson. Radiation off this quark line 
will be underestimated as its now polarized state is not accounted for. 
This is a general problem when embedding electroweak splittings into 
spin--\-averaged parton showers, and we do not suggest a solution 
here\footnote{A solution for this problem may follow some or all of the ideas
  presented in~\cite{Nagy:2008eq,Hoeche:2011fd}.}. 
In the present case, however, it is only a minor effect due to the small 
radiation probabilities and does not effect the outcome of this analysis.

As can be seen, the couplings of the longitudinal modes are derived through 
Goldstone boson equivalence, therefore, they only couple to the fermion mass.
In the case of light jet production studied here they can therefore be 
neglected to a very good approximation.  The $W^\pm$ and $Z$ boson masses 
enter via the splitting kinematics, using the recoil scheme the construction of
the splitting kinematics detailed in~\cite{Hoeche:2009xc,Carli:2010cg}.
All $W^\pm$ and $Z$ bosons produced are decayed immediately and all decay 
channels are considered. Although this neglects additional radiative 
branchings of the type $W^\pm\to W^\pm\gamma$, $W^\pm\to W^\pm Z$ or 
$Z\to W^\pm W^\mp$, it ensures that the color singlet 
$q$-$\bar q^{(\prime)}$ pair produced in the hadronic decay modes starts 
its evolution at the correct scale of $m_W$ or $m_Z$, respectively. 
Higgs radiation, can also be neglected to a very good approximation, similar
to the longitudinal boson polarization modes.  

The factors $f_W$, $f_Z$ and $f_h$, which have been added by hand in 
Eq.\ \eqref{eq:def_K}, are used in the analysis to modify the coupling 
strength of the different electroweak bosons to fermions.  Their Standard 
Model values of course are all unity, $f_W=f_Z=f_h=1$.  For the present study 
it has been found that contributions from the radiation of $Z$ or Higgs 
bosons to all observables investigated are negligible. We therefore set $f=f_W$ 
and $f_Z=f_h=0$ in the analysis of Secs.\ \ref{sec:analysis} and 
\ref{sec:results}.

% \subsection{Electroweak Sudakov effects in dijet production}
% 
% To judge the quality of the approximations made in deriving the 
% electroweak splitting functions and the incorporation of these 
% inherently spin-dependent objects into an spin-avaraged parton shower 
% the case of dijet production is considered.
% 
% 
% {\bf MS: Compare numbers against \cite{Dittmaier:2012kx}, both EW shower 
% only and including QCD shower. Maybe also compare to analytical Sudakovs of 
% \cite{Denner:2000jv,Bell:2010gi}.}
% 
% \begin{figure}[t]
%   \includegraphics[width=0.32\textwidth]{figures/dijet/placeholder}\hfill
%   \includegraphics[width=0.32\textwidth]{figures/dijet/placeholder}\hfill
%   \includegraphics[width=0.32\textwidth]{figures/dijet/placeholder}\\
%   \includegraphics[width=0.32\textwidth]{figures/dijet/placeholder}\hfill
%   \includegraphics[width=0.32\textwidth]{figures/dijet/placeholder}\hfill
%   \includegraphics[width=0.32\textwidth]{figures/dijet/placeholder}
%   \caption{
%             Comparison of the suppression of the dijet cross section 
%             through electroweak effects, calculated either through 
%             an exact EW NLO calculation \cite{Dittmaier:2012kx} or 
%             the Sudakov factors of the parton shower approximation.
%             \label{fig:ewps-vs-nlo}
%           }
% \end{figure}

\section{W reconstruction in dijet events} \label{sec:analysis}

To select events of interest, we group all visible final state particles 
with $p_{T}>0.5$ GeV and $\left|\eta\right|<5.0$ into 
cells of size $\Delta \eta \times \Delta \phi = 0.1 \times 0.1$ to 
account for the granularity of the detector. We identify an isolated electron 
or muon with $p_T>25$ GeV and $|\eta_l | <2.5$ if the 
hadronic energy deposit within a cone of radius $R=0.2$ is less than 
$10 \%$ of the lepton candidate's transverse energy. 
After removing the isolated leptons from the calorimeter cells we use 
the remaining visible final state of the above selection to construct jets.
For triggering on jets we use the anti-$k_\mr{T}$ 
algorithm~\cite{Cacciari:2008gp}.  
We then apply a pre-selection trigger requiring one fat jet with $R=1.5$ and 
$p_{T_J}>200$ GeV.  At this point we consider two cases: if there are no 
isolated leptons we perform a hadronic $W$ reconstruction and if there is 
exactly one lepton we perform a 
leptonic $W$ reconstruction detailed below. In both analyses additional 
conditions on the fat jets are applied, requiring at least two such jets 
with $p_{T_J}>$ 500, 750 or 1000 GeV. These additional cuts will force the 
radiated $W$\!s successively further into the collinear region where 
the approximations of Sec.\ \ref{sec:ewshower} are valid.
Finally, we infer the missing transverse momentum vector from the sum of 
all transverse momenta of visible final state particles with 
$\left|\eta\right|<5.0$. We cluster jets using \Fastjet 
\cite{Cacciari:2011ma} and analyze the hadronic final state using 
\Rivet~\cite{Buckley:2010ar}. 

\subsection{Hadronic Analysis} \label{sec:hadana}

To reconstruct a hadronically decaying $W$ in dijet events, we study methods 
which use either the mass of the $W$ boson or the distribution of the 
radiation emitted off the $W$ decay products, so-called jet shape observables. 
In general we find that subjet-based reconstruction methods perform better 
than jet shape observables and are more efficient in separating the $W$ decay 
products from other hadronic activity in the event. However, applying a jet 
shape observable in combination with a subjet-based mass reconstruction 
technique gives the best performance.

We aim to reconstruct the $W$ boson in the three different kinematic 
regimes $p_{T_W} \gg m_W$, $p_{T_W} > m_W$  and $p_{T_W} \simeq m_W$. 
The first two methods employ jet sub-structure techniques aimed at 
boosted objects, and the last is an event-wide $W$ search:
\begin{enumerate}[(A)]
\item \label{mhard}
      To select a boosted $W$ boson, we recluster the fat jet constituents 
      with a Cambridge/Aachen jet algorithm (C/A)~\cite{Dokshitzer:1997in} 
      with $R=0.5$ and require $p_{T\iota} > 200$ GeV for every subjet 
      $\iota$. Then we apply the BDRS algorithm~\cite{Butterworth:2008iy} 
      to the second hardest of these subjets.  We accept the $W$ candidate if 
      the filtered mass $m_\text{BDRS}\in\left[74,90\right]$ GeV.  
\item \label{mmed}
      Just like in method~\ref{mhard} we recluster the fat jet constituents 
      into C/A subjets but now using $R=0.3$ and $p_{T\iota}>20$ GeV. We call 
      this set of subjets microjets.  Since the quark generated in the hard 
      process carries the bulk of the energy, we discard the hardest microjet. 
      In case a $W$ boson is emitted off a quark, the $W$ decay products 
      are likely to have larger transverse momentum than gluon 
      emissions~\cite{Christiansen:2014kba}. Therefore, if the 
      decay is symmetric and boosted enough, most of the time the 
      second and third microjets are due to the $W$ decay products. 
      We then choose the combined four-momentum of the second and 
      third microjets as a $W$ candidate and their invariant mass 
      $m_{23}$ as a discriminating variable. We find that the mass 
      window which leaves the best signal to background ratio, 
      while taking into account detector resolution effects, is 
      $m_{23}\in\left[70,86\right]$ GeV. If $m_{23}$ is within this 
      mass window, we accept the $W$ candidate.
\item \label{msoft}
      Since the $W$ boson is likely to be emitted at a relatively 
      large angle with respect to the initial high-$p_T$ quark, we 
      recluster the full event into small anti-$k_T$ jets of radius 
      $R=0.4$ with $p_{T_j} > 40$ GeV. We require at least five jets 
      to accept the event. The hardest two jets in the event originate 
      from partons produced in the hard interaction, so we discard 
      them. We use the remaining jets in the event to form pairs, 
      which combine to an invariant mass $m^2_{kl}=(p_{j_k} + p_{j_l})^2$. 
      We apply additional restrictions on the jet pairing to avoid 
      biasing pure QCD events. At \LHC energies even QCD radiation 
      can occur at high virtuality, producing enough mass to match 
      the $W$. If we use all jets within the event, the probability 
      to find a pair with invariant mass within $10$ GeV around 
      $m_{W}$ in a generic event and miss-tag a $W$ boson is 
      non-negligible. To avoid this bias, only jets 
      $j_k$, $k \in [3, 6]$, participate in the pairing algorithm. 
      Additionally, we do not include $m_{34}$ because $j_3$ or 
      $j_4$ is likely to be induced by QCD radiation. This leaves 
      $m_{3l}$ and $m_{4l}$ as $W$ mass candidates, where 
      $l \in [5,6]$ depending on the event multiplicity. We tag a 
      $W$ boson if the candidate pair mass 
      $m_{kl}\in\left[70,86\right]$ GeV cut. If several combinations 
      of jet pairings are within this window, we choose the pair with 
      smallest $\Delta m =  \left|m_{kl}-m_{W}\right|_\mr{min}$ and 
      label the pair mass $m_\mr{min}$. 
\end{enumerate}

\begin{figure}[t]
  \includegraphics[width=.3\linewidth]{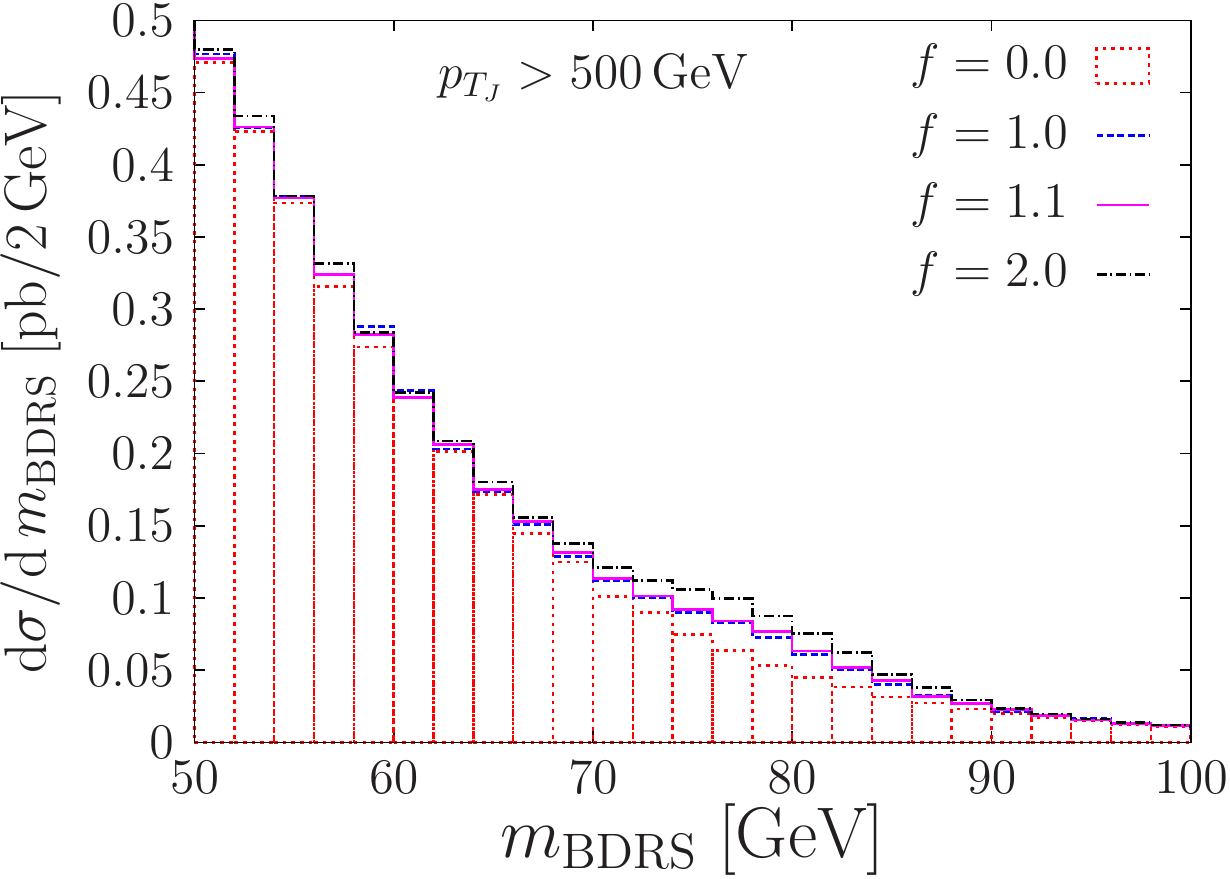}\hspace*{10pt}
  \includegraphics[width=.3\linewidth]{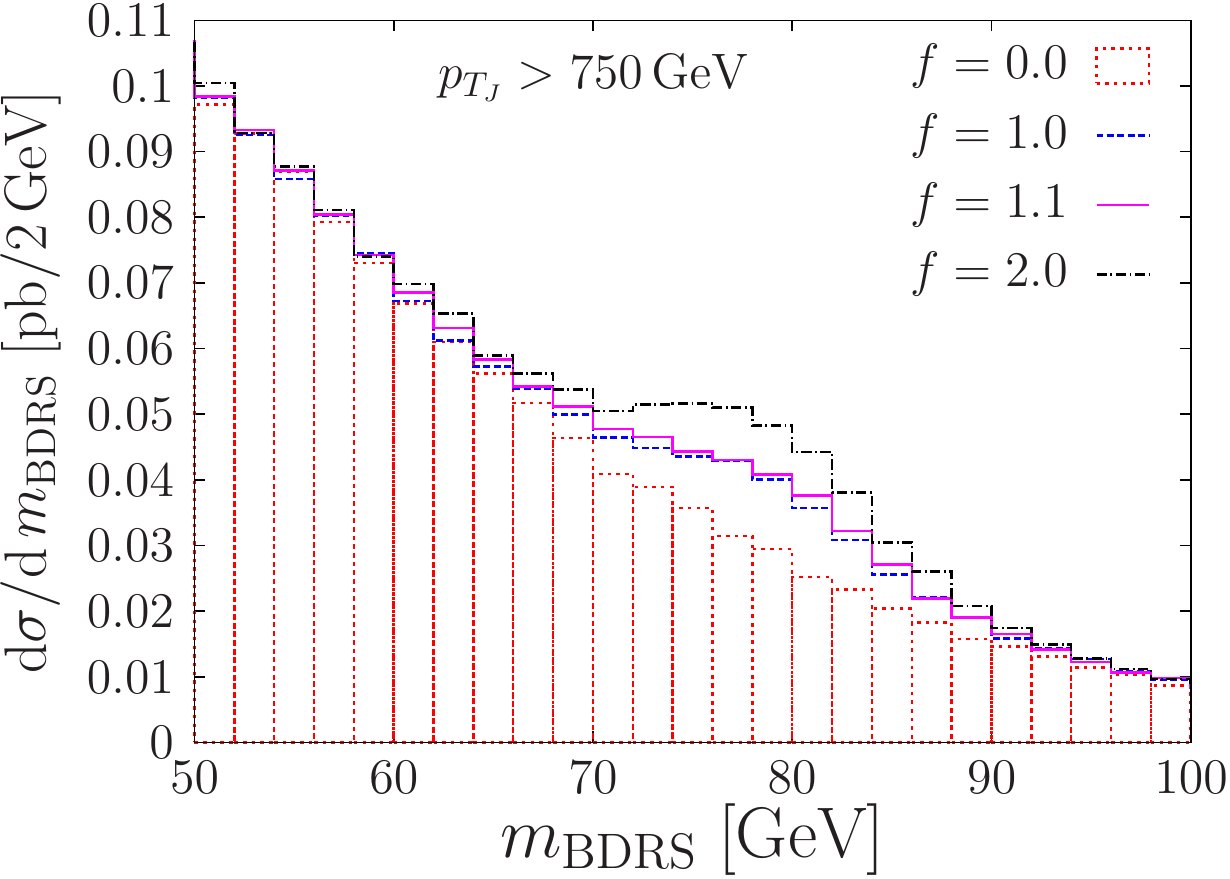}\hspace*{10pt}
  \includegraphics[width=.3\linewidth]{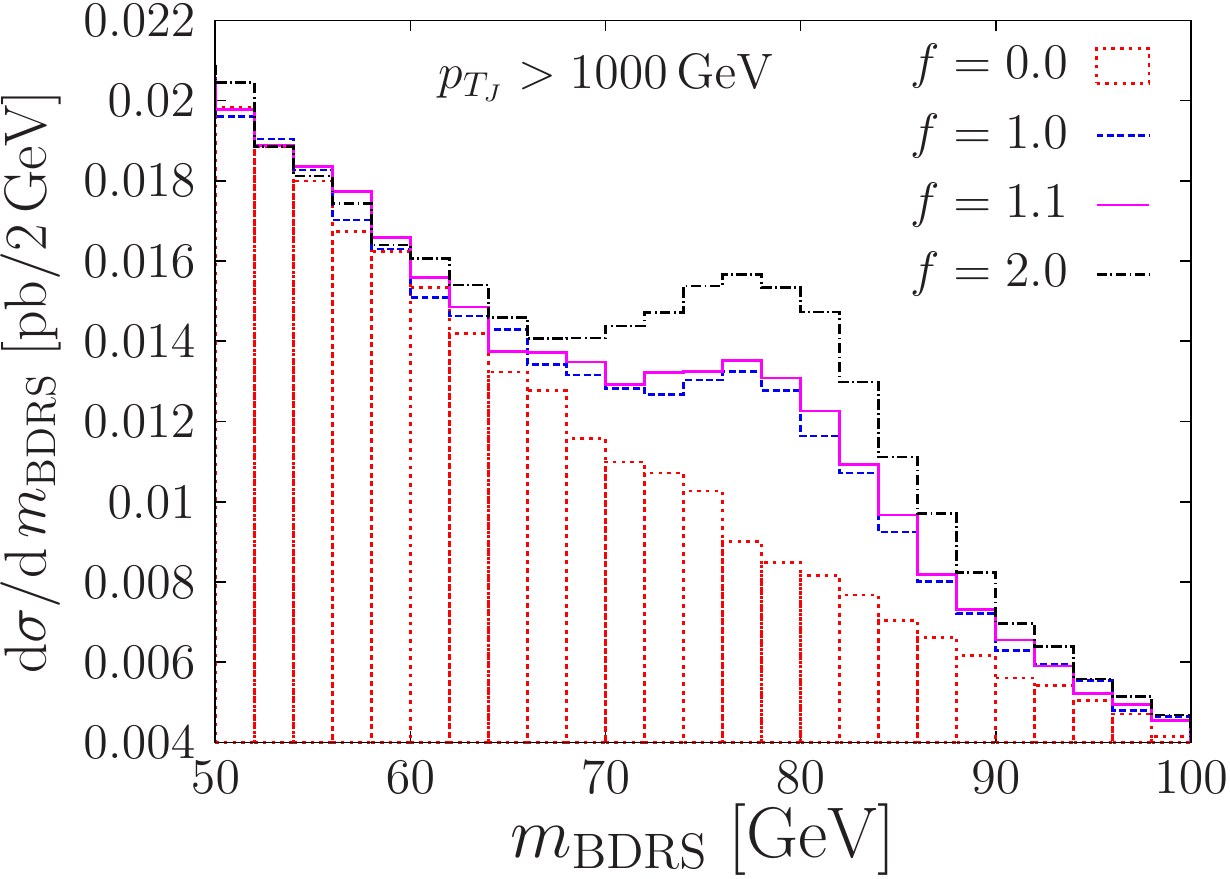}
  \caption{
	    $W$ candidate mass distribution using method~\ref{mhard} 
	    for $p_{T_J} > 500$ (left), $750$ (center) and 
	    $1000$ (right) GeV.
	    \label{fig:mbdrsR5}
	  }
\end{figure}

%%%% method hard
Method~\ref{mhard} is most sensitive if the $W$ boson is highly boosted. 
Because the angular separation of the double decay products is 
$\Delta R_{ab} \simeq 2m_W / p_{T_W}$ a subjet radius of $R=0.5$ implies 
$p_{T_W} \gtrsim 300$ GeV. In Fig.~\ref{fig:mbdrsR5} we show the results 
of method~\ref{mhard} for three different fat jet $p_T$ selections. For 
the two free parameters ($\mu$,~$y_\mathrm{cut}$) of the BDRS method we 
follow the suggestion of \cite{Butterworth:2008iy} using ($0.54$,~$0.13$) 
for $200 < p_{T\iota} < 500$ and ($0.72$,~$0.09$) for $500 < p_{T\iota} $. 
The plots show an excess around $m_\mr{BDRS}=80$ GeV. Higher multiplicative 
factors $f$ increase the EW radiation rate, resulting in larger excesses. 
Moreover, the $W$ reconstruction is more successful as the 
fat jets have larger transverse momenta. This is not surprising because 
more energetic quarks emit boosted $W$ bosons more frequently. This allows 
the BDRS method to tag them more efficiently.  

\begin{figure}[t]
  \includegraphics[width=.3\linewidth]{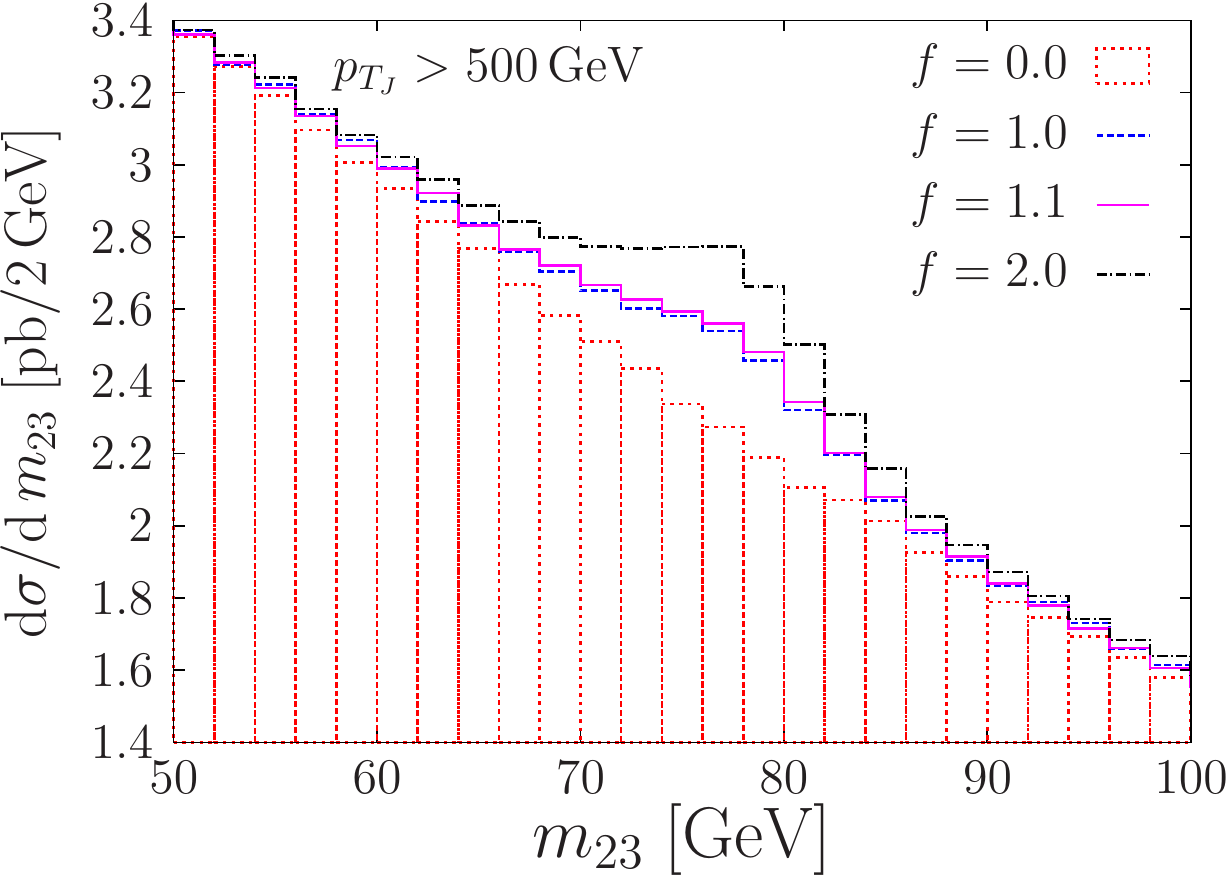}\hspace*{10pt}
  \includegraphics[width=.3\linewidth]{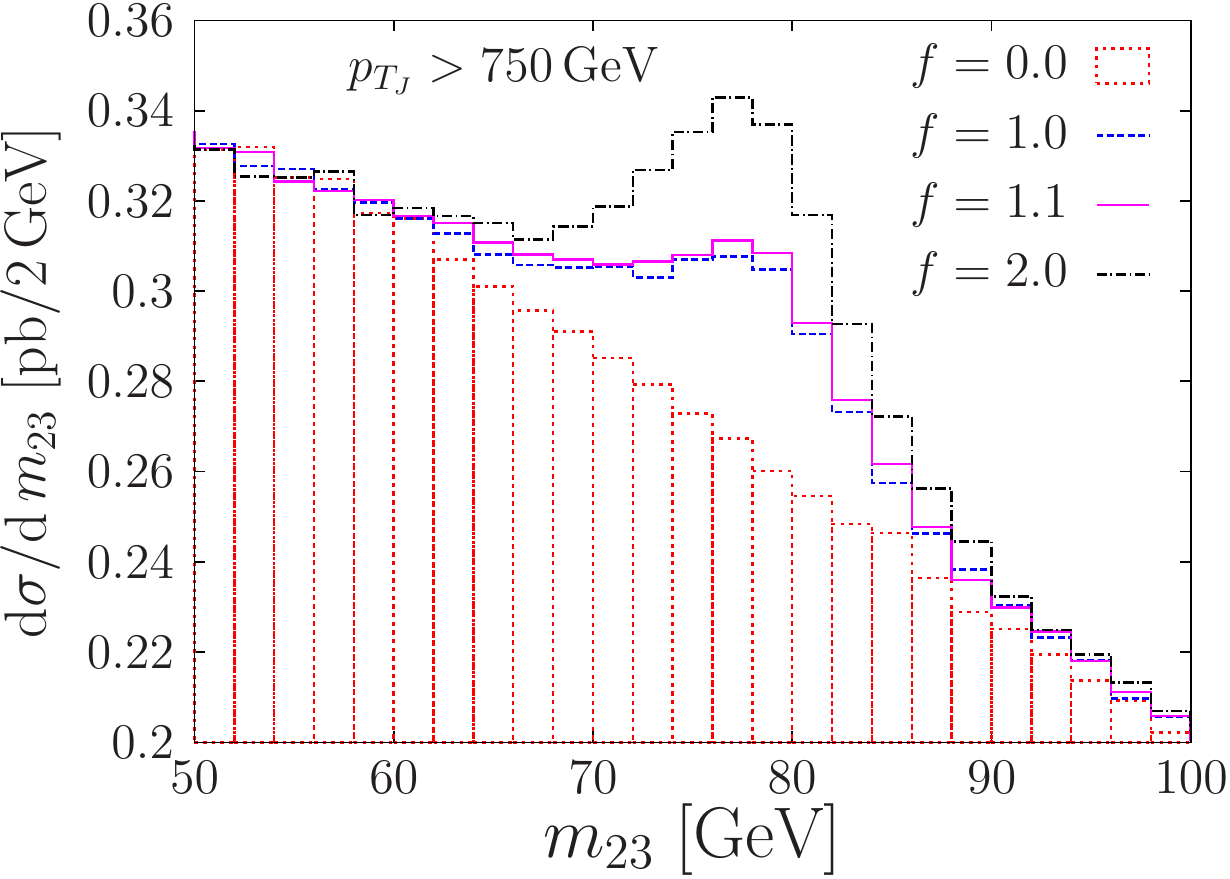}\hspace*{10pt}
  \includegraphics[width=.3\linewidth]{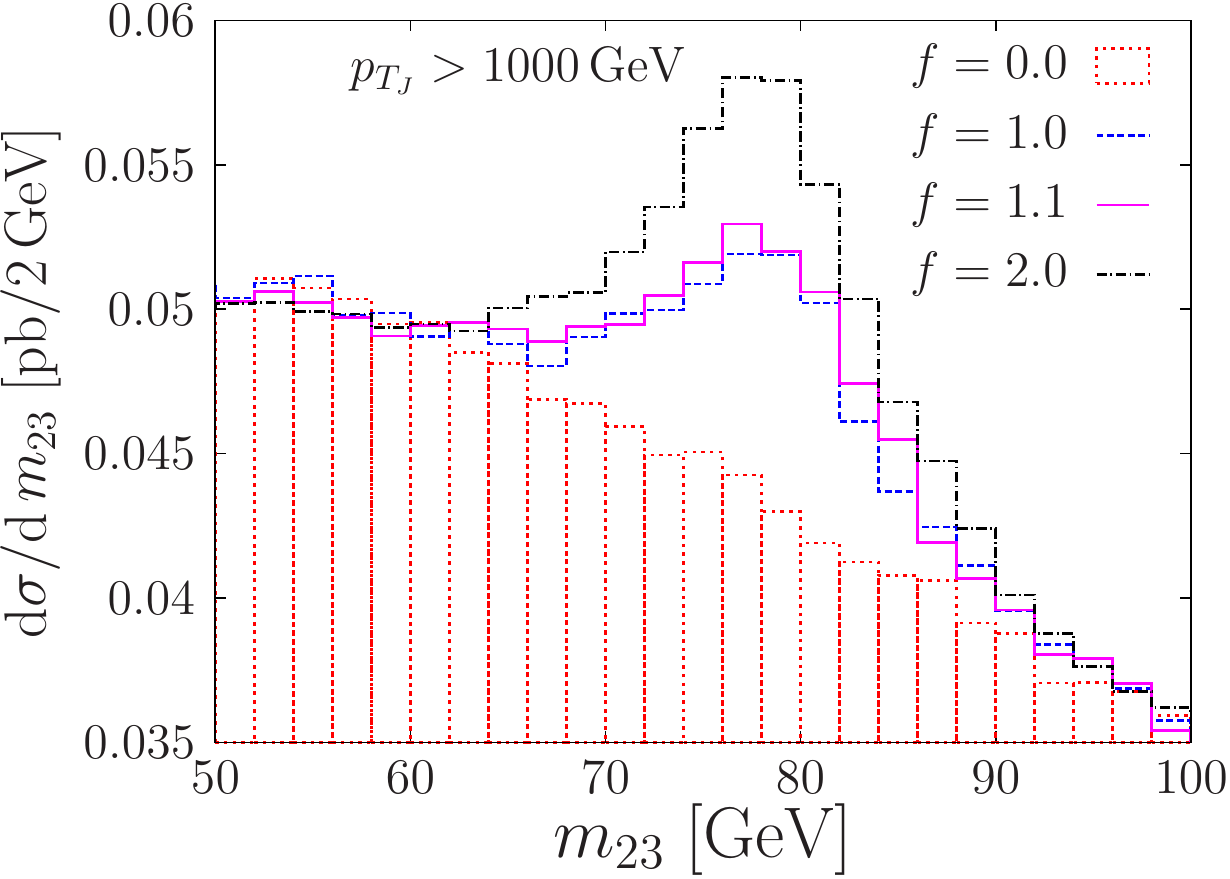}
  \caption{
	    $W$ candidate mass distribution based on microjets 
	    $\iota_{2}$ and $\iota_{3}$ as described in method 
	    \ref{mmed} for $p_{T_J} > 500$ (left), $750$ (center) 
	    and $1000$ (right) GeV.
	    \label{fig:msj23}
	  }
\end{figure}

%%%% method med
Fig.~\ref{fig:msj23} shows the $W$ candidate mass distribution for 
different emission rate hypotheses using method~\ref{mmed}. We observe 
the same tendency here as with method~\ref{mhard}. Higher quark energy 
results in more frequent emissions of $W$ bosons and their decay 
products form microjets two and three within the fat jet. Note the 
different scales and lowest points on the $y$-axes of 
Fig.~\ref{fig:mbdrsR5} and \ref{fig:msj23}. Although the $W$ mass peak 
seem more pronounced in Fig.~\ref{fig:msj23}, closer inspection reveals 
that $S/B$ is similar as for method 1.

\begin{figure}[t]
  \includegraphics[width=.3\linewidth]{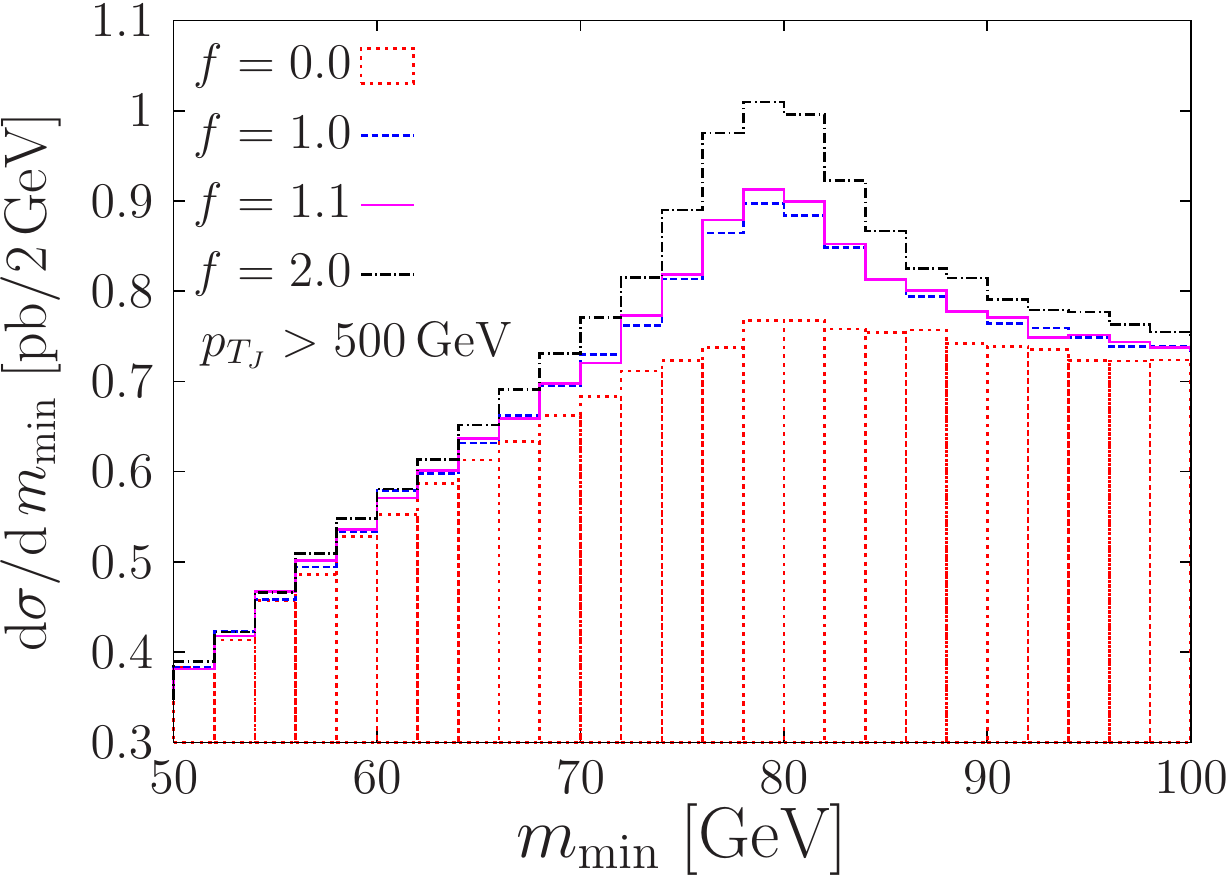}\hspace*{10pt}
  \includegraphics[width=.3\linewidth]{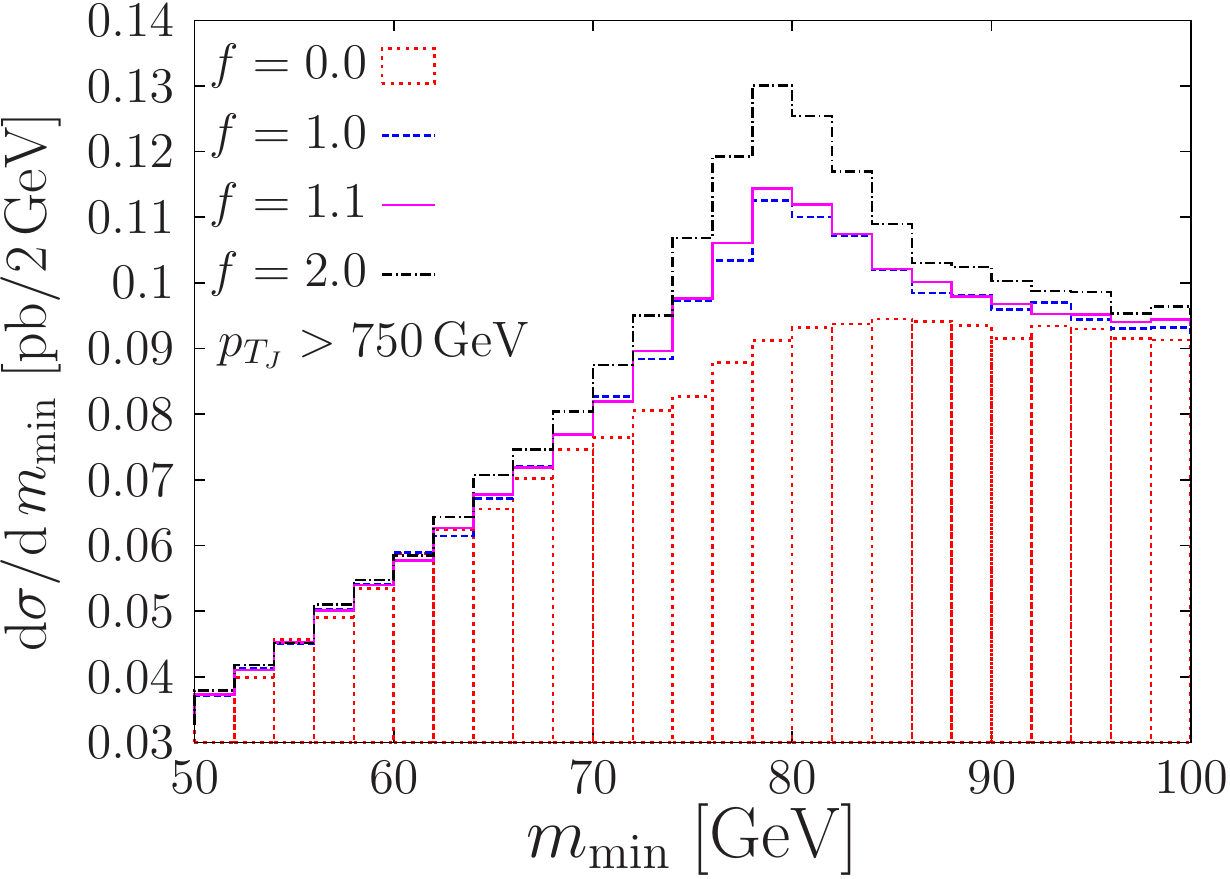}\hspace*{10pt}
  \includegraphics[width=.3\linewidth]{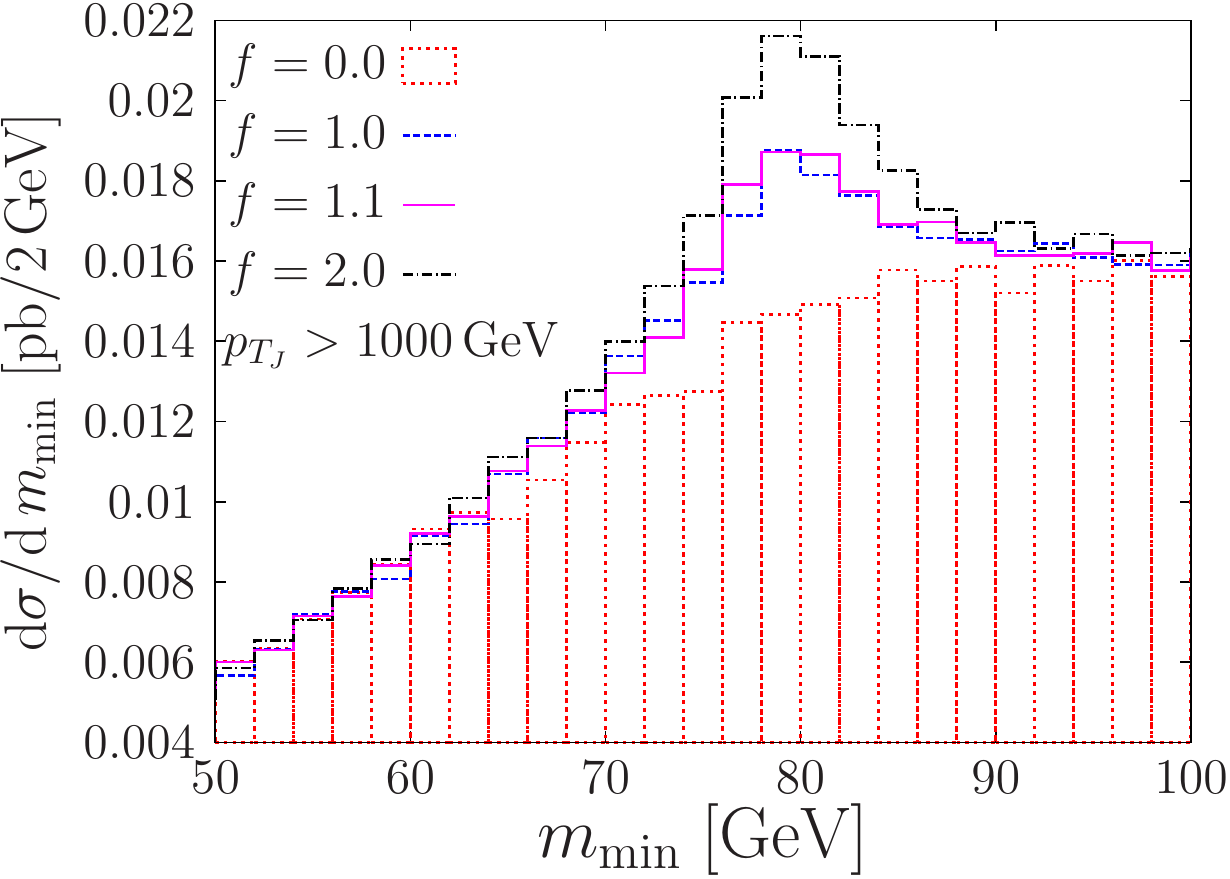}
  \caption{
	    $W$ candidate mass distribution based on method~\ref{msoft} 
	    for $p_{T_J} > 500$ (left), $750$ (center) and $1000$ (right) 
	    GeV.
	    \label{fig:mWmin}
	  }
\end{figure}

%%%%%%%%%% method low
Fig.~\ref{fig:mWmin} shows the mass distribution resulting from 
method~\ref{msoft}. In contrast to methods 1 and 2 the signal to 
background ratio within the excess region does not change drastically 
with an increasing fat jet $p_T$ selection cut.

We find that combining methods~\ref{mhard} and \ref{mmed} with jet shape 
observables, i.e. $n$-subjettiness $\tau_n$ \cite{Thaler:2010tr} and 
ellipticity $\hat{t}$ (Appendix~\ref{App:A}), can improve on the $W$ 
boson identification.

\begin{figure}[t!]
  \centering
  \includegraphics[width=.3\linewidth]{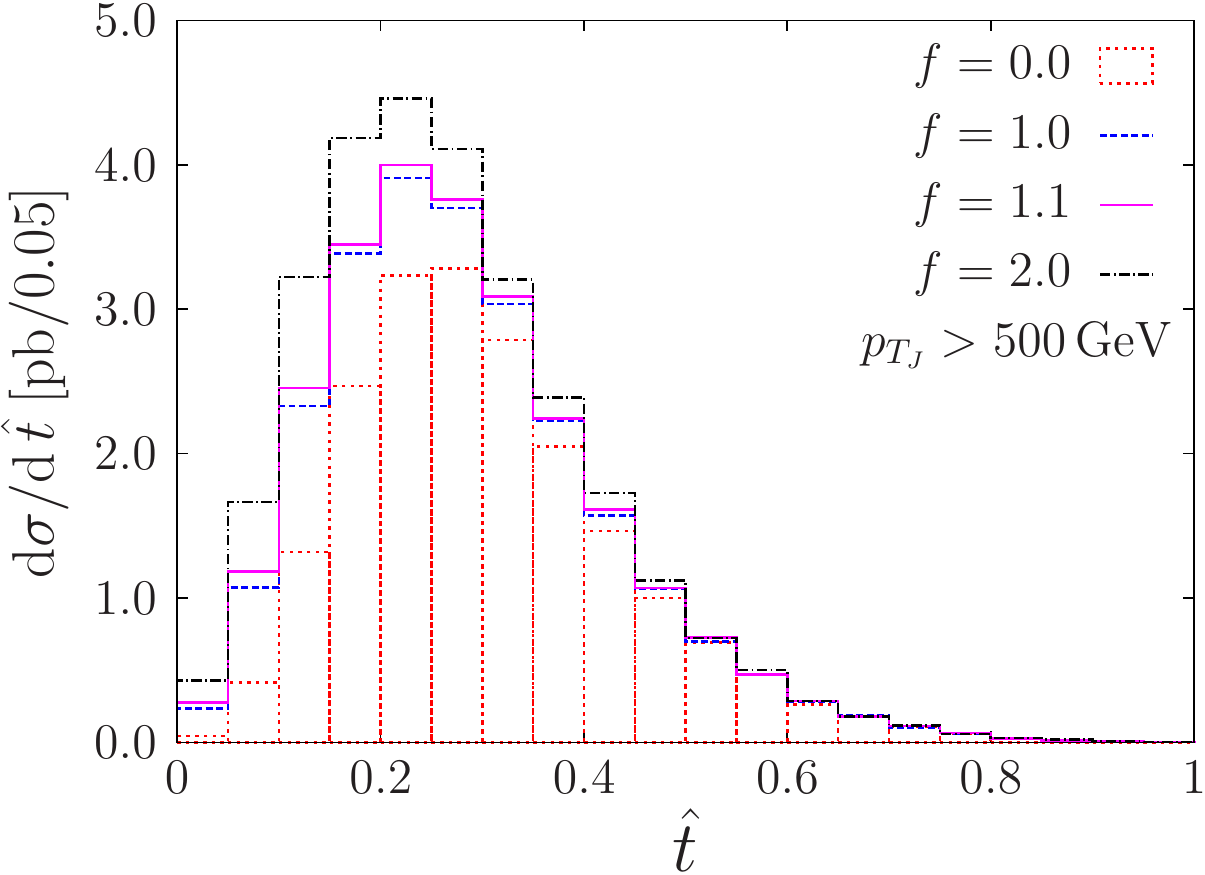}\hspace*{10pt}
  \includegraphics[width=.3\linewidth]{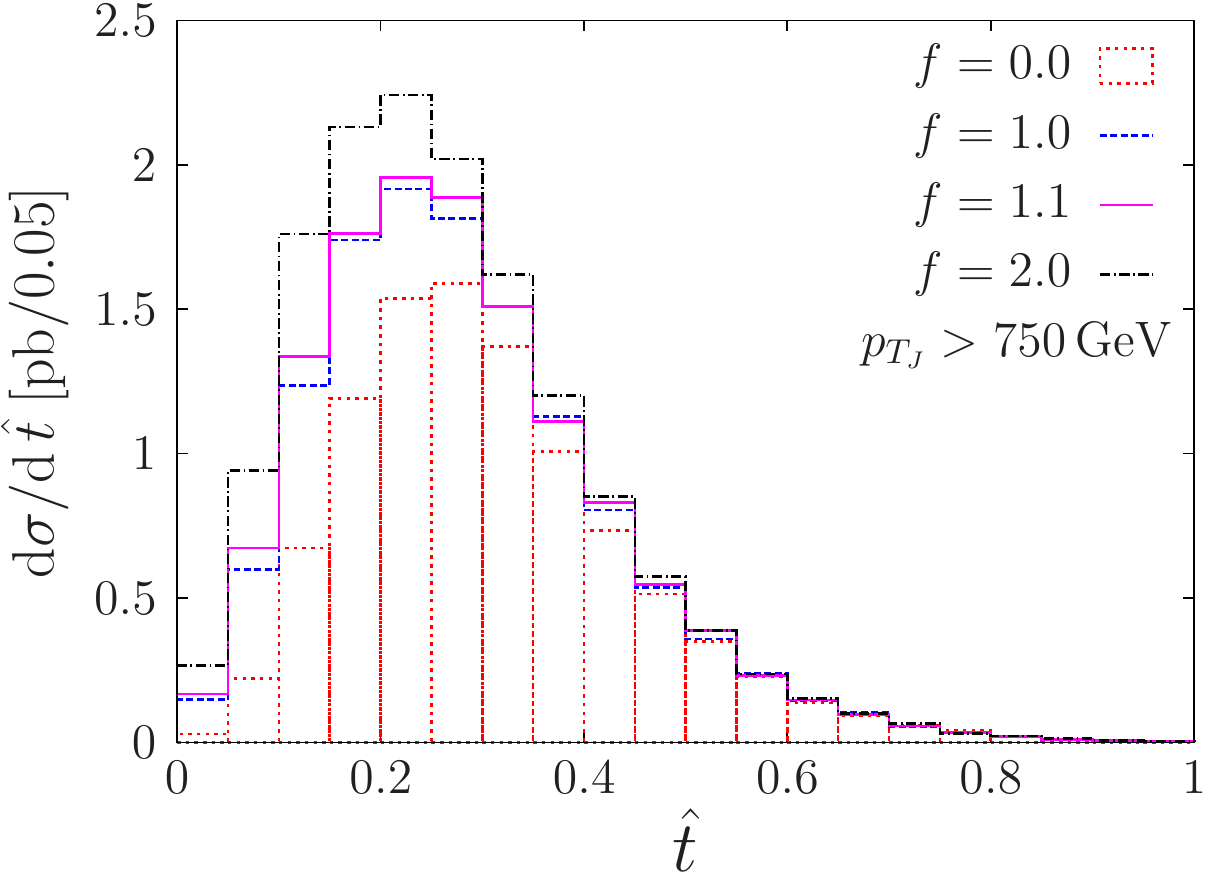}\hspace*{10pt}
  \includegraphics[width=.3\linewidth]{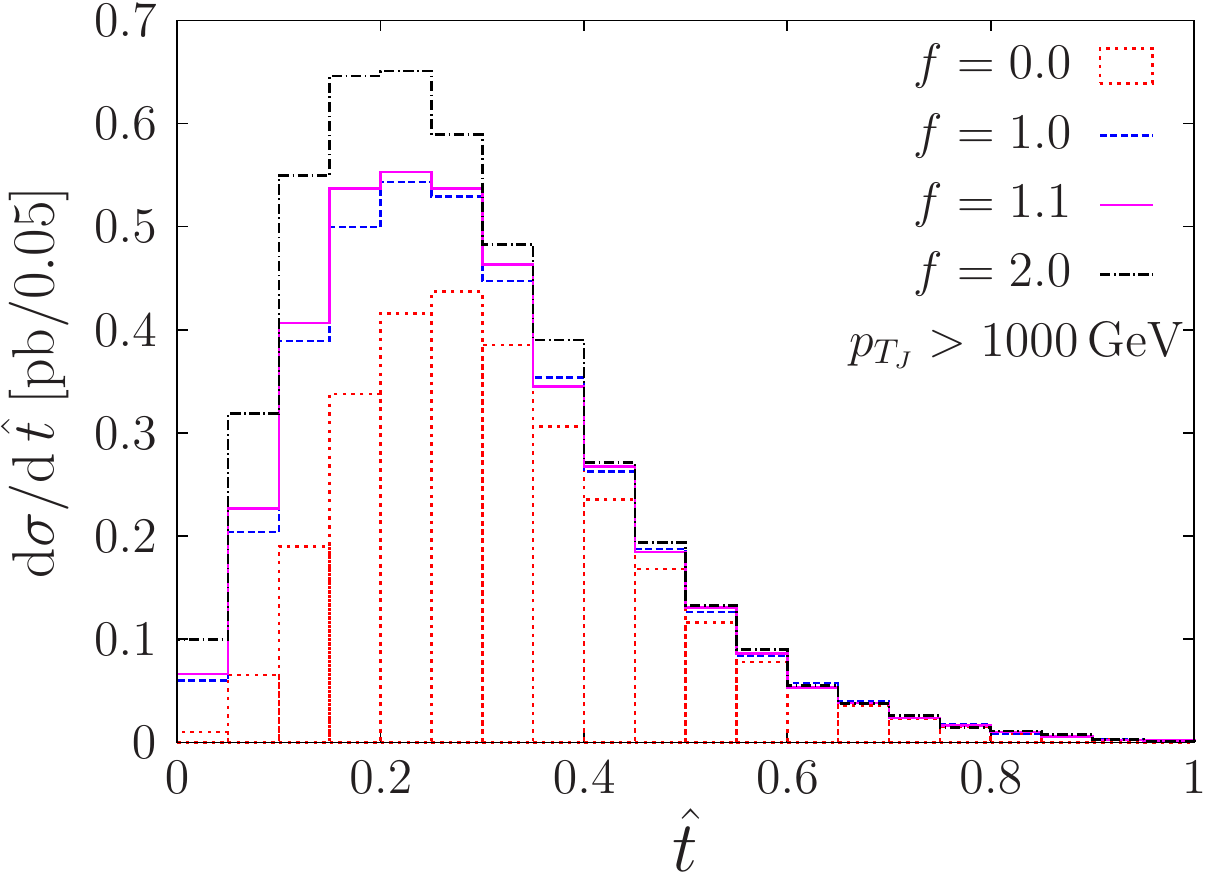}\\[2ex]
  \includegraphics[width=.3\linewidth]{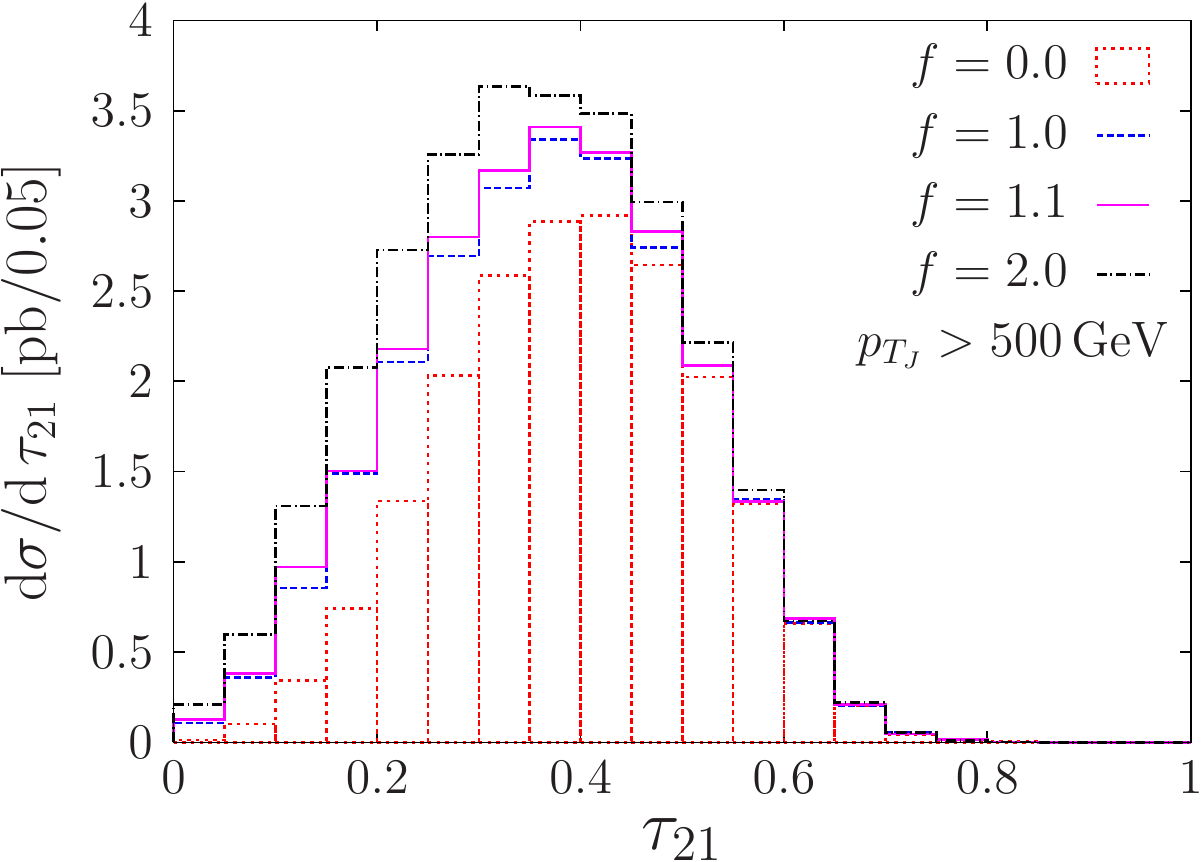}\hspace*{10pt}
  \includegraphics[width=.3\linewidth]{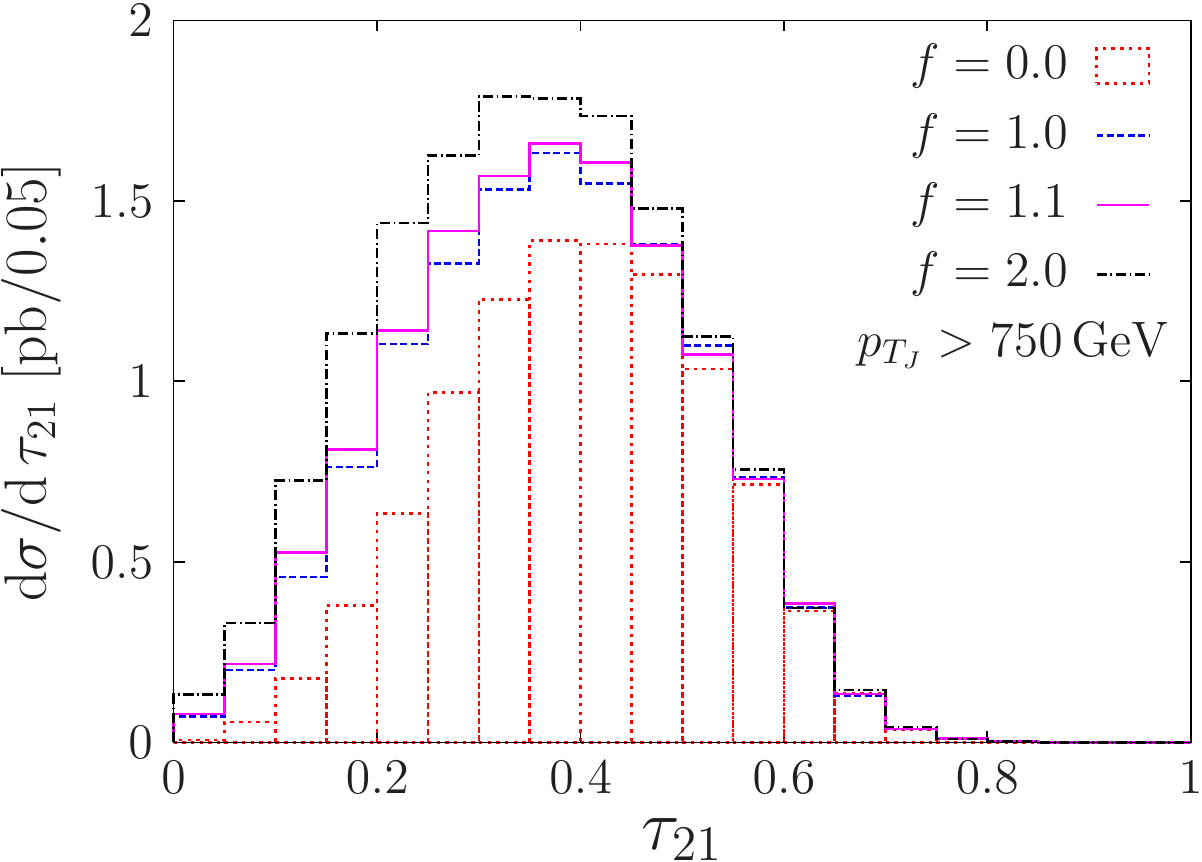}\hspace*{10pt}
  \includegraphics[width=.3\linewidth]{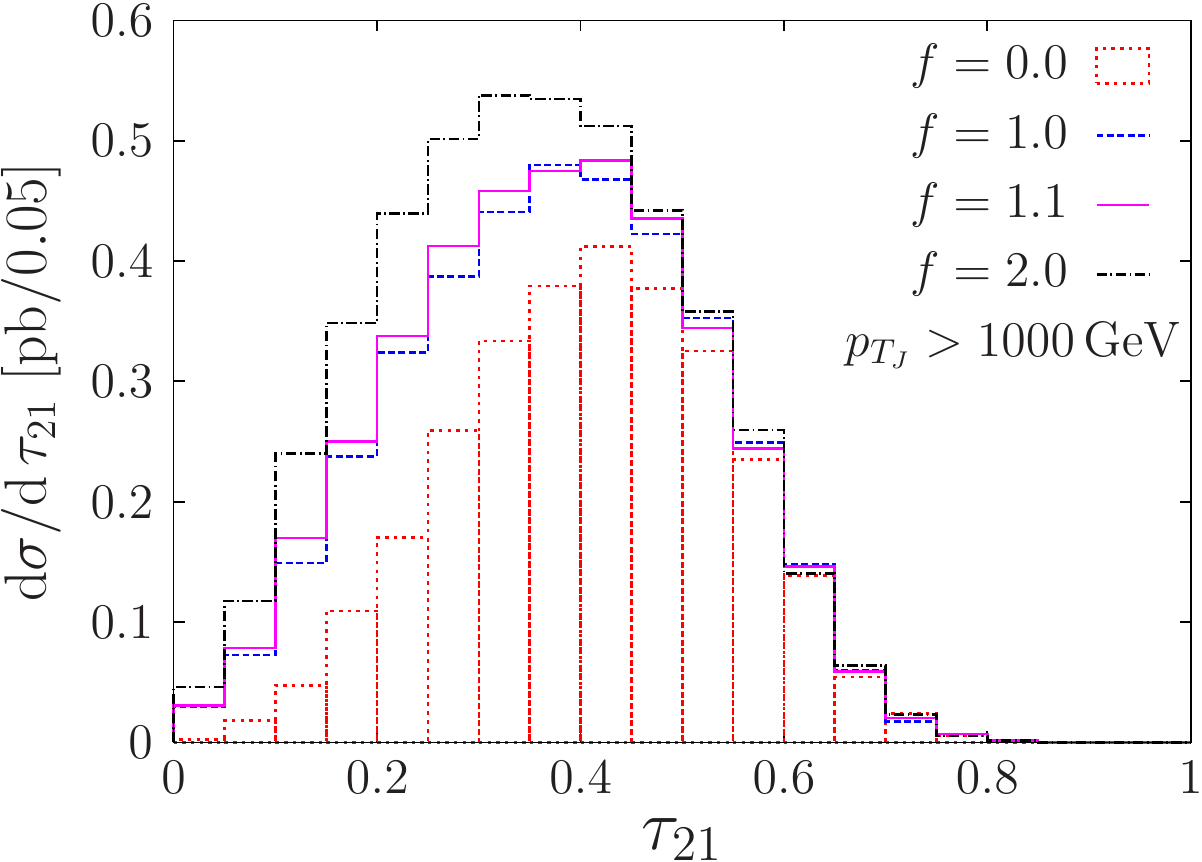}
  \caption{
	    Ellipticity $\hat{t}$ (top row) and $\tau_{21}$ (bottom row) 
	    distributions calculated using constituents of $W$ 
	    candidates identified with method~\ref{mhard} for 
	    $p_{T_J} > 500$ (left), $750$ (center) and $1000$ (right) GeV.
	    \label{fig:Shapes}
	  }
\end{figure}

We measure these observables using the constituents of the successfully 
reconstructed $W$ with methods \ref{mhard} and \ref{mmed}. Ellipticity 
and $\tau_{21} = \tau_2 / \tau_1$ achieve the best results when 
applied on the second hardest boosted subjet of radius $R=0.5$ and mass 
$m_\mr{BDRS} \in \left[74,90\right]$ GeV. In Fig.~\ref{fig:Shapes} we 
show the two distributions $\hat{t}$ (top row) and $\tau_{21}$ (bottom row).

We find that the total cross section substantially increases with $f$. 
This reflects the fact that we only use subjets that pass selection 
method~\ref{mhard}. Just as importantly, the shape of the distributions 
also changes as $f$ is varied. The peak region of the distribution of 
both jet shapes shifts to smaller values. 

We construct ellipticity in such a way that, if the bulk of the jet 
radiation in the transverse plane is along a single line, the value of 
the jet shape observable is small. In contrast, a more circular 
distribution of radiation results in a large $\hat{t}$. A symmetric 
two-body decay of a color singlet resonance, such as 
$W\rightarrow q\overline{q}'$, gives rise to two clusters of comparable 
energies and consecutive QCD emissions in the region between them. This 
energy profile is one-dimensional, therefore the hadronic $W$ final 
state particles will have a small ellipticity. On the other hand, a 
gluon (the main source of background) has color connections to other 
particles and is less likely to form a one-dimensional radiation pattern 
in the transverse plane. Therefore, the signal and background 
ellipticity distributions are shifted with respect to each other. 

The reason behind the shift in $\tau_{21}$ is of similar nature. By 
definition $\tau_{n+1}\leq \tau_n$ for any distribution of particles. 
However, if the radiation forms two well separated clusters 
$\tau_2 \ll \tau_1$. If a jet does not have two pronounced clusters, 
$\tau_2\lesssim \tau_1$. Thus $\tau_{21}$ tends to be smaller for a 
$W$ than for a QCD jet.

\subsection{Leptonic Analysis} \label{sec:lepana}

We assume at this stage that the event has already passed the 
tagging criteria of a single isolated lepton with transverse momentum 
$p_{T_l} > 25$ GeV and $|\eta_l| < 2.5$. A leptonically decaying $W$ gives 
rise to a substantial amount of missing transverse energy. We 
therefore require $\slashed{E}_{T} > 50$ GeV.

\begin{figure}[t]
  \includegraphics[width=.3\linewidth]{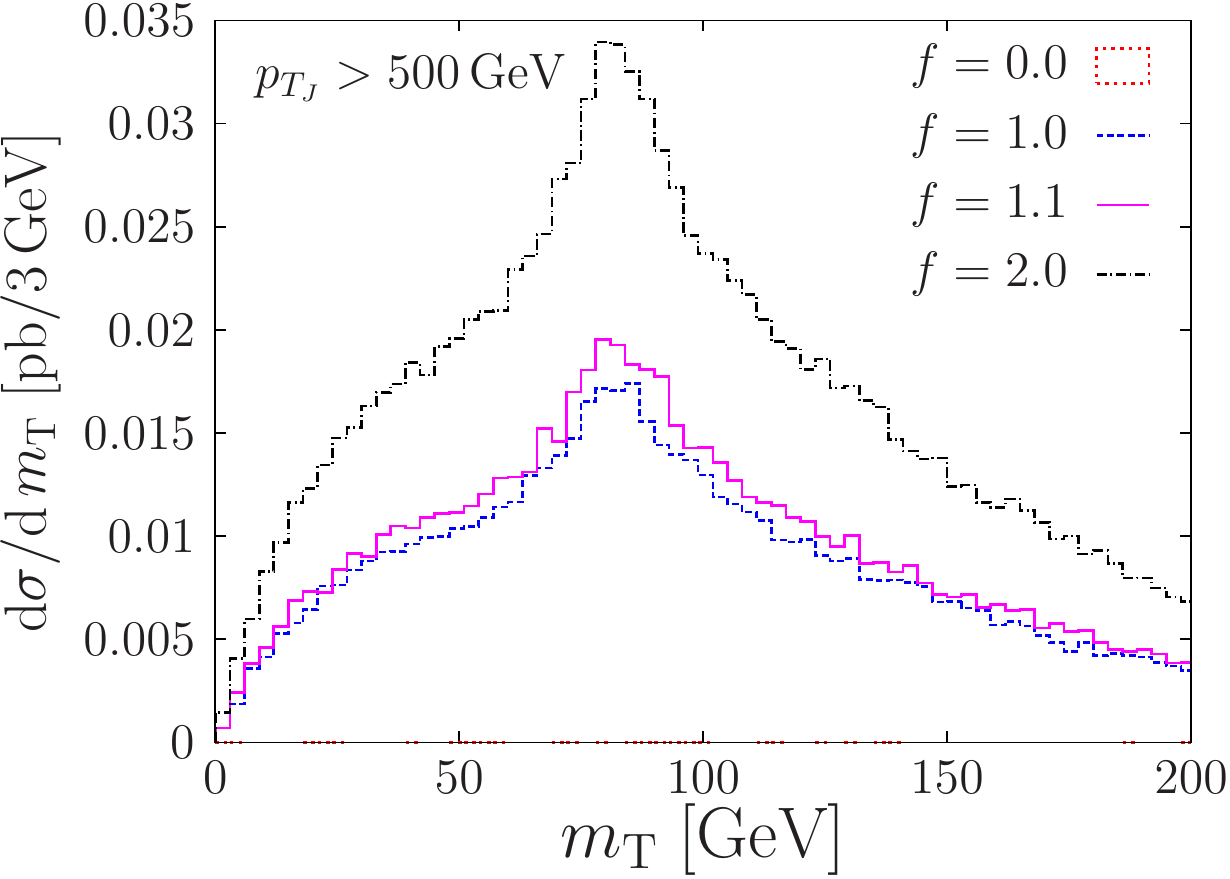}
  \includegraphics[width=.3\linewidth]{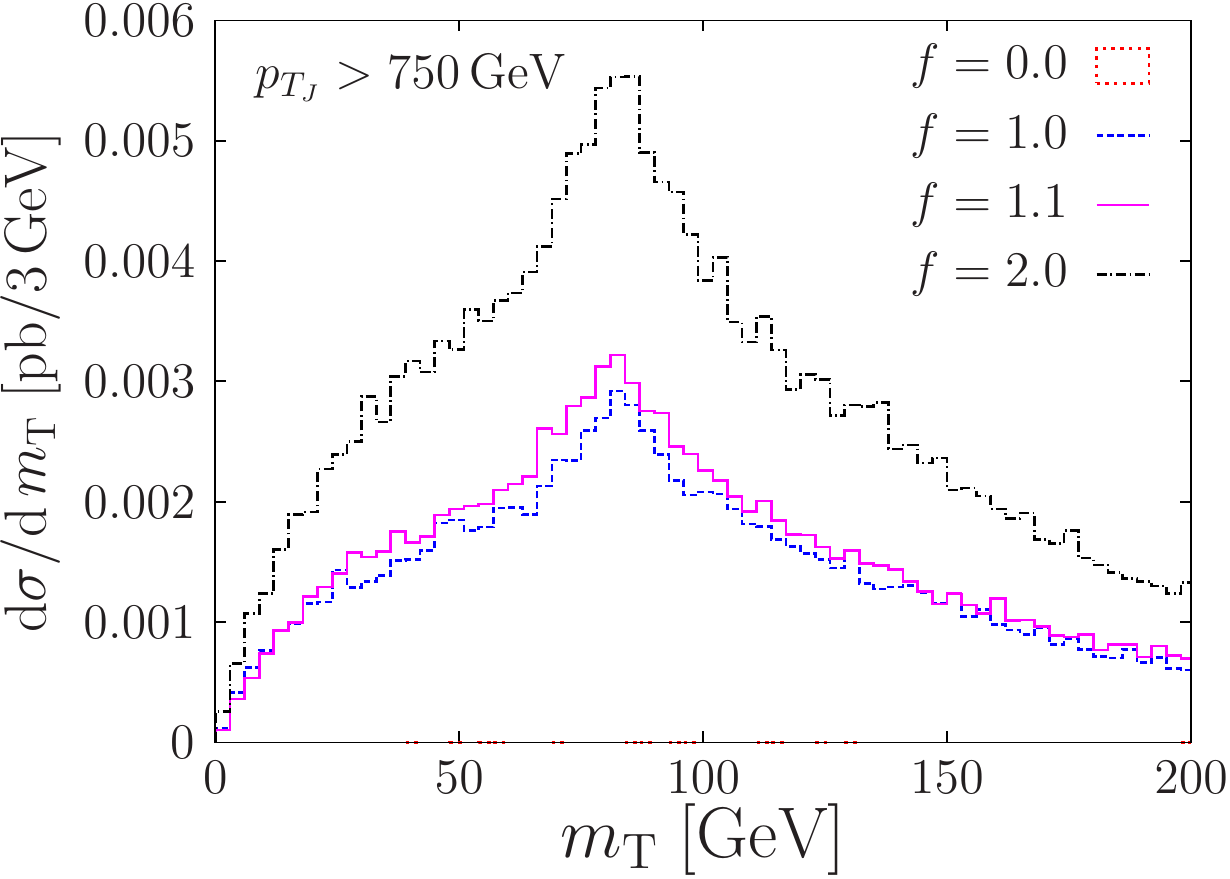}
  \includegraphics[width=.3\linewidth]{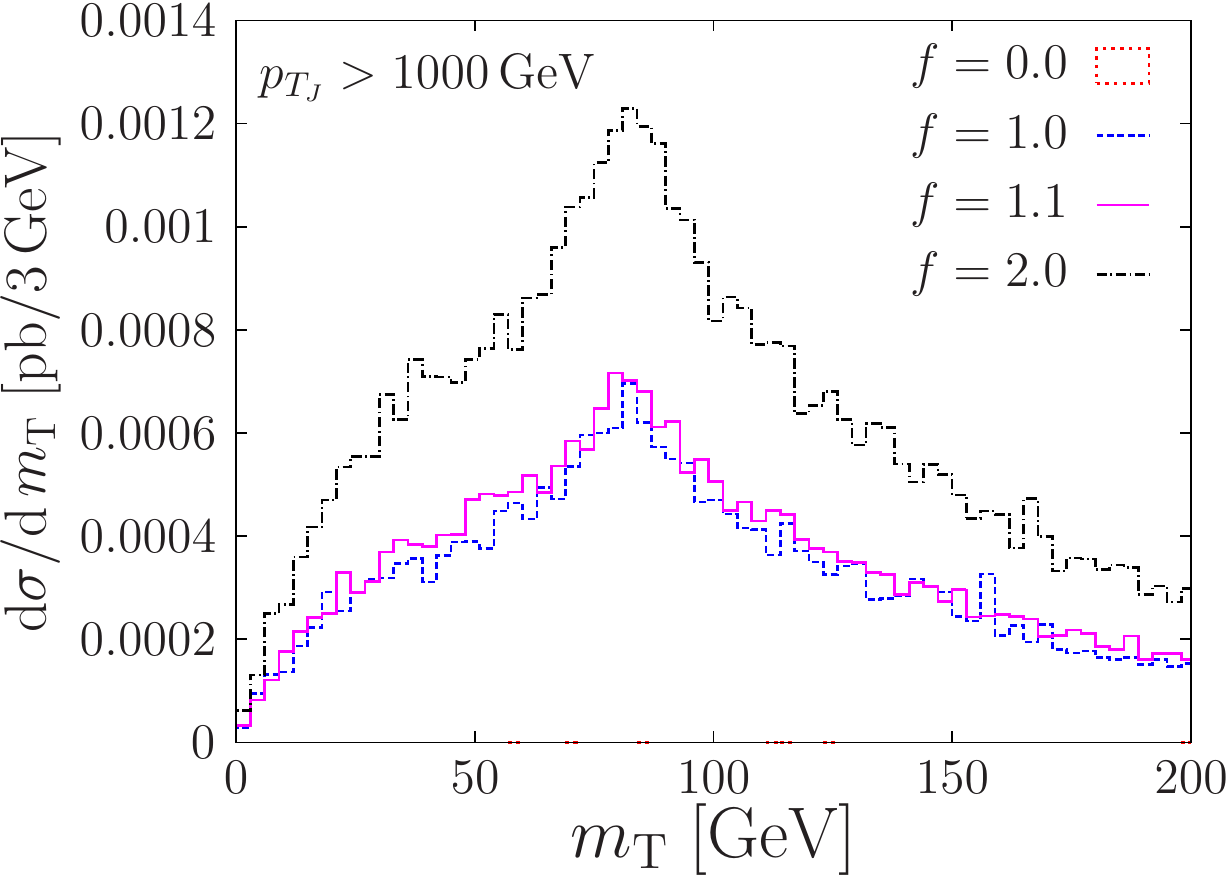}
  \caption{
	    Transverse mass of the leptonic $W$ candidate $m_\mr{T}$ 
	    for $p_{T_J} > 500$ (left), $750$ (center) and $1000$ (right).
	    \label{fig:mTW}
	  }
\end{figure}

To reconstruct the leptonic $W$ we define its transverse mass as 
\begin{equation}\label{eq:mT}
  m_\mr{T} = \sqrt{2E_{T_l} \slashed{E}_{T}\left(1-\cos\theta \right)} ,
\end{equation}
where $\theta$ is the angle between the missing energy vector and the 
isolated lepton. Fig \ref{fig:mTW} shows a pronounced peak for $m_\mr{T}$ 
as defined in Eq.~\ref{eq:mT} in the mass window $\left[60,100\right]$ 
GeV. The $W$ candidate is accepted if the transverse mass of the 
missing momentum and lepton momentum system falls in the aforementioned 
bin. See Table \ref{tab:xseclep} for the cross section of accepted 
events. We see there is virtually no tagged events when $f=0$. Therefore, 
this method provides a perfect QCD rejection. The sharpness of the 
leptonic $W$ peak slightly broadens as the fat jet selection cut becomes 
more restrictive. The lepton isolation criteria might restrict lepton 
tagging as the $W$ is more likely to be emitted at a small angle from 
the parent quark, thereby getting more radiation within the $R=0.2$ 
isolation radius, described in the beginning of this section. A more 
flexible mini-isolation criterion~\cite{Rehermann:2010vq} might help 
recover sensitivity in the highly boosted regime. 

\section{Measuring \texorpdfstring{\protect{$W$}}{W} boson emission rates} \label{sec:results}

We display the cross sections for a $W$ boson emitted in light dijet production at various stages of the analysis in Tabs.\ \ref{tab:xsecpT}, 
\ref{tab:xsechad}, and \ref{tab:xseclep}. The 
first column always refers to the modification factor $f=f_W$ of the 
Sudakov factor as defined in Eq.~\eqref{eq:def_K}. The first table, Tab.\ 
\ref{tab:xsecpT}, shows the cross sections after the different event 
selection criteria for the hadronic and the leptonic analyses. The first 
column for each analysis states the cross section after requiring exactly 
zero or one isolated lepton in the event, labeled $n_l=0$ and $n_l=1$, 
respectively. At this stage only the presence of one fat jet with 
$p_{T_J}>$ 200 GeV is required. The three remaining columns then detail the effects 
of additionally applying minimum fat jet transverse momentum requirements of 
$p_{T_J}>$ 500, 750 and 1000 GeV, as described in Sec.~\ref{sec:analysis}.
For $f=0$, assuming pure QCD evolution, no $W$ bosons are emitted by the 
jets. Then, only rarely leptons from meson or baryon decays are accepted 
as isolated, resulting in cross sections of $\mathcal{O}(1)$ fb or less 
for the leptonic analysis. 

Tab.~\ref{tab:xsechad} then further details the cross sections 
remaining after additionally applying mass window cuts on each 
mass-related variable for the three different methods of the hadronic 
analysis as described in Sec.~\ref{sec:hadana}. The three minimum fat 
jet transverse momenta involved in the sample preparation and the 
requirements on their subjets effected during their subsequent reclustering 
strongly affect the cross sections left after applying methods \ref{mhard}, 
\ref{mmed} and \ref{msoft}. As the subjets used in method \ref{mhard} 
have $p_{T_\iota}>$ 200 GeV, followed by method~\ref{msoft} with 
$p_{T_\iota}>$ 40 GeV and method~\ref{mmed} with $p_{T_\iota}>$ 20 GeV, we 
observe the cross sections ascending in the same order.
While the cross section of methods~\ref{mmed} and \ref{msoft} 
maintain a constant difference (roughly of factor 3) as the fat jet 
selection cut increases, method~\ref{mhard} has a slower drop in 
cross section with increasing $p_{T_J}$. We attribute this behavior
to the fact that emissions of subjets with $p_{T_\iota}>$ 200 GeV and 
separation $\Delta R \geq 0.5$ are relatively rare if 
$p_{T_J}\gtrsim$ 500 GeV. For example, Tab.~\ref{tab:xsechad} shows that roughly 
$50\%$ of all tagged $W$ candidates with fat jet selection 
$p_{T_J}>500$ GeV are actually emitted when the fat jet has $p_{T_J}>750$ 
GeV. In comparison, almost $90\%$ of all tagged $W$ candidates in 
method~\ref{msoft} with selection cut $p_{T_J}>500$ GeV stem from a 
fat jet with $p_{T_J}<750$ GeV. 
Tab.~\ref{tab:xseclep} is dedicated to the leptonic analysis. It shows the 
cross sections after successively applying the minimum missing transverse 
energy cut in the left column and the transverse mass requirement in the 
right column, as outlined in Sec.~\ref{sec:lepana}.

\begin{table}[t]
  \begin{tabular}{c|c|ccc|c|ccc}
  & \multicolumn{4}{c|}{hadronic} & \multicolumn{4}{c}{leptonic} \\
  \hline
  \multirow{2}{*}{\;\;\;$f$\;\;\;}
      & \multirow{2}{*}{\quad $n_l=0$ \quad} 
      & \multicolumn{3}{c|}{$p_{T_J}$} 
      & \multirow{2}{*}{\quad $n_l=1$ \quad} 
      & \multicolumn{3}{c}{$p_{T_J}$} \\
  & & \;\;\;500 GeV\;\; & \;\;\;750 GeV\;\; & \;\;\;1000 GeV\;\; 
  & & \;\;\;500 GeV\;\; & \;\;\;750 GeV\;\; & \;\;\;1000 GeV\;\; \\
  \hline 
  0   & 2116
      & 551.2 & 59.53 & 10.24 
      & 0.001 
      & 0.002 & 0.0002 & 3$\times10^{-5}$\hspace*{-6pt} \\ 
  1.0 & 2092 
      & 539.1 & 57.74 & 9.856 
      & 23.37 
      & 3.663 & 0.5795 & 0.1286 \\ 
  1.1 & 2090 
      & 537.9 & 57.57 & 9.826 
      & 25.73 
      & 4.056 & 0.6341 & 0.1389 \\ 
  2.0 & 2070 
      & 527.5 & 56.00 & 9.481 
      & 45.71 
      & 7.081 & 1.117\hPO & 0.2439
  \end{tabular}
  \caption{
	    Cross sections of the hadronic and leptonic 
	    analyses in pb. Where applicable a column has three 
	    numbers to account for different fat jet $p_T$ cuts:
	    $p_{T_J}> 500$ (left), $750$ (middle) and 
	    $1000$ (right) GeV.
	    \label{tab:xsecpT}
	  }
\end{table}

\begin{table}[t]
  \begin{tabular}{c|ccc|ccc|ccc}
    & \multicolumn{3}{c|}{method \ref{mhard} ($m_{\mr{BDRS}}\in[74,90]$ GeV)}
    & \multicolumn{3}{c|}{method \ref{mmed} ($m_{23}\in[70,86]$ GeV)} 
    & \multicolumn{3}{c} {method \ref{msoft} ($m_\mr{min}\in[70,86]$ GeV)} \\
  \hline
  \multirow{2}{*}{\;\;\;$f$\;\;\;} 
      & \multicolumn{3}{c|}{$p_{T_J}$} 
      & \multicolumn{3}{c|}{$p_{T_J}$} 
      & \multicolumn{3}{c} {$p_{T_J}$} \\
  & \;\;\;500 GeV\;\; & \;\;\;750 GeV\;\; & \;\;\;1000 GeV\;\; 
  & \;\;\;500 GeV\;\; & \;\;\;750 GeV\;\; & \;\;\;1000 GeV\;\; 
  & \;\;\;500 GeV\;\; & \;\;\;750 GeV\;\; & \;\;\;1000 GeV\;\; \\
  \hline
  0   & 0.9939 & 0.4906 & 0.1447 
      & 35.87 & 4.228 & 0.6943 
      & 11.81 & 1.401 & 0.2255 \\ 
  1.0 & 1.219\hPO & 0.6202 & 0.1923 
      & 38.83 & 4.698 & 0.7890 
      & 13.22 & 1.607 & 0.2643 \\ 
  1.1 & 1.251\hPO & 0.6386 & 0.1977 
      & 39.11 & 4.741 & 0.8000 
      & 13.34 & 1.623 & 0.2661 \\ 
  2.0 & 1.422\hPO & 0.7312 & 0.2286 
      & 41.43 & 5.085 & 0.8584 
      & 14.49 & 1.780 & 0.2939 
  \end{tabular}
  \caption{
	    Cross sections after the three mass reconstruction cuts in 
	    the three different methods for the hadronic analysis in pb. 
	    Each column contains three numbers to account for different 
	    fat jet cuts:
	    $p_{T_J}>500$ (left), $750$ (middle) and $1000$ (right) GeV.
	    \label{tab:xsechad}
	  }
\end{table}

\begin{table}[t!]
  \begin{tabular}{c|ccc|ccc}
      & \multicolumn{3}{c|}{$\slashed{E}_{T}>50$ GeV} 
      & \multicolumn{3}{c}{$m_\mr{T}\in[60,100]$ GeV} \\
  \hline
  \multirow{2}{*}{\;\;\;$f$\;\;\;} 
      & \multicolumn{3}{c|}{$p_{T_J}$} 
      & \multicolumn{3}{c} {$p_{T_J}$} \\
  & \;\;\;500 GeV\;\; & \;\;\;750 GeV\;\; & \;\;\;1000 GeV\;\; 
  & \;\;\;500 GeV\;\; & \;\;\;750 GeV\;\; & \;\;\;1000 GeV\;\; \\
  \hline
  0   & 0.001 & 1$\times10^{-5}$\hspace*{-6pt} & 4$\times10^{-7}$ 
      & 6$\times10^{-5}$\hspace*{-6pt} & 5$\times10^{-6}$ & 1$\times10^{-7}$ \\ 
  1.0 & 2.062 & 0.3481 & 0.07988 
      & 0.5769 & 0.09271 & 0.02156 \\ 
  1.1 & 2.280 & 0.3795 & 0.08654 
      & 0.6402 & 0.1046\hPO & 0.02323 \\ 
  2.0 & 4.000 & 0.6765 & 0.1531\hPO 
      & 1.108\hPO & 0.1830\hPO & 0.04099
  \end{tabular}
  \caption{
	    Cross sections after the $\slashed{E}_{T}>50$ GeV cut and 
	    the $m_\mr{T}$ cut in the leptonic analysis in pb. 
	    Each column contains three numbers 
	    to account for different fat jet cuts:
	    $p_{T_J}>500$ (left), $750$ (middle) and $1000$ (right) GeV.
	    \label{tab:xseclep}
	  }
\end{table}

The three mass reconstruction observables $m_\mr{min}$, $m_{23}$, 
and $m_{\mr{BDRS}}$ in the hadronic analysis of Sec.~\ref{sec:hadana} 
do not suffer from low statistics even in the most boosted kinematic 
region where both fat jets have $p_{T_J}>1$ TeV, although the cross 
sections after cuts are in the sub-picobarn range. This is well within 
the Run 2 capabilities of the LHC, expected to reach an integrated 
luminosity of $\mathcal{L}\approx 100\,\rm{fb^{-1}}$. Therefore, the 
sensitivity in measuring electroweak emissions in the collinear 
approximation is mostly limited by the signal vs background ratio 
achieved with the reconstruction methods and their intrinsic 
experimental uncertainties. The observables we use induce a peak 
structure which allows to perform side-band analyses, reducing 
sensitivity on theoretical uncertainties.

In order to estimate how well the electroweak emissions can be 
measured at the LHC, we use a binned log-likelihood hypothesis 
test when varying the emission-probability modification factor $f$, as documented in \cite{Junk:1999kv} and implemented in the \Root \cite{Brun:1997pa} package 
\TLimit. Throughout we include a flat systematic error added linearly to the 
statistical error for each bin, 
$\sigma_\mr{syst} + \sigma_\mr{stat} = \sigma_\mr{tot}$, when 
performing the likelihood test.  In general the statistical error is dominant at low 
integrated luminosity. However, as the number of events increases, 
the relative statistical error decreases and the fixed statistical 
error can become dominant, i.e. increasing the integrated luminosity 
will not improve the exclusion limit.
We evaluate the LHC's potential to exclude modified $W$ emission rates $(f\neq 1)$ in favor of the Standard Model $(f=1)$. 

\begin{figure}[t]
  \centering
  \includegraphics[width=.32\linewidth]{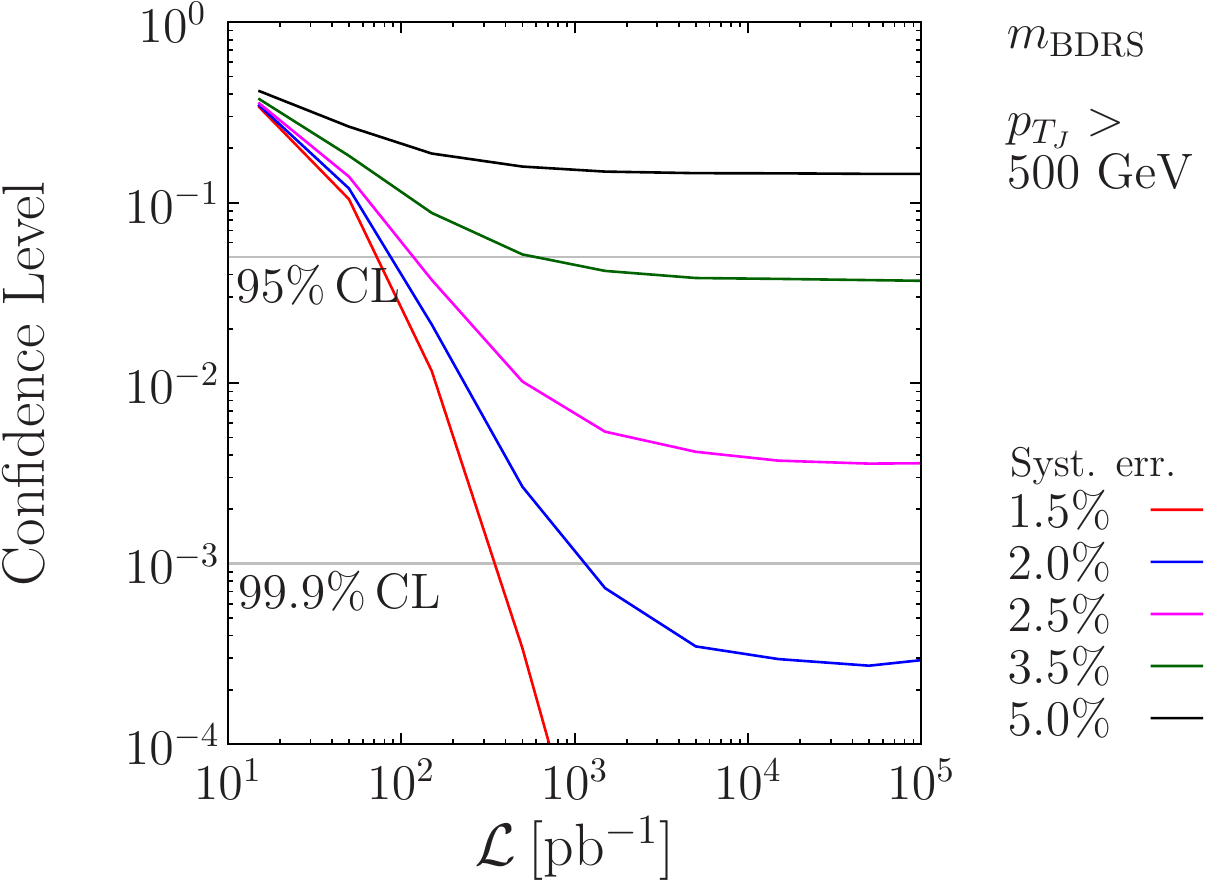}\hfill
  \includegraphics[width=.32\linewidth]{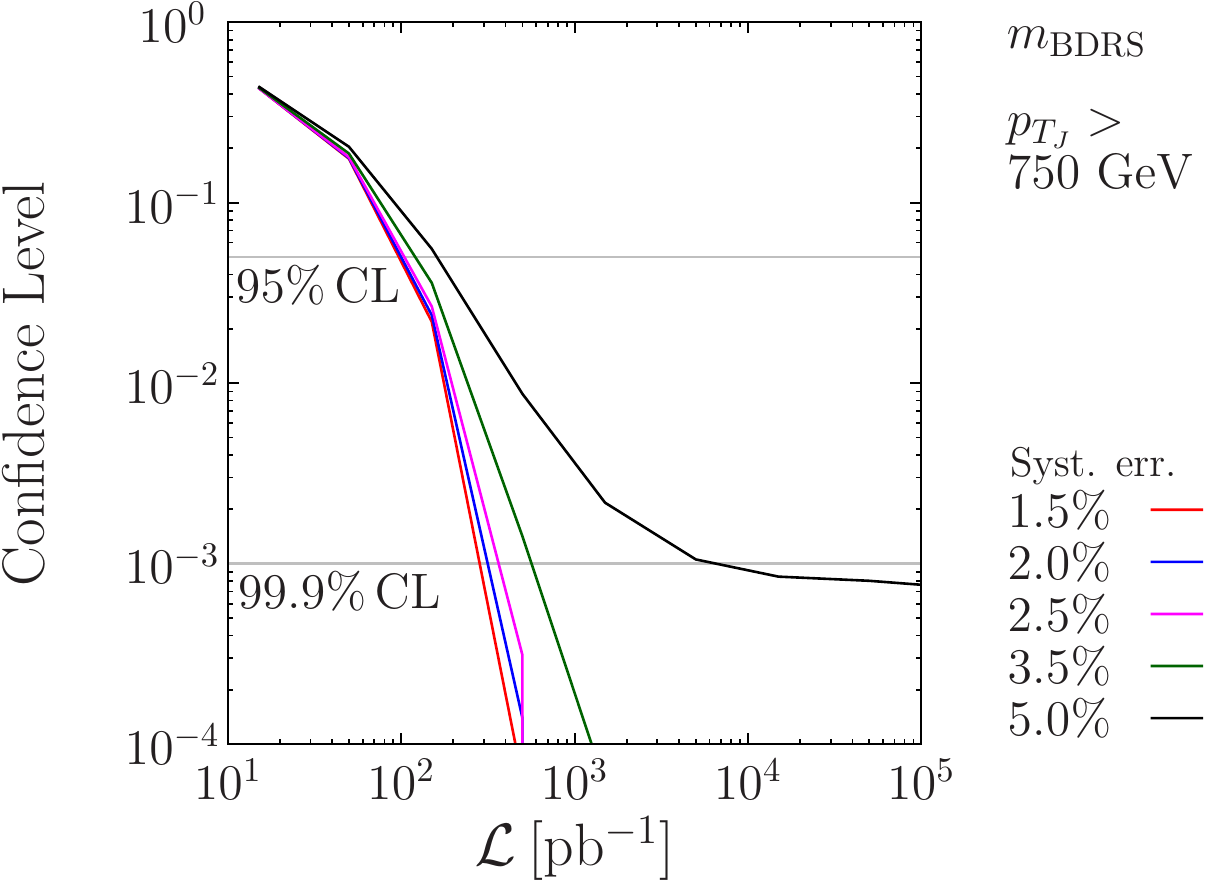}\hfill
  \includegraphics[width=.32\linewidth]{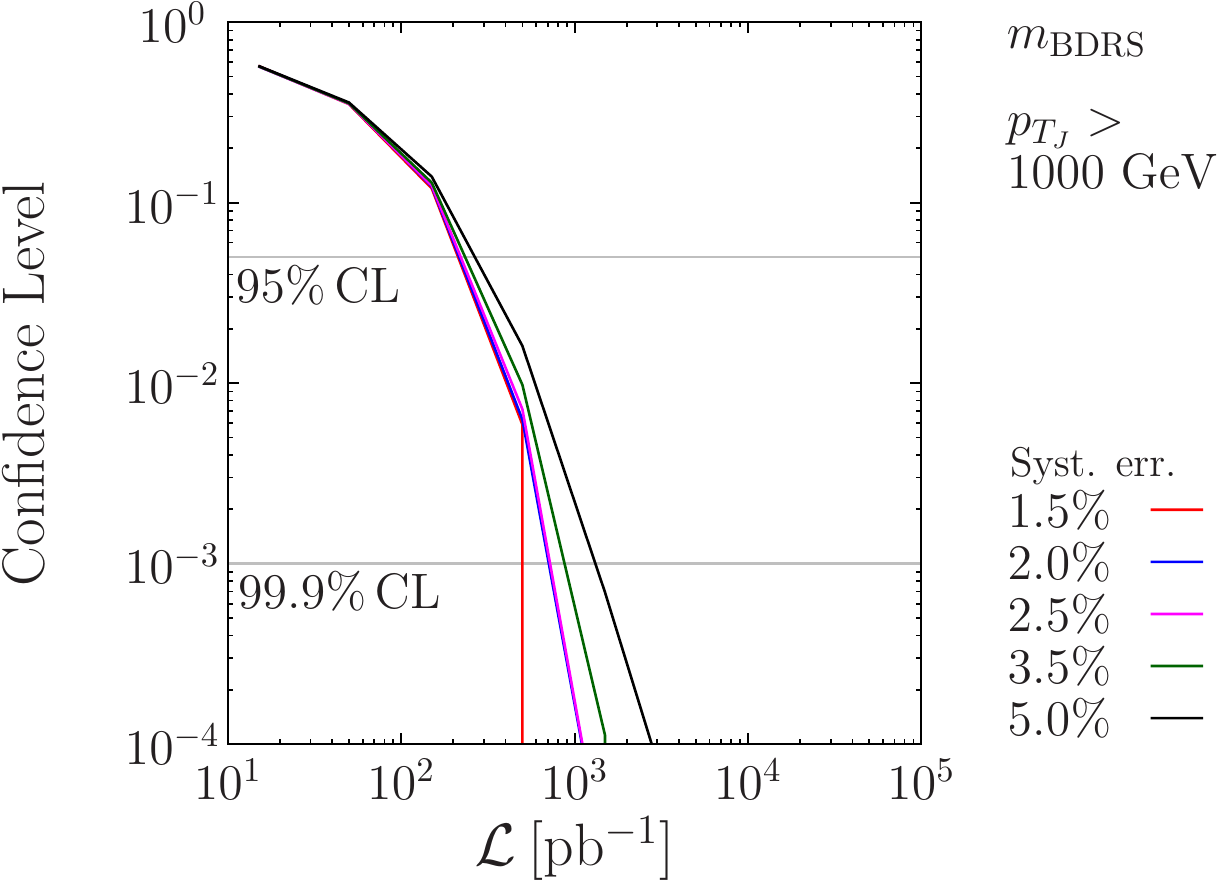}\\[10pt]
  \includegraphics[width=.32\linewidth]{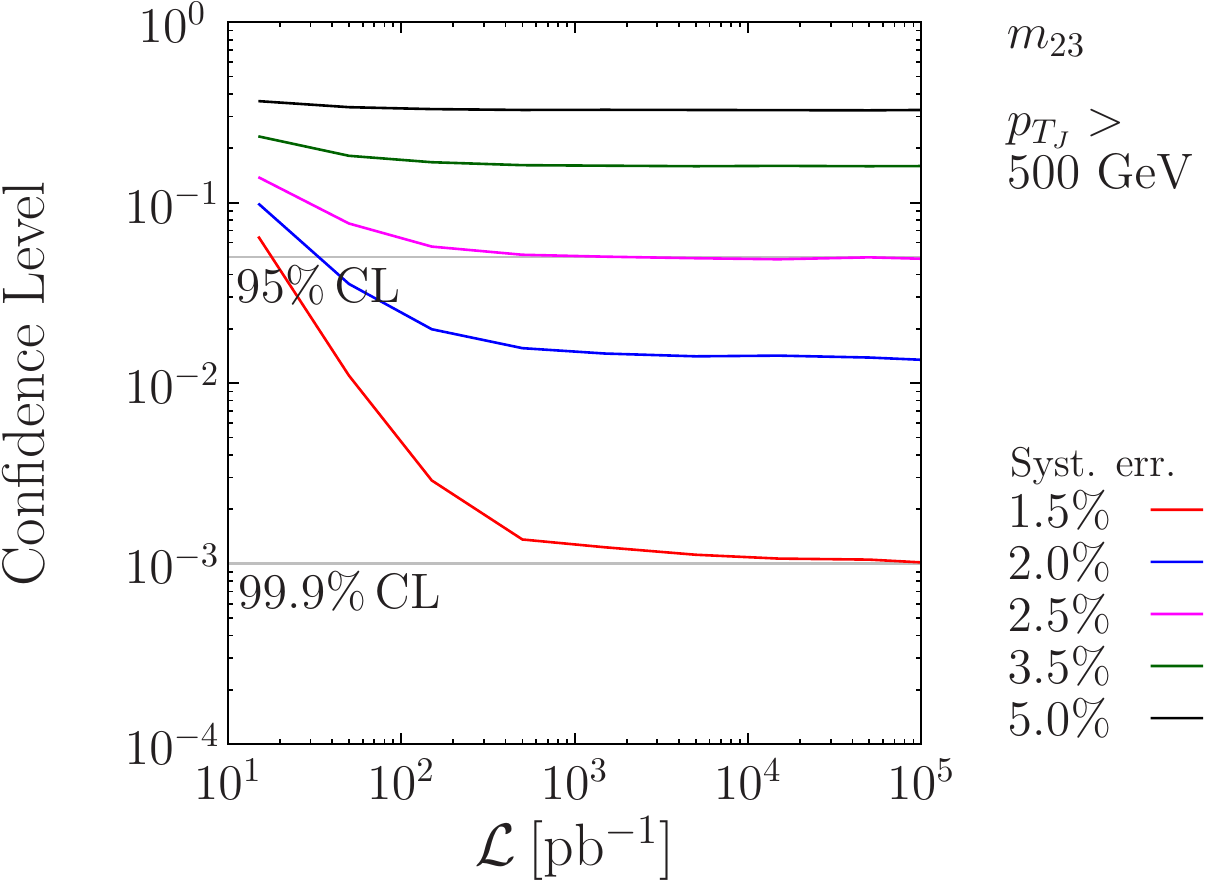}\hfill
  \includegraphics[width=.32\linewidth]{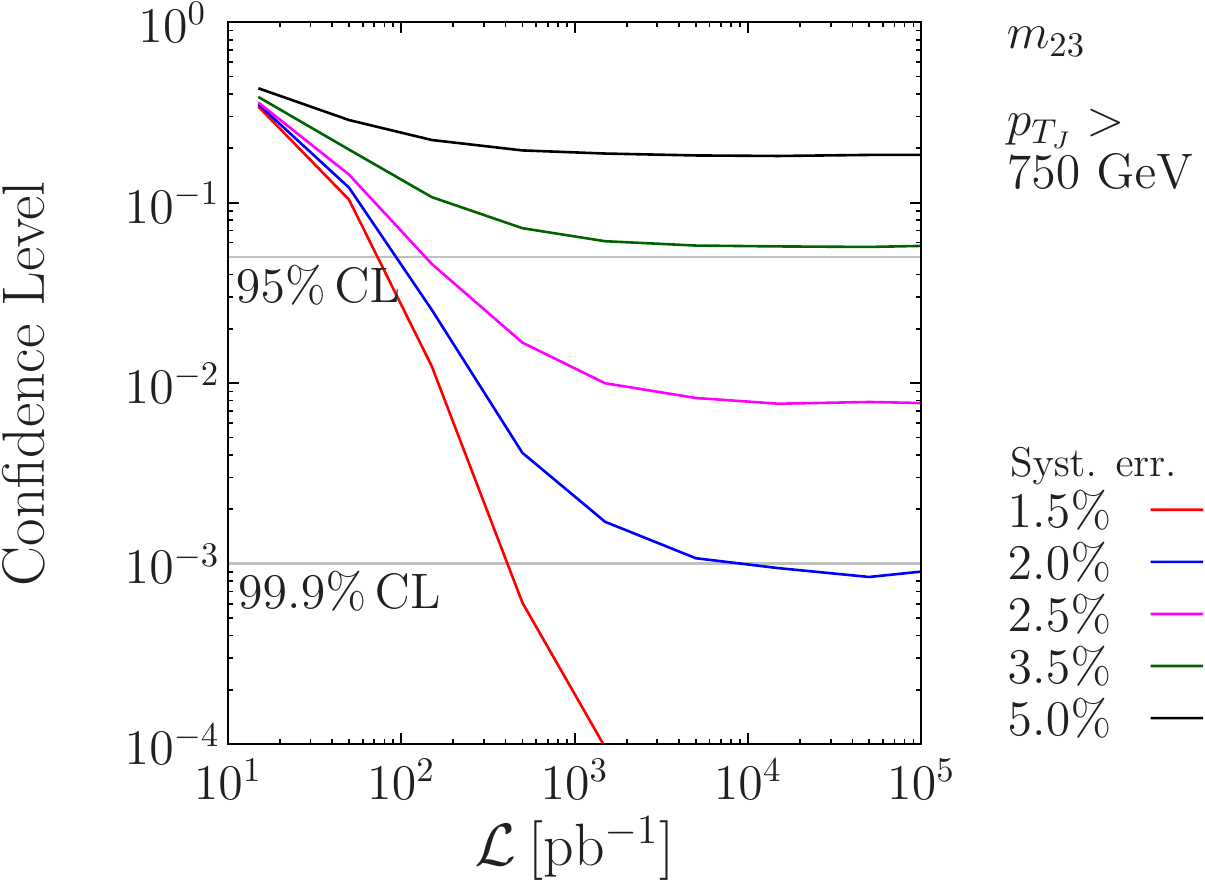}\hfill
  \includegraphics[width=.32\linewidth]{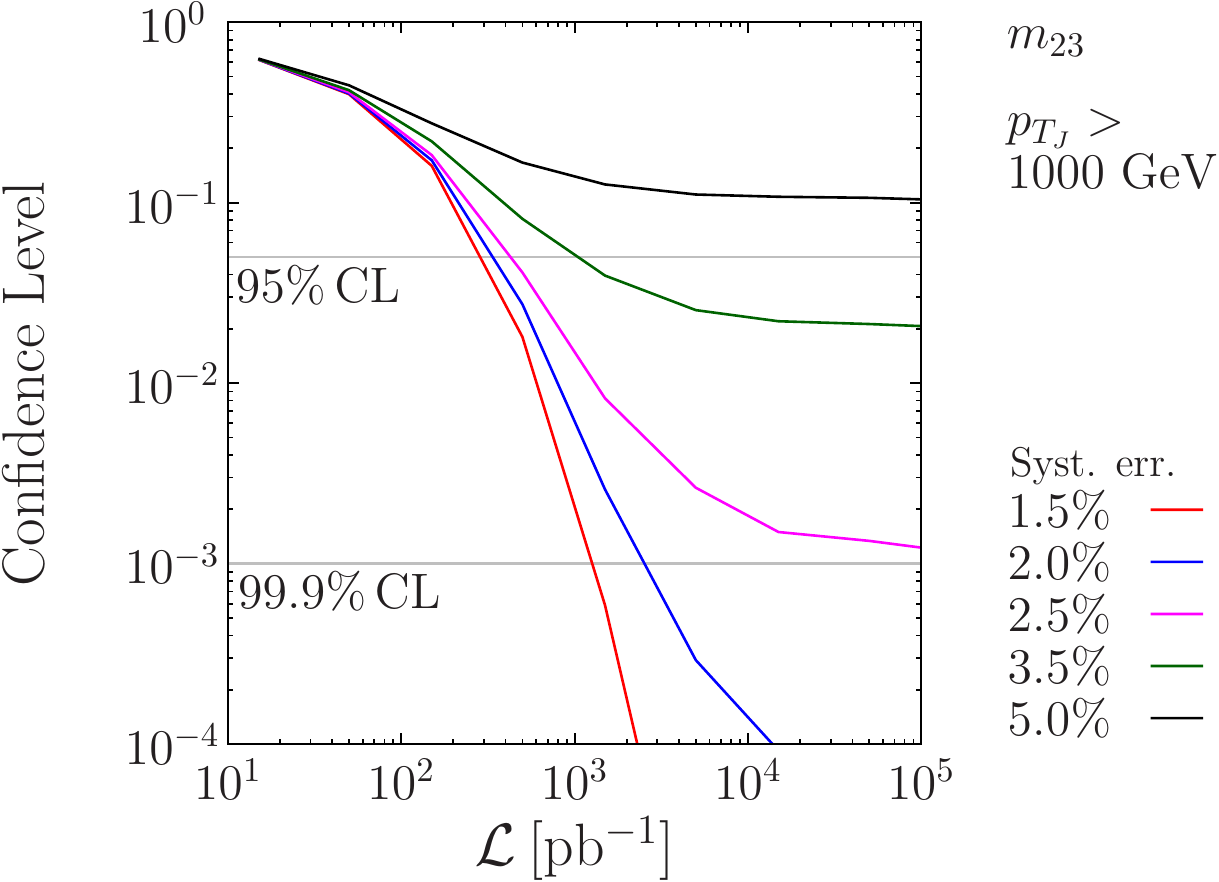}\\[10pt]
  \includegraphics[width=.32\linewidth]{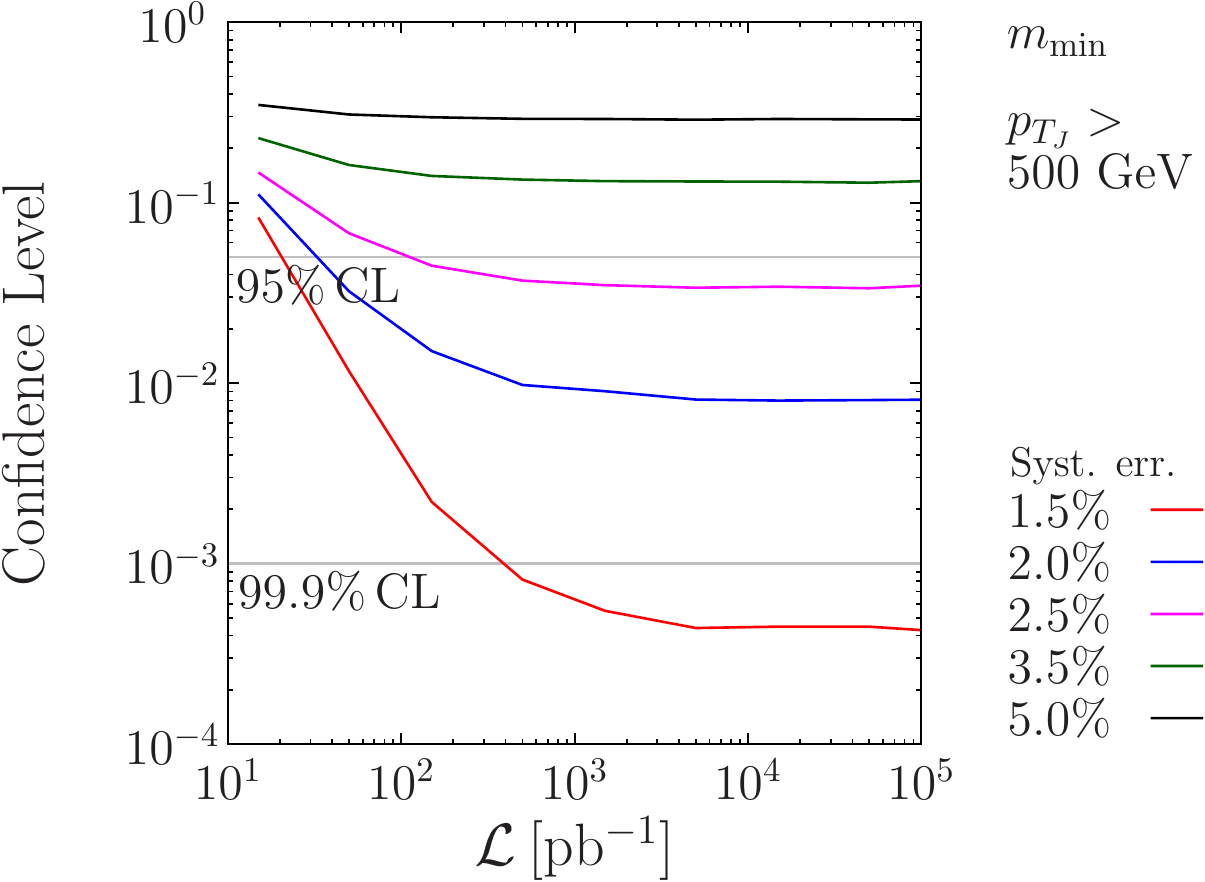}\hfill
  \includegraphics[width=.32\linewidth]{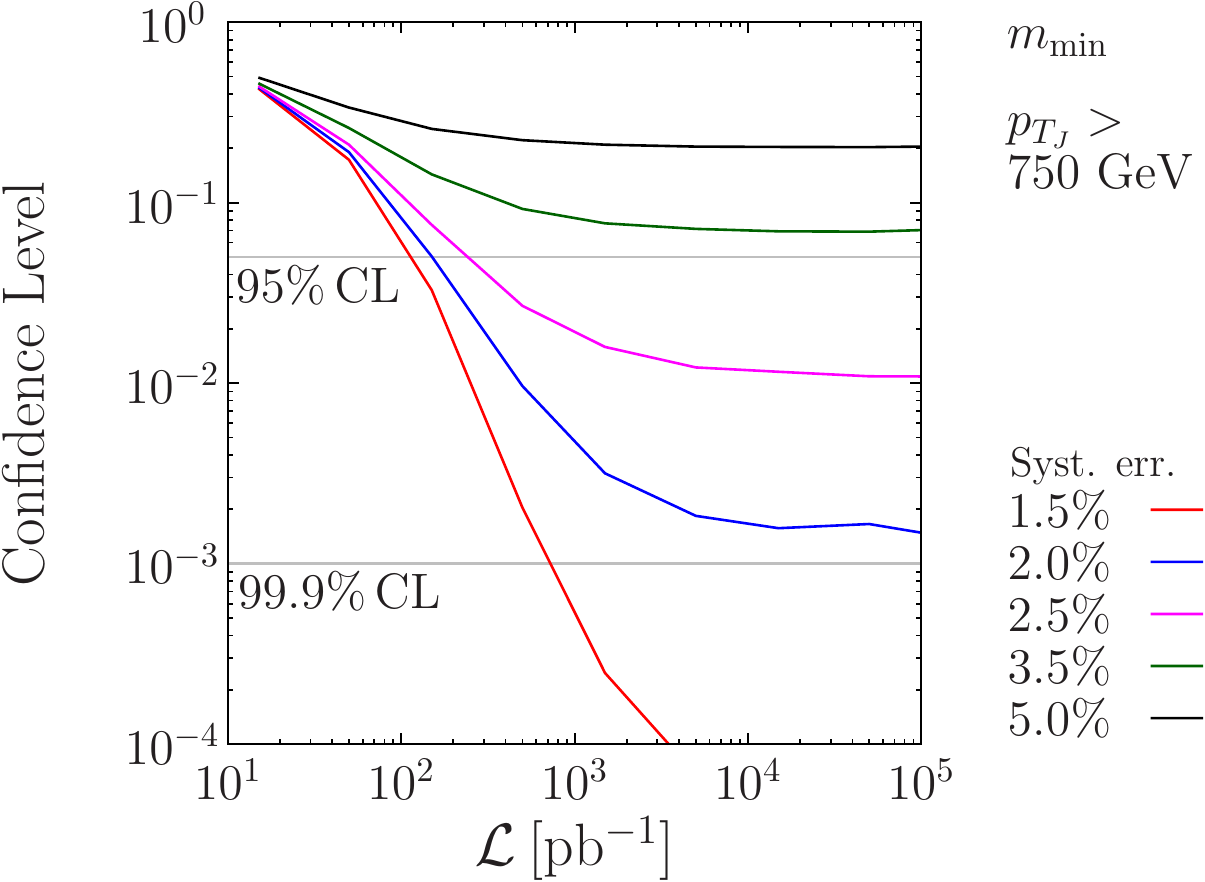}\hfill
  \includegraphics[width=.32\linewidth]{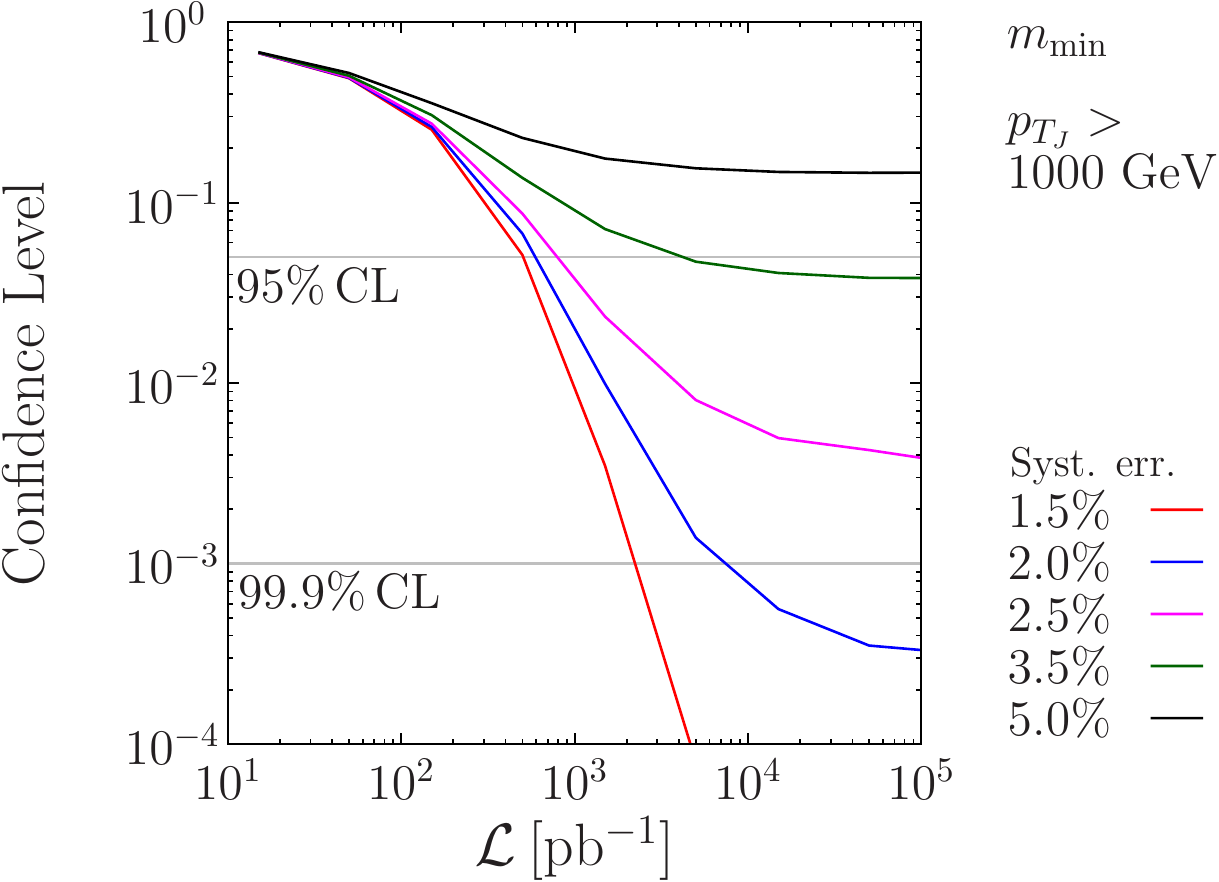}
  \caption{
	    $\mathrm{1-CL_b}$ for the $W$ mass reconstruction through 
	    method \ref{mhard} using $m_\text{BDRS}$ (top row), 
	    method \ref{mmed} using $m_{23}$ (center row), and 
	    method \ref{msoft} using $m_\text{min}$ (bottom row) 
	    of the hadronic analysis for the three different minimum 
	    jet transverse momenta: $p_{T_J} > 500$ (left column), 
	    $750$ (center column) and $1000$ (right column) GeV. 
	    The background corresponds to $f=0$ and signal + background 
	    to $f=1$.
	    \label{fig:CLb}
	  }
\end{figure}

First, we test whether we can observe hadronically decaying $W$ bosons over the large QCD background. Here we use $1-\mathrm{CL_b}$ confidence level (as defined in \cite{Junk:1999kv}) with $f=0$ as background hypothesis. The results in Fig.~\ref{fig:CLb} indicate what systematic error $\sigma_\mr{syst}$ on each mass distribution would allow the exclusion of the QCD-only  
($f=0$) from the combined electroweak and QCD parton shower ($f=1$) 
at $95\%$ and $99.9\%$ CL. Only here we set $f=1$ to be our 
$S+B$ hypothesis. We plot $1-\mathrm{CL_b}$ as 
a function of the integrated luminosity in Fig.~\ref{fig:CLb}. In 
each row we show the background rejection by one of the hadronic mass 
reconstruction methods from Sec.~\ref{sec:hadana}. From left to right 
the fat jet selection cut becomes more stringent. Despite the large 
statistics for $p_{T,J} > 500$ GeV, the variables result in a better $S/B$ and a better sensitivity in excluding $f=0$ in the regimes were the fat jet is harder.  We find that all mass reconstruction methods exclude the 
QCD-only hypothesis in favor of the Standard Model at $95\%$ CL 
with $\sigma_\mr{syst} = 3.5\%$, while using the BDRS method allows 
for a $99.9\%$ exclusion with $\sigma_\mr{syst} = 5\%$. Moreover, 
$m_{\rm{BDRS}}$ gives a far more successful than $99.9\%$ exclusion 
(almost $5\sigma$). Hence, all methods designed for hadronic $W$ 
reconstruction will suffice to disprove a QCD-only hypothesis during 
the second LHC run.

\begin{figure}[t]
  \includegraphics[width=.32\linewidth]{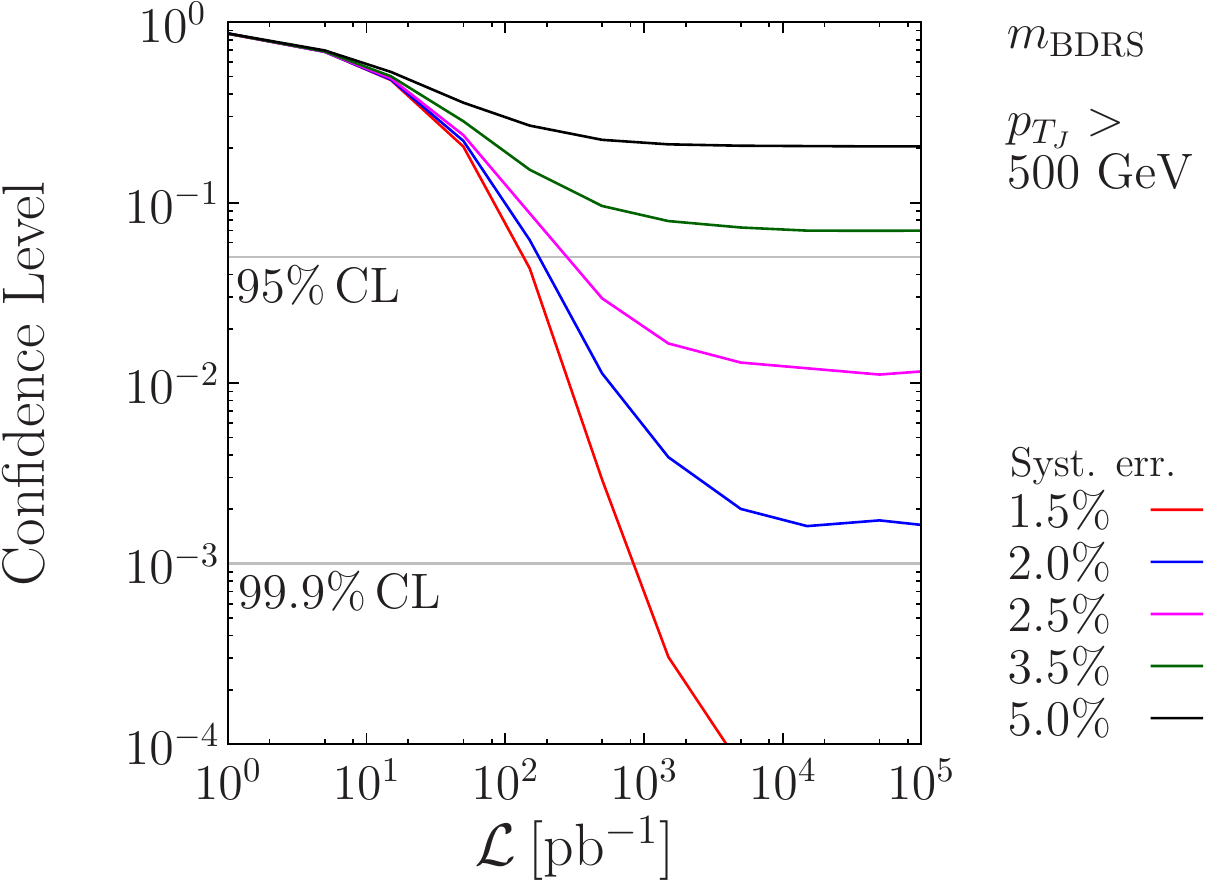}\hfill
  \includegraphics[width=.32\linewidth]{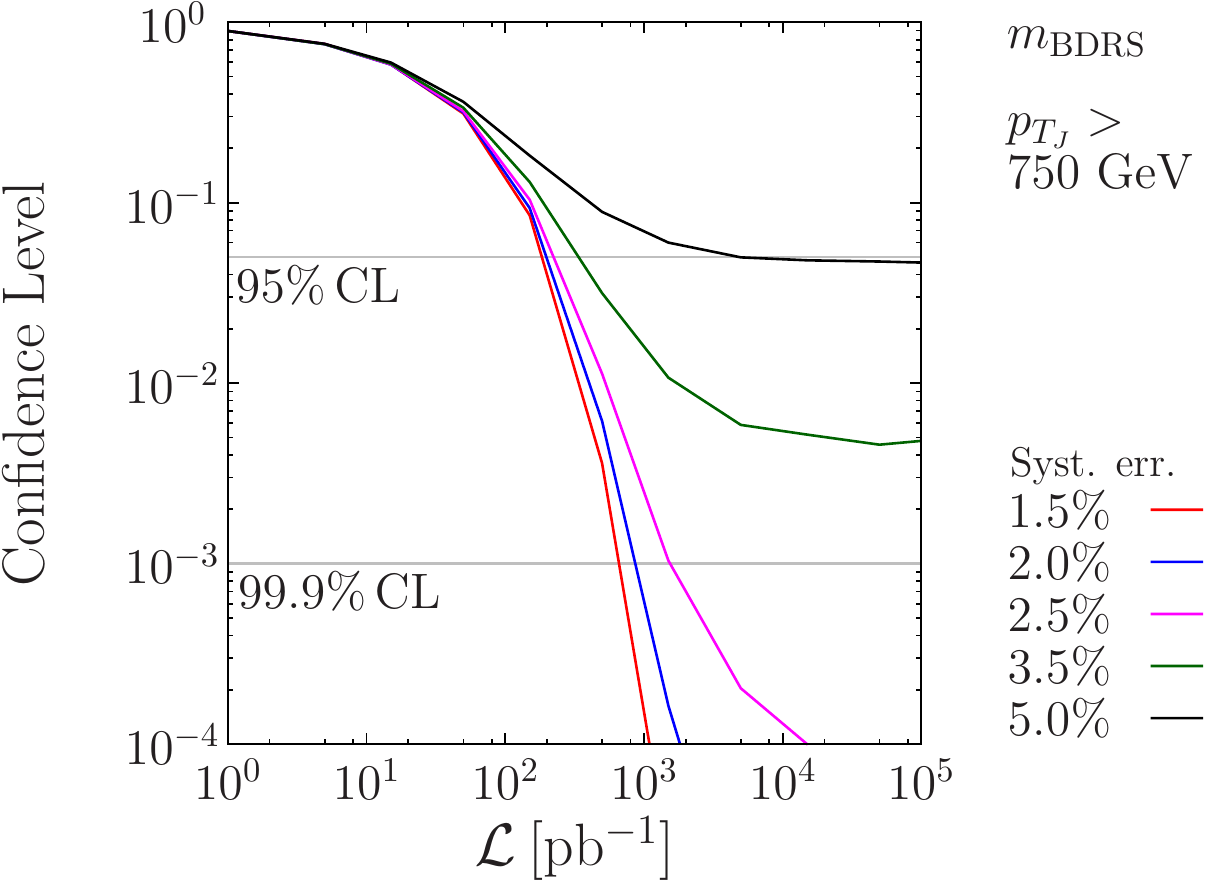}\hfill
  \includegraphics[width=.32\linewidth]{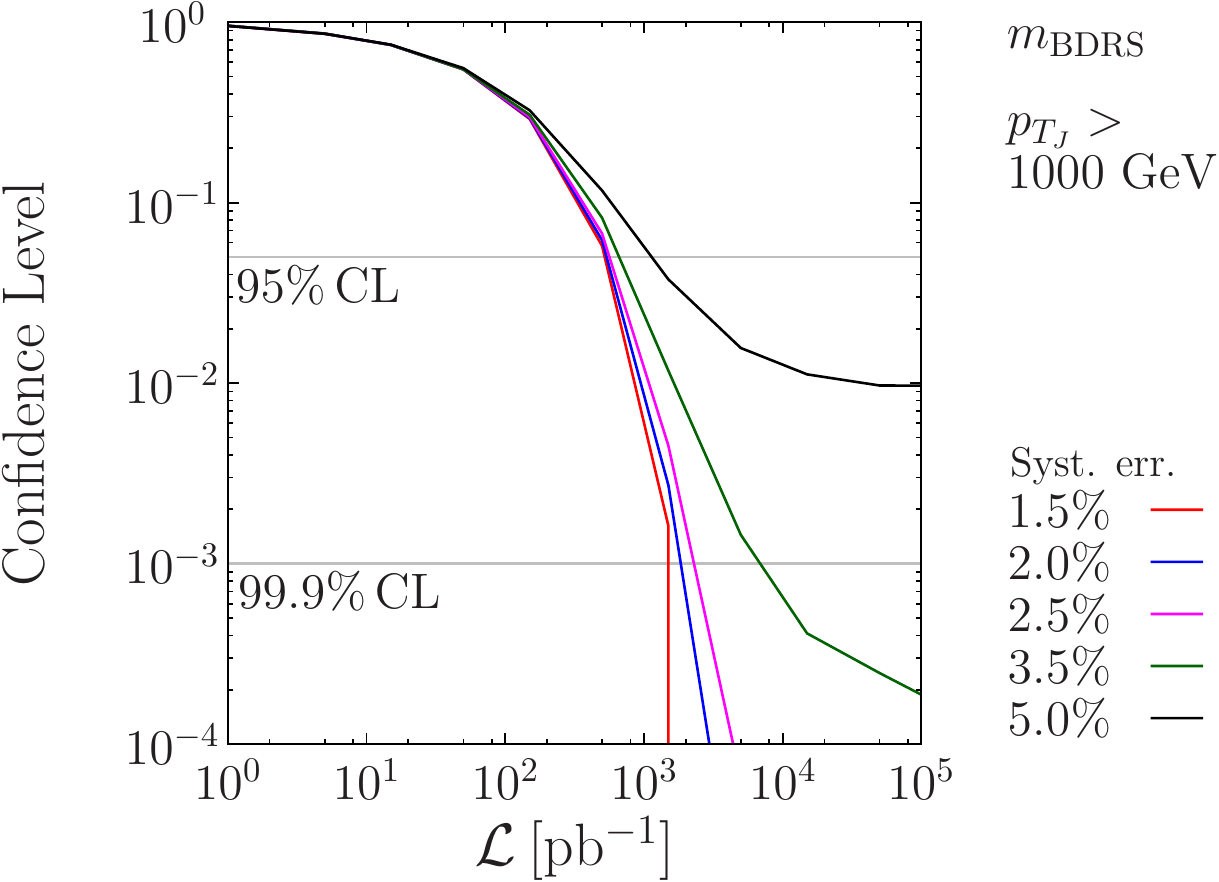}\\[10pt]
  \includegraphics[width=.32\linewidth]{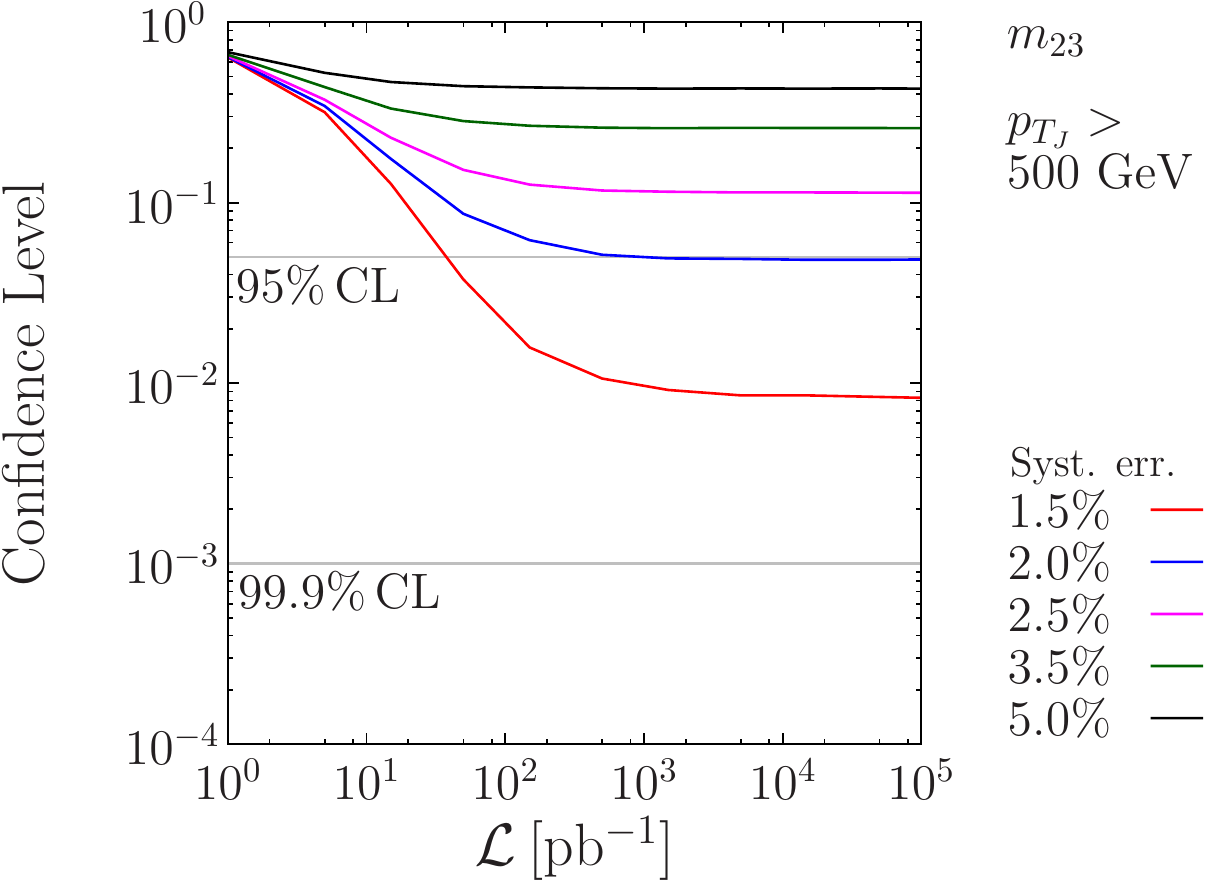}\hfill
  \includegraphics[width=.32\linewidth]{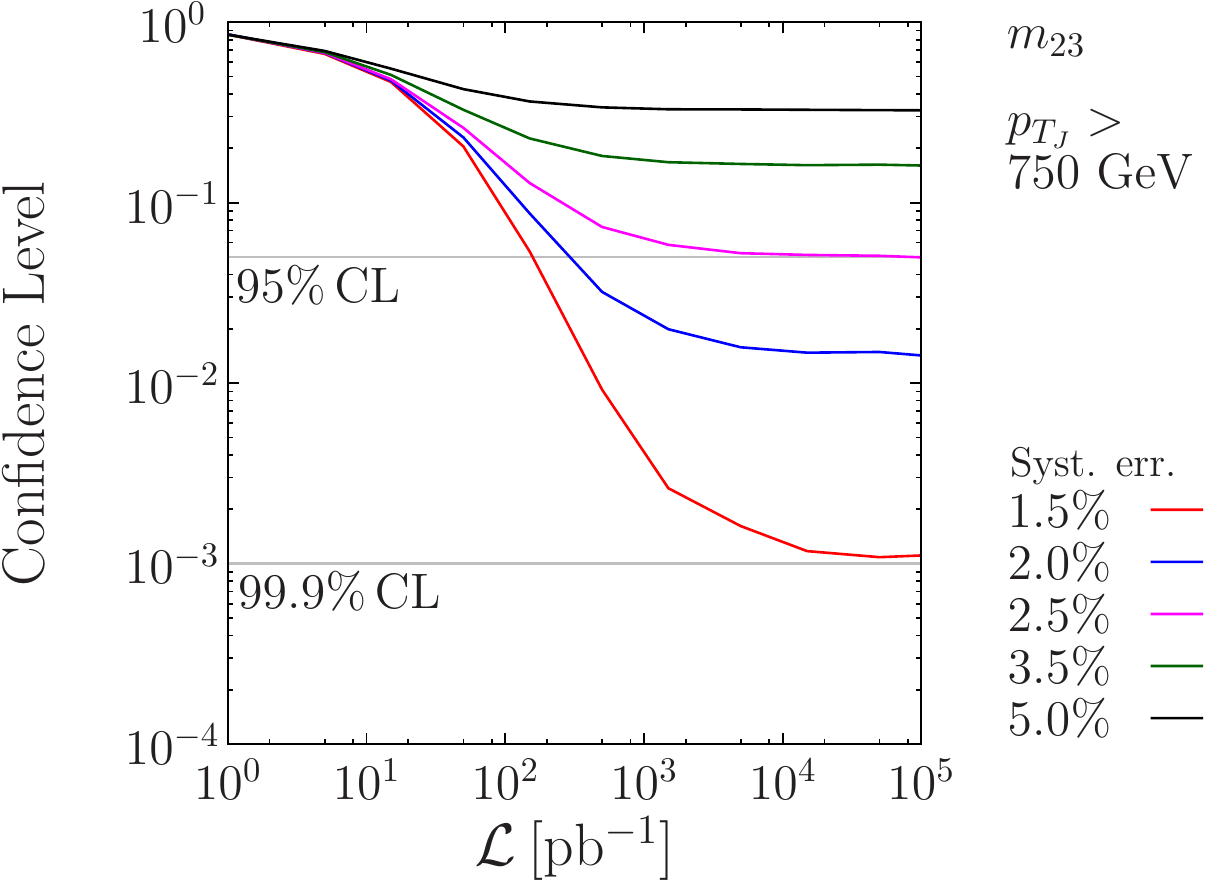}\hfill
  \includegraphics[width=.32\linewidth]{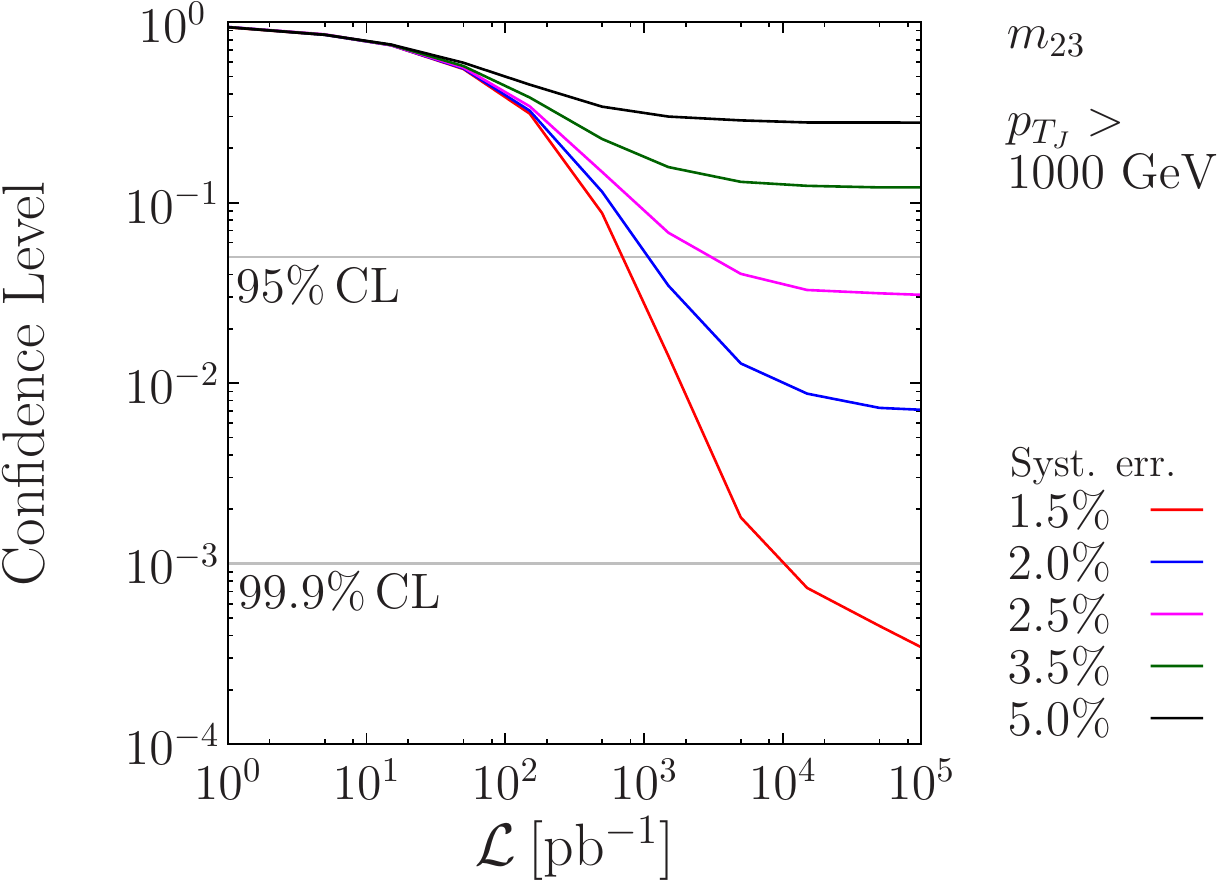}\\[10pt]
  \includegraphics[width=.32\linewidth]{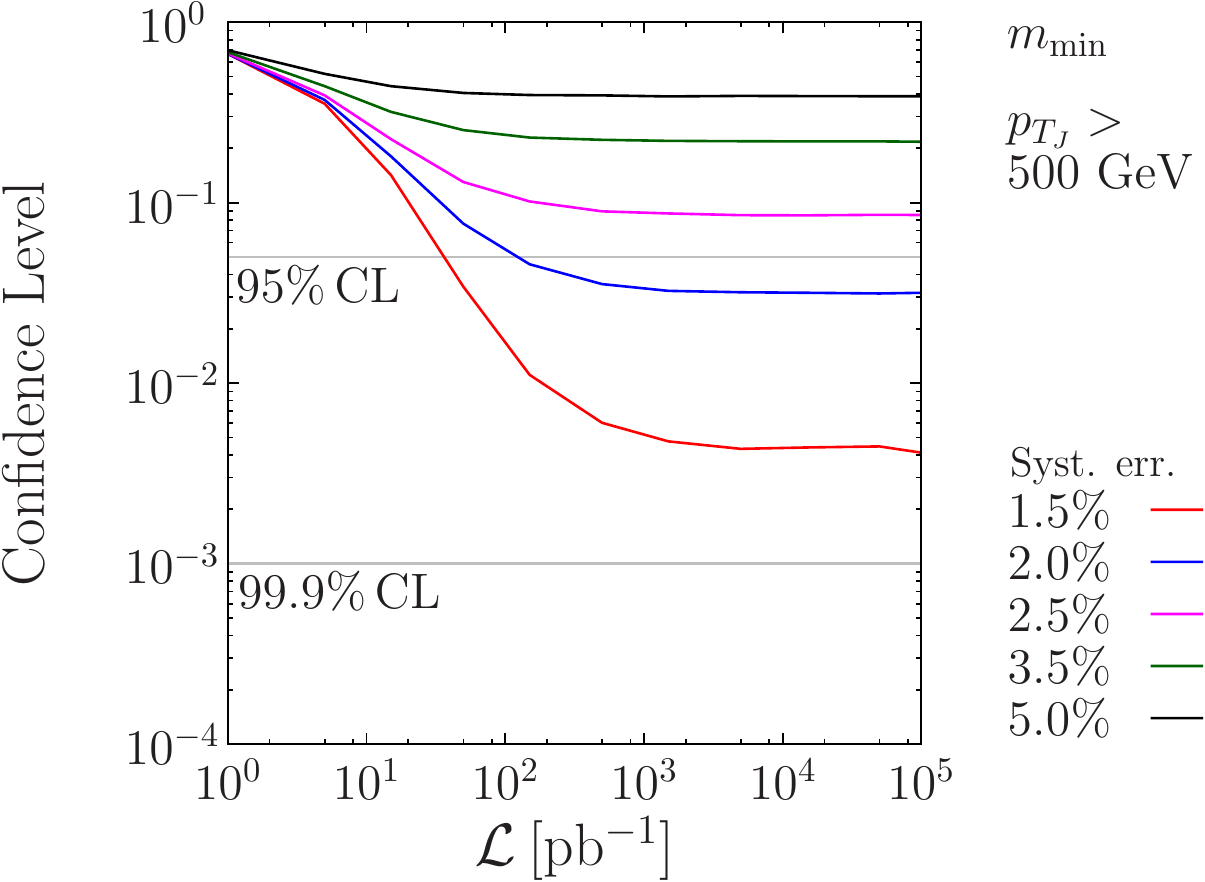}\hfill
  \includegraphics[width=.32\linewidth]{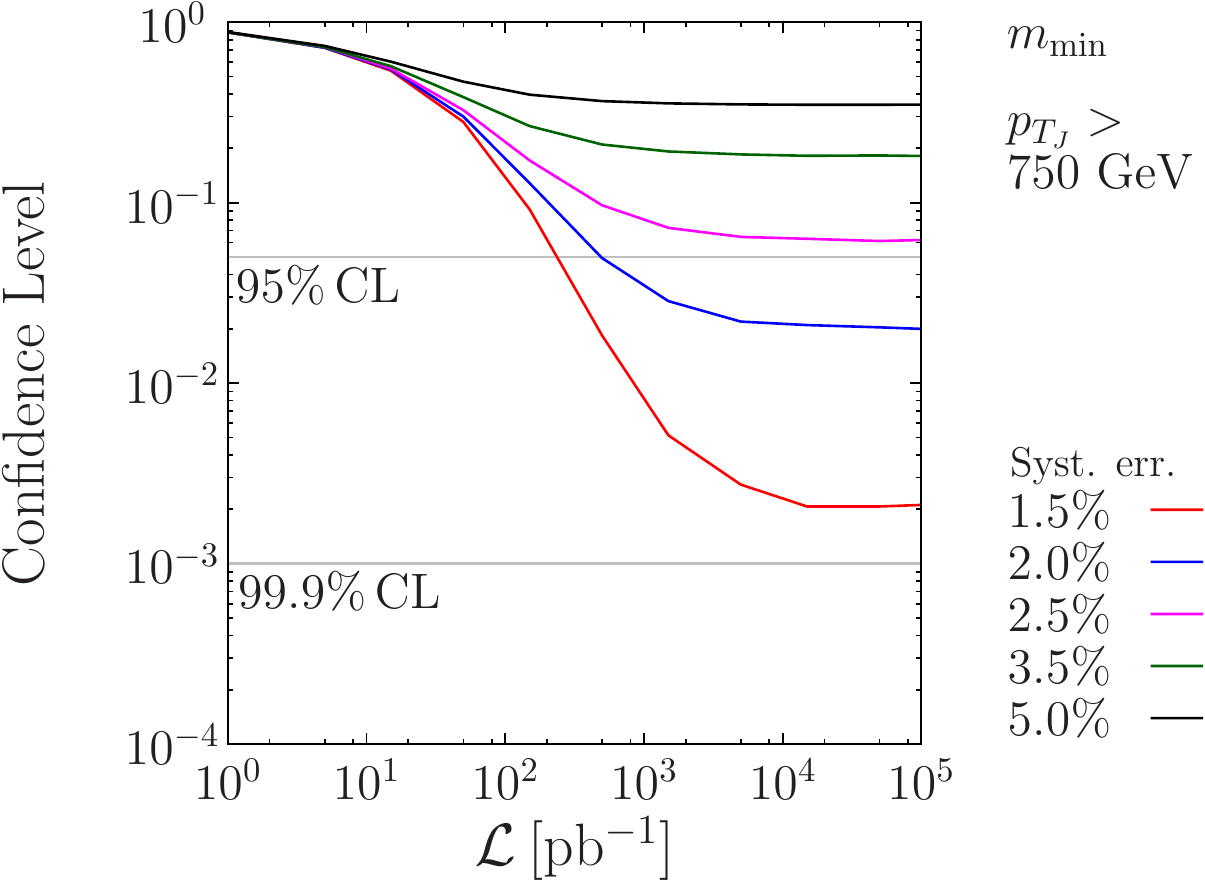}\hfill
  \includegraphics[width=.32\linewidth]{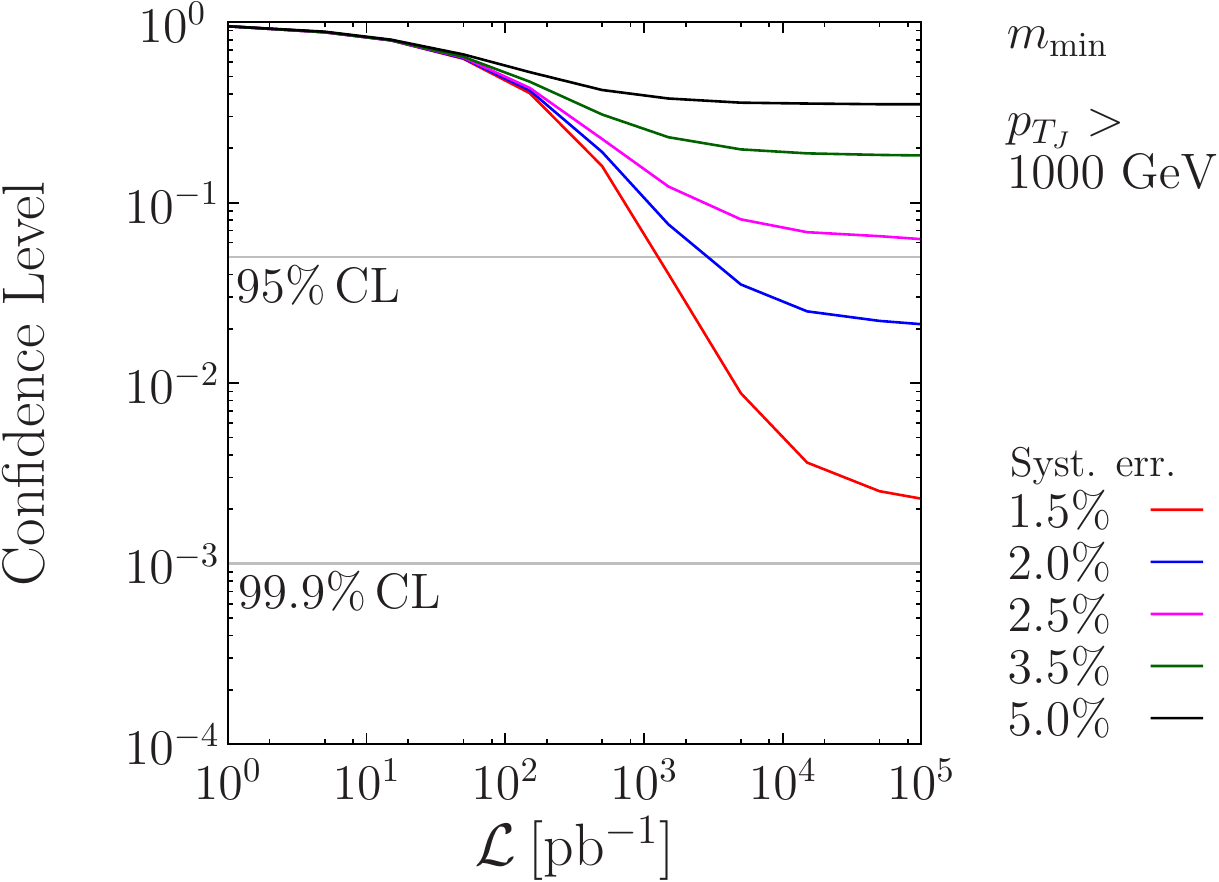}
  \caption{
	    CLs obtained from the $W$ mass reconstruction through 
	    method \ref{mhard} using $m_\text{BDRS}$ (top row), 
	    method \ref{mmed} using $m_{23}$ (center row), and 
	    method \ref{msoft} using $m_\text{min}$ (bottom row) 
	    of the hadronic analysis for the three different minimum 
	    jet transverse momenta: $p_{T_J} > 500$ (left column), 
	    $750$ (center column) and $1000$ (right column) GeV. 
	    The background corresponds to the Standard Model emission 
	    rate ($f=1$) and signal + background to $f=2$.
	    \label{fig:CLs}
	  }
\end{figure}

\begin{figure}[t]
  \centering
  \includegraphics[width=.32\linewidth]{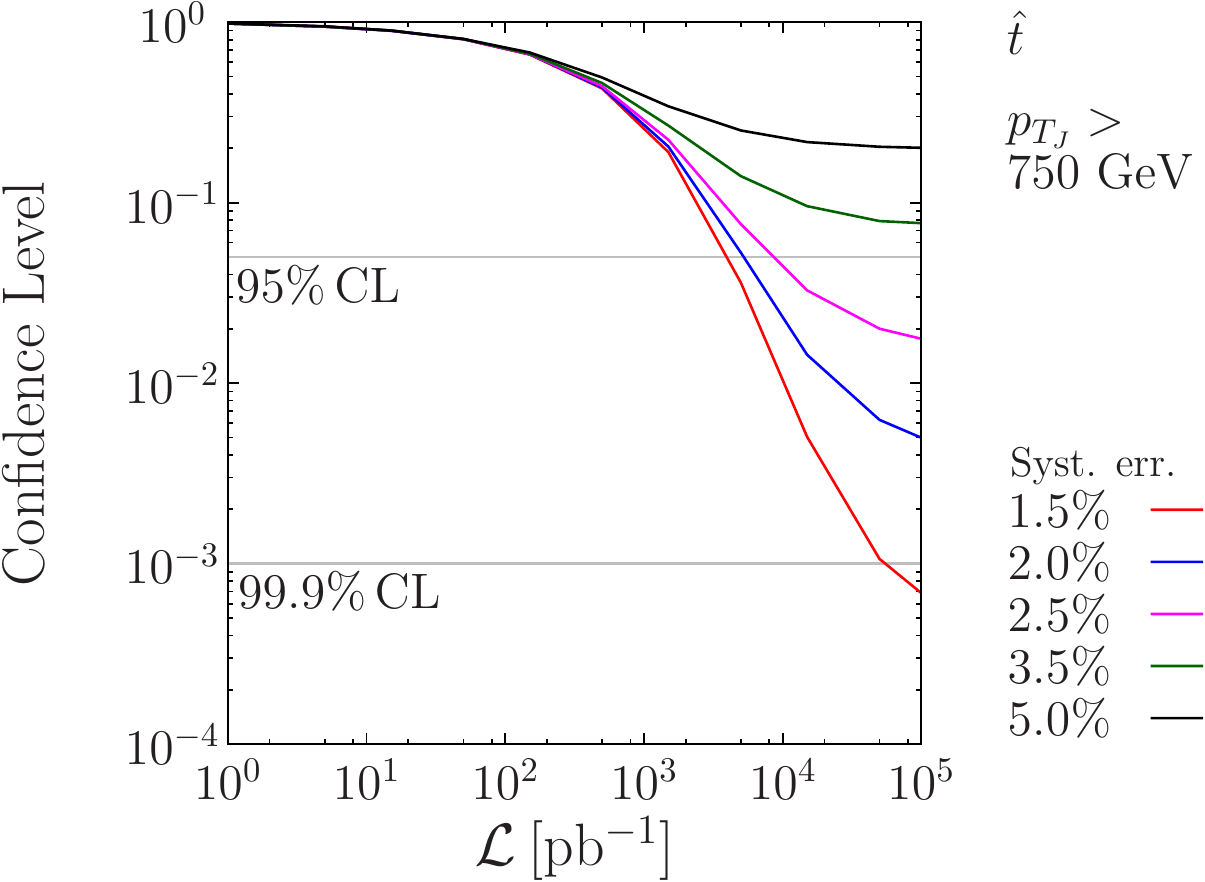}\hspace*{20pt}
  \includegraphics[width=.32\linewidth]{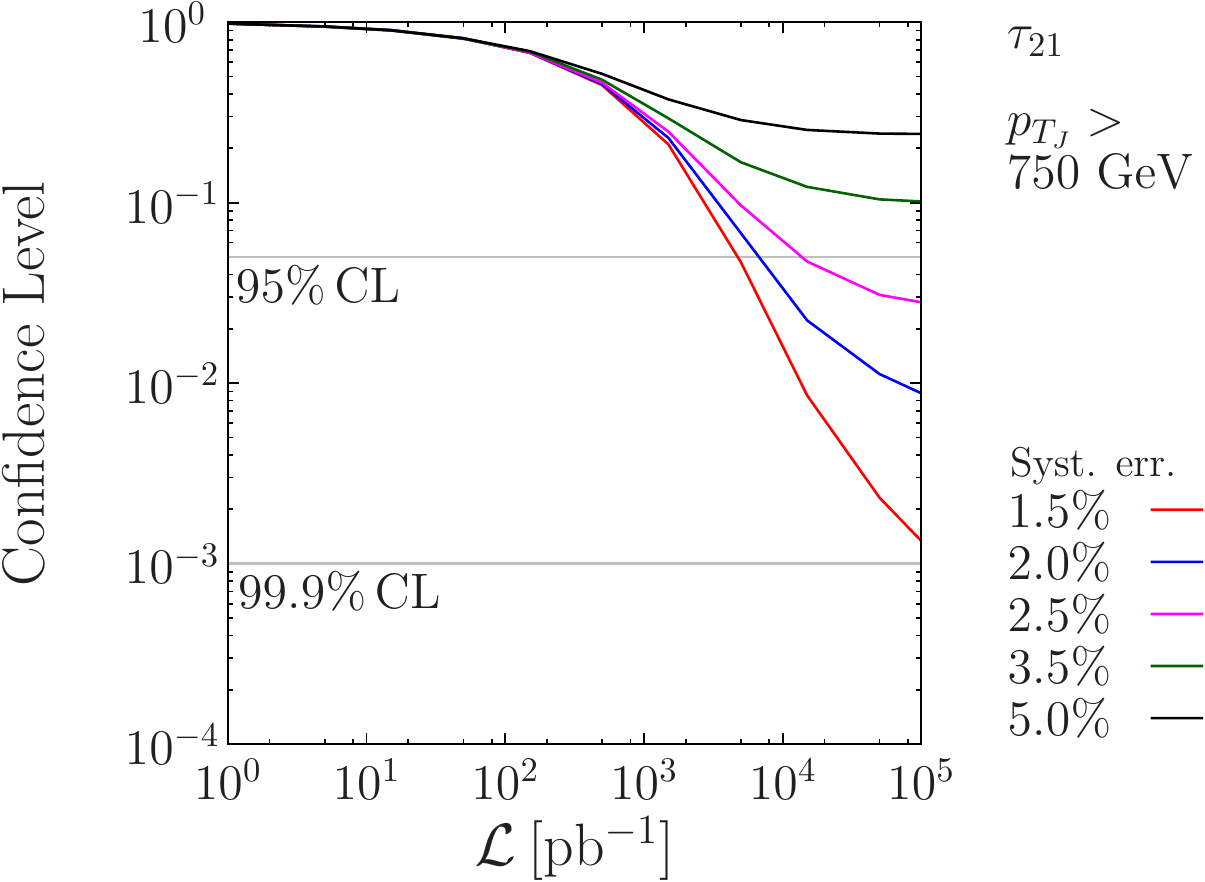}
  \caption{
	    CLs obtained from the ellipticity $\hat{t}$ (left) and 
	    $\tau_{21}$ (right) distributions calculated from the constituents 
            of the $W$ candidates that pass the BDRS cut on the second 
	    boosted subjet. $p_{T_J}>750$ GeV. The background is the SM emission 
	    rate ($f=1$), signal + background sample is $f=1.1$.
	    \label{fig:CLsShapes}
	  }
\end{figure}

\begin{figure}[t]
  \centering
  \includegraphics[width=.32\linewidth]{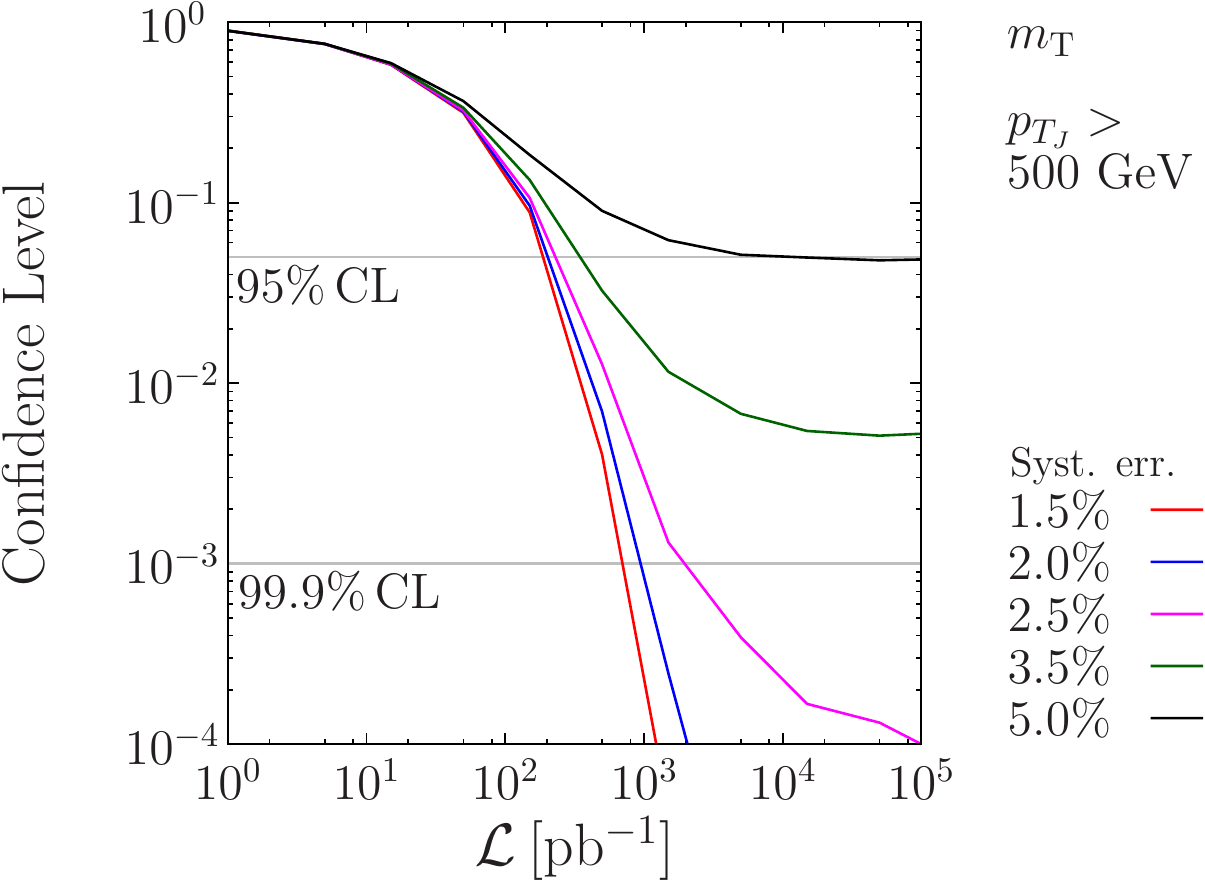}\hfill
  \includegraphics[width=.32\linewidth]{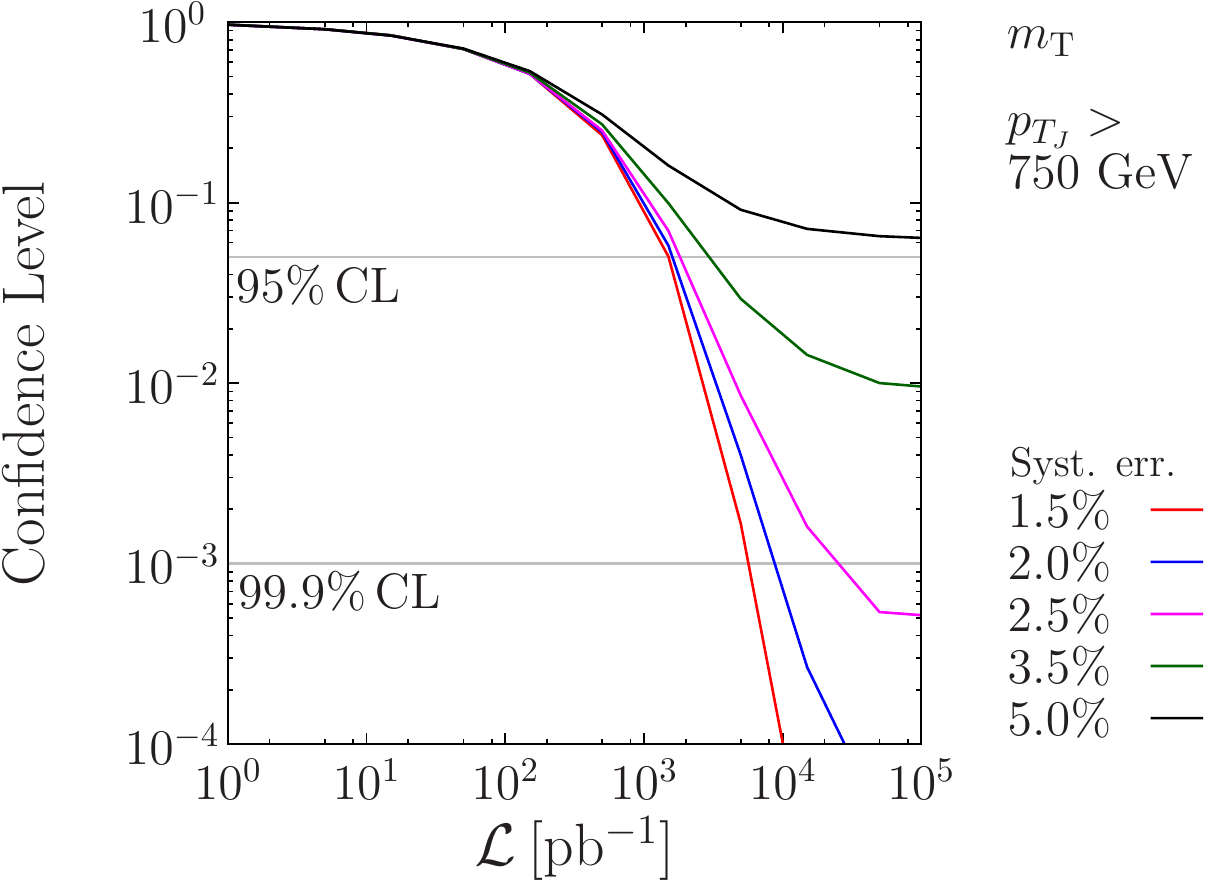}\hfill
  \includegraphics[width=.32\linewidth]{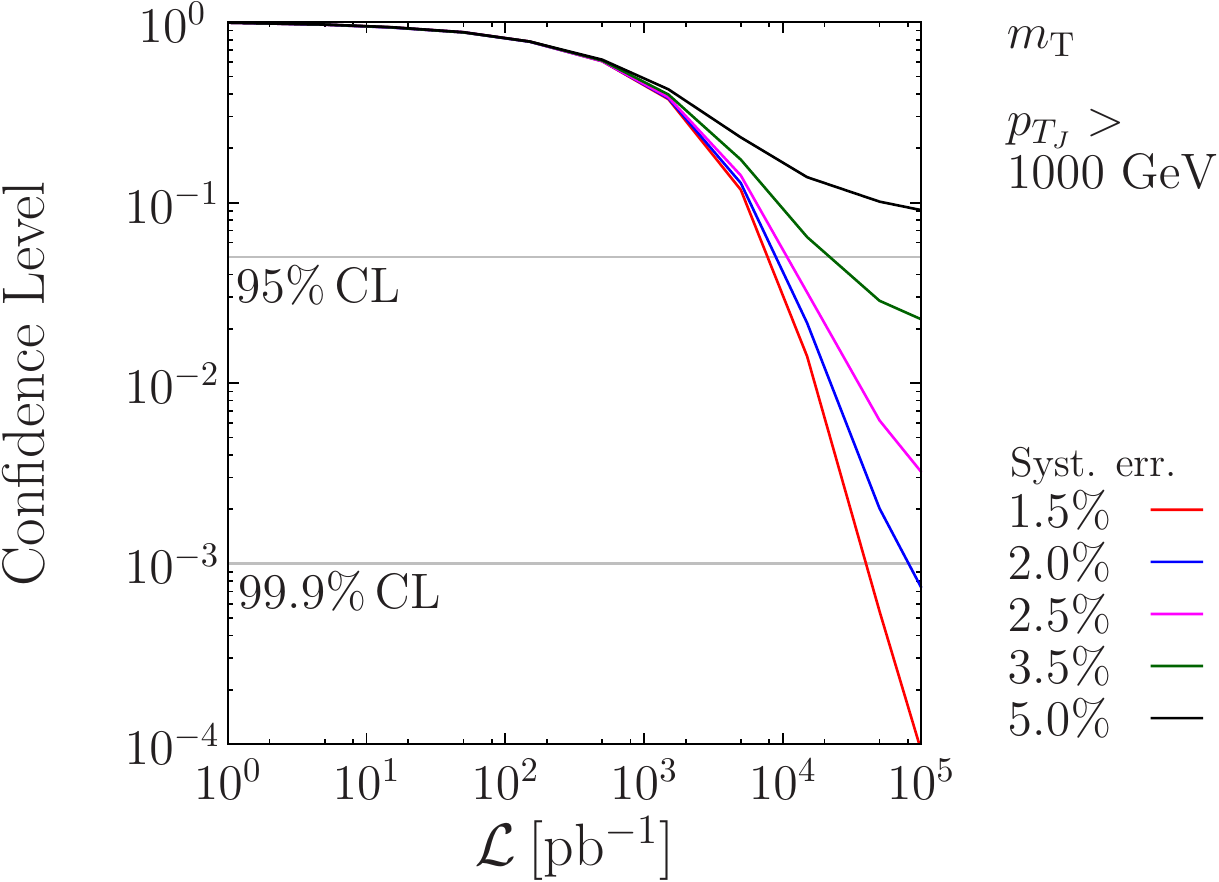}
  \caption{
	    CLs obtained from the $W$ transverse mass $m_\mr{T}$ 
	    reconstruction in the leptonic analysis. The background 
	    sample is the SM emission rate ($f=1$). The signal plus 
	    background sample is $f=1.1$.
	    \label{fig:CLsmT}
	  }
\end{figure}

Second, to evaluate how well one can measure different $W$-emission rates, 
we now use the $\mr{CL_s}$ method and take the Standard Model value $f=1$ as background $B$,
aiming to exclude $S+B$ hypotheses with $f>1$. 
In Figs.~\ref{fig:CLs}-\ref{fig:CLsmT} we show the signal confidence 
levels $\mr{CL_s}$ as a function of the integrated luminosity. In an ideal scenario, without systematic uncertainty, data from 
the LHC will provide sufficient statistics in all methods to 
exclude $f = 1.1$ at $99.9\%$ CL. However, more realistically, 
after adding a systematic error $\sigma_\mr{syst} \geq 1.5\%$ 
$f=1.1$ cannot be excluded anymore with methods \ref{mhard}-\ref{msoft} 
alone. This is not surprising because the $W$ emission rate is only 
increased by $10\%$ with respect to the Standard Model rate, resulting in $S/B \sim \mathcal{O}(1)\%$.

Apparently a more feasible target is $f=2$ for which we show results in 
Fig~\ref{fig:CLs}. We see that $W$ mass 
reconstruction methods \ref{mmed} and \ref{msoft} allow the 
exclusion at $95\%$ CL  with 
$\sigma_\mr{syst}\gtrsim2.5\%$ and $2\%$ respectively. Again, the 
reconstruction involving mass drop and filtering fares better. Using 
$m_{\mr{BDRS}}$, it is possible to achieve $S/B \simeq 20\%$ with the 
$p_{T_J}>1$ TeV selection. This allows to exclude the signal to $95\%$ 
with $5\%$ systematic uncertainty and to $99.9\%$ with $3.5\%$ 
systematic uncertainty.

Consequently, to be able to exclude $f=1.1$ we need to increase $S/B$ by improving on the $W$ 
reconstruction. The combination of the jet shapes 
discussed in Sec.~\ref{sec:hadana}, i.e. $\hat{t}$ and $\tau_{21}$, 
with method \ref{mhard} allows to boost the sensitivity of the 
likelihood ratio. Fig.~\ref{fig:CLsShapes} shows the signal confidence 
level using $\hat{t}$ (left) and $\tau_{21}$ (right) in 
combination with method \ref{mhard}. A fat jet transverse momentum 
cut $p_{T_J}>750$ yields the best result for two reasons: the $W$s 
are boosted enough for the jet shapes to perform well and the cross 
section is large enough to accumulate sufficient statistics during 
Run 2 of the LHC. The combination of method \ref{mhard} with either 
of the jet shapes can exclude the signal to $95\%$ CL with $2.5\%$ 
systematic uncertainty, while $\sigma_\mr{syst} = 1.5\%$ would allow 
to exclude it in favor of the Standard Model to $3 \sigma$. What allows the 
discrimination is the strong rejection of QCD emissions with low jet 
shape values. Still, this additional step is not enough to 
exclude $f=1.1$ at $95\%$ CL given a $5\%$ systematic 
uncertainty. 

Therefore, in order to test small deviations from the Standard Model 
emission rate, a $W$ reconstruction with small statistical 
and systematic uncertainties is needed. The leptonic analysis outlined 
in Sec.~\ref{sec:lepana} can be of use. Even though the cross sections for 
$f=1$ and $f=1.1$ differ only by a few femtobarns, i.e.\ for 
$p_{T,J}>1~\mathrm{TeV}$, the high luminosity or run 2 will provide sufficient 
statistics. A clear advantage of the leptonic reconstruction is the improved 
$S/B \simeq 10\%$ compared to $S/B\simeq1\%$ for the hadronic mass 
reconstruction. This quantitative difference is due to the fact that the 
lepton analysis completely rejects the QCD background. Thus, changing the 
$W$-emission rate by $\mathcal{O}(10)\%$ directly translates to 
$S/B \simeq 10\%$ between the most similar hypotheses, namely $f=1$ and 
$f=1.1$. The large $S/B$ ratio 
achieved by $m_\mr{T}$ allows us to exclude the latter to $95\%$ CL 
with a systematic uncertainty of $5\%$, and to $99.9\%$ with $2.5\%$ error. 
Curiously, the discriminating power of the leptonic analysis diminishes 
with increasing fat jet transverse momentum cut. We attribute it to the 
fact that here the $S/B$ ratio does not improve with more boosted jets 
and to the reduced isolated tagging efficiency explained in 
Sec.~\ref{sec:lepana}. Even though the hadronic decay comes with a larger 
cross section, the improved background rejection in the leptonic analysis 
allows for a better discrimination of non-Standard Model emission rates.

\section{Conclusion} \label{sec:conclusion}
We propose leptonic and hadronic $W$-tagging strategies for bosons produced 
in association with a hard jet. Hadronic $W$ reconstruction relies on jet 
substructure techniques and a subsequent use of jet shapes to separate $W$ 
jets from the overwhelming QCD background. We compare the Standard Model with 
increased $W$ emission rates. Using a binned log-likelihood approach while 
varying "systematic" uncertainties we find that electroweak emission rates 
can be measured at the LHC. 

Hadronic $W$ mass reconstruction methods are capable of distinguishing 
differences in the cross section comparable with the SM value, i.e. 
$\left|\sigma_{f\cdot\mr{SM}} - \sigma_\mr{SM} \right| / \sigma_\mr{SM} \simeq 1$. 
In other words, hadronic mass reconstructions can exclude the $f=0$ and 
$f=2$ hypotheses in favor of the SM ($f=1$) with a $5\%$ systematic 
uncertainty up to $95\%$ CL.  The best subjet-based reconstruction method we 
tested, $m_\mr{BDRS}$, can exclude $f=0$ to $99.9\%$ with 
$\sigma_\mr{syst} = 5\%$ and $f=2$ with $\sigma_\mr{syst} = 3.5\%$. 

However, if 
$\left|\sigma_{f\cdot\mr{SM}} - \sigma_\mr{SM} \right| / \sigma_\mr{SM} \simeq 0.1$, 
i.e.\ $f=1.1$, no hadronic mass reconstruction can discriminate signal from 
background given a non-negligible systematic uncertainty. To this end, we 
calculate jet shapes of the subjets which pass the mass cuts given in 
Sec.~\ref{sec:analysis}. The jet shape distributions we find most useful are 
ellipticity $\hat{t}$ and $\tau_{21}$ applied on boosted subjets with 
$m_\mr{BDRS} \in \left[74,90\right]$ GeV. Binned log-likelihood analyses of 
these distributions allow a $95\%$ CL exclusion of the $f=1.1$ hypothesis with 
$\sigma_\mr{syst} =2.5\%$.  With a smaller error of $1.5\%$ we can exclude 
$f=1.1$ with $99.9\%$ CL. The statistical error at this stage amounts to 
$\mathcal{O}(1)\%$. This suggests the possibility that in the high luminosity 
run, we would be able to exclude the $f=1.1$ hypothesis with an even larger 
systematic uncertainty, or if the systematics are well understood, probe 
smaller values of 
$\left|\sigma_{f\cdot\mr{SM}} - \sigma_\mr{SM} \right| / \sigma_\mr{SM}$.
While this seems a futile exercise -- we essentially know the coupling of
$W$ bosons to quarks -- this kind of measurement will offer an opportunity
for the determination of electroweak Sudakov effects.

The simple leptonic analysis rejects the background more successfully. It 
benefits from the large cross section of the low jet $p_T$ cut. Given a 
systematic uncertainty $\sigma_\mr{syst}\leq 2.5\%$, a cut on the transverse 
mass $m_\mr{T}$ of a leptonic $W$ candidate will exclude $f=1.1$ up to $99.9\%$.
Moreover, it can probe 
$\left|\sigma_{f\cdot\mr{SM}} - \sigma_\mr{SM} \right| / \sigma_\mr{SM} < 0.1$.

\section*{Acknowledgments}

MS acknowledges support 
by the Research Executive Agency (REA) of the European Union under the Grant 
Agreement number PITN-GA-2010-264564 (LHCPhenoNet) and PITN-GA-2012-315877 
(MCnet). MS further gratefully acknowledges fruitful discussion with A.\ 
Denner, kindly supplying the work with F.\ Hebenstreit, A.\ Huss for 
supplying the results of \cite{Dittmaier:2012kx} to cross-check our 
implementation.  FK and MS are grateful for useful discussions with 
S.\ H\"oche and other members of the \Sherpa team. PP would like to acknowledge
the support provided by the Science and Technology Facilities Council (STFC).

\appendix
\section{Ellipticity} \label{App:A}

To calculate the ellipticity $\hat{t}$ of a jet we define the 
particles' three-momentum components $\mathbf{k}_{Ti}$ transverse 
to the jet it is part of. Thus, it is defined in the plane transverse 
to the momentum $\mathbf{p}_J=\sum^{}_{i}\mathbf{p}_i$, where 
$\mathbf{p}_i$ are the three-momenta of the jet constituents, as
\begin{equation}
  \mathbf{k}_{Ti}
  \;=\;\mathbf{p}_i - \left( \mathbf{p}_J \cdot \mathbf{p}_i \right) 
                      \frac{\mathbf{p}_J}{\left| \mathbf{p}_J \right|^2}\;.
\end{equation} 
While we take $\mathbf{p}_J$ to be the thrust axis, we calculate 
thrust major $T_\mr{maj}$ and and thrust minor $T_\mr{min}$ using 
the $\mathbf{k}_{Ti}$ as input
\begin{equation} 
  T_\mr{maj}
  \;=\;\max\limits_{\mathbf{n}_\mr{maj}}\;
       \frac{\sum_{i} \left|\mathbf{k}_{Ti} \cdot \mathbf{n}_\mr{maj} \right|}
            {\sum_{i} \left| \mathbf{p}_{Ti} \right|}
  \qquad\text{and}\qquad
  T_\mr{min}
  \;=\;\frac{\sum_{i} \left|\mathbf{k}_{Ti} \cdot \mathbf{n}_\mr{min} \right|}
            {\sum_{i} \left| \mathbf{p}_{Ti} \right|}\;,
\end{equation}
where $\mathbf{n}_\mr{maj}^2=\mathbf{n}_\mr{min}^2=1$, 
$\mathbf{n}_\mr{min} \cdot \mathbf{n}_\mr{maj} =0$ and 
$\mathbf{n}_\mr{min} \cdot \mathbf{p}_J =0$. We then define 
the ellipticity to be the ratio
\begin{equation} 
  \hat{t}\;=\;\frac{T_\mr{min}}{T_\mr{maj}}\;.
\end{equation}
The thereby defined ellipticity makes a distinction between different 
jet topologies. The two limiting cases are either when the radiation 
of the jet is distributed homogeneously within the jet cone (the energy 
profile in the jet transverse plane is a circle), leading to 
$T_\mr{maj} = T_\mr{min}$ and the ellipticity $\hat{t} = 1$, or when 
the radiation is two-dimensional (the energy profile in the jet 
transverse plane is one-dimensional), leading to $T_\mr{min} = 0$ 
and $T_\mr{maj}>0$ and the ellipticity $\hat{t} = 0$.

\bibliography{references}

\begin{thebibliography}{69}
\expandafter\ifx\csname natexlab\endcsname\relax\def\natexlab#1{#1}\fi
\expandafter\ifx\csname bibnamefont\endcsname\relax
  \def\bibnamefont#1{#1}\fi
\expandafter\ifx\csname bibfnamefont\endcsname\relax
  \def\bibfnamefont#1{#1}\fi
\expandafter\ifx\csname citenamefont\endcsname\relax
  \def\citenamefont#1{#1}\fi
\expandafter\ifx\csname url\endcsname\relax
  \def\url#1{\texttt{#1}}\fi
\expandafter\ifx\csname urlprefix\endcsname\relax\def\urlprefix{URL }\fi
\providecommand{\bibinfo}[2]{#2}
\providecommand{\eprint}[2][]{\url{#2}}

\bibitem[{\citenamefont{Aad et~al.}(2012)}]{Aad:2012tfa}
\bibinfo{author}{\bibfnamefont{G.}~\bibnamefont{Aad}} \bibnamefont{et~al.}
  (\bibinfo{collaboration}{ATLAS Collaboration}), \bibinfo{journal}{Phys.Lett.}
  \textbf{\bibinfo{volume}{B716}}, \bibinfo{pages}{1} (\bibinfo{year}{2012}),
  \eprint{1207.7214}.

\bibitem[{\citenamefont{Chatrchyan et~al.}(2012)}]{Chatrchyan:2012ufa}
\bibinfo{author}{\bibfnamefont{S.}~\bibnamefont{Chatrchyan}}
  \bibnamefont{et~al.} (\bibinfo{collaboration}{CMS Collaboration}),
  \bibinfo{journal}{Phys.Lett.} \textbf{\bibinfo{volume}{B716}},
  \bibinfo{pages}{30} (\bibinfo{year}{2012}), \eprint{1207.7235}.

\bibitem[{\citenamefont{Butterworth et~al.}(2008)\citenamefont{Butterworth,
  Davison, Rubin, and Salam}}]{Butterworth:2008iy}
\bibinfo{author}{\bibfnamefont{J.~M.} \bibnamefont{Butterworth}},
  \bibinfo{author}{\bibfnamefont{A.~R.} \bibnamefont{Davison}},
  \bibinfo{author}{\bibfnamefont{M.}~\bibnamefont{Rubin}}, \bibnamefont{and}
  \bibinfo{author}{\bibfnamefont{G.~P.} \bibnamefont{Salam}},
  \bibinfo{journal}{Phys.Rev.Lett.} \textbf{\bibinfo{volume}{100}},
  \bibinfo{pages}{242001} (\bibinfo{year}{2008}), \eprint{0802.2470}.

\bibitem[{\citenamefont{Krohn et~al.}(2010)\citenamefont{Krohn, Thaler, and
  Wang}}]{Krohn:2009th}
\bibinfo{author}{\bibfnamefont{D.}~\bibnamefont{Krohn}},
  \bibinfo{author}{\bibfnamefont{J.}~\bibnamefont{Thaler}}, \bibnamefont{and}
  \bibinfo{author}{\bibfnamefont{L.-T.} \bibnamefont{Wang}},
  \bibinfo{journal}{JHEP} \textbf{\bibinfo{volume}{1002}}, \bibinfo{pages}{084}
  (\bibinfo{year}{2010}), \eprint{0912.1342}.

\bibitem[{\citenamefont{Ellis et~al.}(2010)\citenamefont{Ellis, Vermilion, and
  Walsh}}]{Ellis:2009me}
\bibinfo{author}{\bibfnamefont{S.~D.} \bibnamefont{Ellis}},
  \bibinfo{author}{\bibfnamefont{C.~K.} \bibnamefont{Vermilion}},
  \bibnamefont{and} \bibinfo{author}{\bibfnamefont{J.~R.} \bibnamefont{Walsh}},
  \bibinfo{journal}{Phys.Rev.} \textbf{\bibinfo{volume}{D81}},
  \bibinfo{pages}{094023} (\bibinfo{year}{2010}), \eprint{0912.0033}.

\bibitem[{\citenamefont{Kim}(2011)}]{Kim:2010uj}
\bibinfo{author}{\bibfnamefont{J.-H.} \bibnamefont{Kim}},
  \bibinfo{journal}{Phys.Rev.} \textbf{\bibinfo{volume}{D83}},
  \bibinfo{pages}{011502} (\bibinfo{year}{2011}), \eprint{1011.1493}.

\bibitem[{\citenamefont{Thaler and Van~Tilburg}(2011)}]{Thaler:2010tr}
\bibinfo{author}{\bibfnamefont{J.}~\bibnamefont{Thaler}} \bibnamefont{and}
  \bibinfo{author}{\bibfnamefont{K.}~\bibnamefont{Van~Tilburg}},
  \bibinfo{journal}{JHEP} \textbf{\bibinfo{volume}{1103}}, \bibinfo{pages}{015}
  (\bibinfo{year}{2011}), \eprint{1011.2268}.

\bibitem[{\citenamefont{Soper and Spannowsky}(2011)}]{Soper:2011cr}
\bibinfo{author}{\bibfnamefont{D.~E.} \bibnamefont{Soper}} \bibnamefont{and}
  \bibinfo{author}{\bibfnamefont{M.}~\bibnamefont{Spannowsky}},
  \bibinfo{journal}{Phys.Rev.} \textbf{\bibinfo{volume}{D84}},
  \bibinfo{pages}{074002} (\bibinfo{year}{2011}), \eprint{1102.3480}.

\bibitem[{\citenamefont{Ellis et~al.}(2012)\citenamefont{Ellis, Hornig, Roy,
  Krohn, and Schwartz}}]{Ellis:2012sn}
\bibinfo{author}{\bibfnamefont{S.~D.} \bibnamefont{Ellis}},
  \bibinfo{author}{\bibfnamefont{A.}~\bibnamefont{Hornig}},
  \bibinfo{author}{\bibfnamefont{T.~S.} \bibnamefont{Roy}},
  \bibinfo{author}{\bibfnamefont{D.}~\bibnamefont{Krohn}}, \bibnamefont{and}
  \bibinfo{author}{\bibfnamefont{M.~D.} \bibnamefont{Schwartz}},
  \bibinfo{journal}{Phys.Rev.Lett.} \textbf{\bibinfo{volume}{108}},
  \bibinfo{pages}{182003} (\bibinfo{year}{2012}), \eprint{1201.1914}.

\bibitem[{\citenamefont{Soper and Spannowsky}(2013)}]{Soper:2012pb}
\bibinfo{author}{\bibfnamefont{D.~E.} \bibnamefont{Soper}} \bibnamefont{and}
  \bibinfo{author}{\bibfnamefont{M.}~\bibnamefont{Spannowsky}},
  \bibinfo{journal}{Phys.Rev.} \textbf{\bibinfo{volume}{D87}},
  \bibinfo{pages}{054012} (\bibinfo{year}{2013}), \eprint{1211.3140}.

\bibitem[{\citenamefont{Ferroglia et~al.}(2013)\citenamefont{Ferroglia, Pecjak,
  and Yang}}]{Ferroglia:2013zwa}
\bibinfo{author}{\bibfnamefont{A.}~\bibnamefont{Ferroglia}},
  \bibinfo{author}{\bibfnamefont{B.~D.} \bibnamefont{Pecjak}},
  \bibnamefont{and} \bibinfo{author}{\bibfnamefont{L.~L.} \bibnamefont{Yang}},
  \bibinfo{journal}{JHEP} \textbf{\bibinfo{volume}{1309}}, \bibinfo{pages}{032}
  (\bibinfo{year}{2013}), \eprint{1306.1537}.

\bibitem[{\citenamefont{Rubin et~al.}(2010)\citenamefont{Rubin, Salam, and
  Sapeta}}]{Rubin:2010xp}
\bibinfo{author}{\bibfnamefont{M.}~\bibnamefont{Rubin}},
  \bibinfo{author}{\bibfnamefont{G.~P.} \bibnamefont{Salam}}, \bibnamefont{and}
  \bibinfo{author}{\bibfnamefont{S.}~\bibnamefont{Sapeta}},
  \bibinfo{journal}{JHEP} \textbf{\bibinfo{volume}{1009}}, \bibinfo{pages}{084}
  (\bibinfo{year}{2010}), \eprint{1006.2144}.

\bibitem[{\citenamefont{Ciafaloni and Comelli}(1999)}]{Ciafaloni:1998xg}
\bibinfo{author}{\bibfnamefont{P.}~\bibnamefont{Ciafaloni}} \bibnamefont{and}
  \bibinfo{author}{\bibfnamefont{D.}~\bibnamefont{Comelli}},
  \bibinfo{journal}{Phys.Lett.} \textbf{\bibinfo{volume}{B446}},
  \bibinfo{pages}{278} (\bibinfo{year}{1999}), \eprint{hep-ph/9809321}.

\bibitem[{\citenamefont{Kuhn et~al.}(2000)\citenamefont{Kuhn, Penin, and
  Smirnov}}]{Kuhn:1999nn}
\bibinfo{author}{\bibfnamefont{J.~H.} \bibnamefont{Kuhn}},
  \bibinfo{author}{\bibfnamefont{A.}~\bibnamefont{Penin}}, \bibnamefont{and}
  \bibinfo{author}{\bibfnamefont{V.~A.} \bibnamefont{Smirnov}},
  \bibinfo{journal}{Eur.Phys.J.} \textbf{\bibinfo{volume}{C17}},
  \bibinfo{pages}{97} (\bibinfo{year}{2000}), \eprint{hep-ph/9912503}.

\bibitem[{\citenamefont{Fadin et~al.}(2000)\citenamefont{Fadin, Lipatov,
  Martin, and Melles}}]{Fadin:1999bq}
\bibinfo{author}{\bibfnamefont{V.~S.} \bibnamefont{Fadin}},
  \bibinfo{author}{\bibfnamefont{L.}~\bibnamefont{Lipatov}},
  \bibinfo{author}{\bibfnamefont{A.~D.} \bibnamefont{Martin}},
  \bibnamefont{and} \bibinfo{author}{\bibfnamefont{M.}~\bibnamefont{Melles}},
  \bibinfo{journal}{Phys.Rev.} \textbf{\bibinfo{volume}{D61}},
  \bibinfo{pages}{094002} (\bibinfo{year}{2000}), \eprint{hep-ph/9910338}.

\bibitem[{\citenamefont{Ciafaloni et~al.}(2000)\citenamefont{Ciafaloni,
  Ciafaloni, and Comelli}}]{Ciafaloni:2000df}
\bibinfo{author}{\bibfnamefont{M.}~\bibnamefont{Ciafaloni}},
  \bibinfo{author}{\bibfnamefont{P.}~\bibnamefont{Ciafaloni}},
  \bibnamefont{and} \bibinfo{author}{\bibfnamefont{D.}~\bibnamefont{Comelli}},
  \bibinfo{journal}{Phys.Rev.Lett.} \textbf{\bibinfo{volume}{84}},
  \bibinfo{pages}{4810} (\bibinfo{year}{2000}), \eprint{hep-ph/0001142}.

\bibitem[{\citenamefont{Hori et~al.}(2000)\citenamefont{Hori, Kawamura, and
  Kodaira}}]{Hori:2000tm}
\bibinfo{author}{\bibfnamefont{M.}~\bibnamefont{Hori}},
  \bibinfo{author}{\bibfnamefont{H.}~\bibnamefont{Kawamura}}, \bibnamefont{and}
  \bibinfo{author}{\bibfnamefont{J.}~\bibnamefont{Kodaira}},
  \bibinfo{journal}{Phys.Lett.} \textbf{\bibinfo{volume}{B491}},
  \bibinfo{pages}{275} (\bibinfo{year}{2000}), \eprint{hep-ph/0007329}.

\bibitem[{\citenamefont{Denner and Pozzorini}(2001)}]{Denner:2000jv}
\bibinfo{author}{\bibfnamefont{A.}~\bibnamefont{Denner}} \bibnamefont{and}
  \bibinfo{author}{\bibfnamefont{S.}~\bibnamefont{Pozzorini}},
  \bibinfo{journal}{Eur.Phys.J.} \textbf{\bibinfo{volume}{C18}},
  \bibinfo{pages}{461} (\bibinfo{year}{2001}), \eprint{hep-ph/0010201}.

\bibitem[{\citenamefont{Accomando et~al.}(2002)\citenamefont{Accomando, Denner,
  and Pozzorini}}]{Accomando:2001fn}
\bibinfo{author}{\bibfnamefont{E.}~\bibnamefont{Accomando}},
  \bibinfo{author}{\bibfnamefont{A.}~\bibnamefont{Denner}}, \bibnamefont{and}
  \bibinfo{author}{\bibfnamefont{S.}~\bibnamefont{Pozzorini}},
  \bibinfo{journal}{Phys.Rev.} \textbf{\bibinfo{volume}{D65}},
  \bibinfo{pages}{073003} (\bibinfo{year}{2002}), \eprint{hep-ph/0110114}.

\bibitem[{\citenamefont{Beenakker and Werthenbach}(2002)}]{Beenakker:2001kf}
\bibinfo{author}{\bibfnamefont{W.}~\bibnamefont{Beenakker}} \bibnamefont{and}
  \bibinfo{author}{\bibfnamefont{A.}~\bibnamefont{Werthenbach}},
  \bibinfo{journal}{Nucl.Phys.} \textbf{\bibinfo{volume}{B630}},
  \bibinfo{pages}{3} (\bibinfo{year}{2002}), \eprint{hep-ph/0112030}.

\bibitem[{\citenamefont{Denner et~al.}(2003)\citenamefont{Denner, Melles, and
  Pozzorini}}]{Denner:2003wi}
\bibinfo{author}{\bibfnamefont{A.}~\bibnamefont{Denner}},
  \bibinfo{author}{\bibfnamefont{M.}~\bibnamefont{Melles}}, \bibnamefont{and}
  \bibinfo{author}{\bibfnamefont{S.}~\bibnamefont{Pozzorini}},
  \bibinfo{journal}{Nucl.Phys.} \textbf{\bibinfo{volume}{B662}},
  \bibinfo{pages}{299} (\bibinfo{year}{2003}), \eprint{hep-ph/0301241}.

\bibitem[{\citenamefont{Denner and Pozzorini}(2005)}]{Denner:2004iz}
\bibinfo{author}{\bibfnamefont{A.}~\bibnamefont{Denner}} \bibnamefont{and}
  \bibinfo{author}{\bibfnamefont{S.}~\bibnamefont{Pozzorini}},
  \bibinfo{journal}{Nucl.Phys.} \textbf{\bibinfo{volume}{B717}},
  \bibinfo{pages}{48} (\bibinfo{year}{2005}), \eprint{hep-ph/0408068}.

\bibitem[{\citenamefont{Denner et~al.}(2008)\citenamefont{Denner, Jantzen, and
  Pozzorini}}]{Denner:2008yn}
\bibinfo{author}{\bibfnamefont{A.}~\bibnamefont{Denner}},
  \bibinfo{author}{\bibfnamefont{B.}~\bibnamefont{Jantzen}}, \bibnamefont{and}
  \bibinfo{author}{\bibfnamefont{S.}~\bibnamefont{Pozzorini}},
  \bibinfo{journal}{JHEP} \textbf{\bibinfo{volume}{0811}}, \bibinfo{pages}{062}
  (\bibinfo{year}{2008}), \eprint{0809.0800}.

\bibitem[{\citenamefont{Baur}(2007)}]{Baur:2006sn}
\bibinfo{author}{\bibfnamefont{U.}~\bibnamefont{Baur}},
  \bibinfo{journal}{Phys.Rev.} \textbf{\bibinfo{volume}{D75}},
  \bibinfo{pages}{013005} (\bibinfo{year}{2007}), \eprint{hep-ph/0611241}.

\bibitem[{\citenamefont{Bell et~al.}(2010)\citenamefont{Bell, Kuhn, and
  Rittinger}}]{Bell:2010gi}
\bibinfo{author}{\bibfnamefont{G.}~\bibnamefont{Bell}},
  \bibinfo{author}{\bibfnamefont{J.}~\bibnamefont{Kuhn}}, \bibnamefont{and}
  \bibinfo{author}{\bibfnamefont{J.}~\bibnamefont{Rittinger}},
  \bibinfo{journal}{Eur.Phys.J.} \textbf{\bibinfo{volume}{C70}},
  \bibinfo{pages}{659} (\bibinfo{year}{2010}), \eprint{1004.4117}.

\bibitem[{\citenamefont{Dittmaier et~al.}(2012)\citenamefont{Dittmaier, Huss,
  and Speckner}}]{Dittmaier:2012kx}
\bibinfo{author}{\bibfnamefont{S.}~\bibnamefont{Dittmaier}},
  \bibinfo{author}{\bibfnamefont{A.}~\bibnamefont{Huss}}, \bibnamefont{and}
  \bibinfo{author}{\bibfnamefont{C.}~\bibnamefont{Speckner}},
  \bibinfo{journal}{JHEP} \textbf{\bibinfo{volume}{1211}}, \bibinfo{pages}{095}
  (\bibinfo{year}{2012}), \eprint{1210.0438}.

\bibitem[{\citenamefont{K{\"u}hn et~al.}(2013)\citenamefont{K{\"u}hn, Scharf,
  and Uwer}}]{Kuhn:2013zoa}
\bibinfo{author}{\bibfnamefont{J.}~\bibnamefont{K{\"u}hn}},
  \bibinfo{author}{\bibfnamefont{A.}~\bibnamefont{Scharf}}, \bibnamefont{and}
  \bibinfo{author}{\bibfnamefont{P.}~\bibnamefont{Uwer}}
  (\bibinfo{year}{2013}), \eprint{1305.5773}.

\bibitem[{\citenamefont{Bierweiler et~al.}(2013)\citenamefont{Bierweiler,
  Kasprzik, and K{\"u}hn}}]{Bierweiler:2013dja}
\bibinfo{author}{\bibfnamefont{A.}~\bibnamefont{Bierweiler}},
  \bibinfo{author}{\bibfnamefont{T.}~\bibnamefont{Kasprzik}}, \bibnamefont{and}
  \bibinfo{author}{\bibfnamefont{J.~H.} \bibnamefont{K{\"u}hn}},
  \bibinfo{journal}{JHEP} \textbf{\bibinfo{volume}{1312}}, \bibinfo{pages}{071}
  (\bibinfo{year}{2013}), \eprint{1305.5402}.

\bibitem[{\citenamefont{Mangano et~al.}(2003)\citenamefont{Mangano, Moretti,
  Piccinini, Pittau, and Polosa}}]{Mangano:2002ea}
\bibinfo{author}{\bibfnamefont{M.~L.} \bibnamefont{Mangano}},
  \bibinfo{author}{\bibfnamefont{M.}~\bibnamefont{Moretti}},
  \bibinfo{author}{\bibfnamefont{F.}~\bibnamefont{Piccinini}},
  \bibinfo{author}{\bibfnamefont{R.}~\bibnamefont{Pittau}}, \bibnamefont{and}
  \bibinfo{author}{\bibfnamefont{A.~D.} \bibnamefont{Polosa}},
  \bibinfo{journal}{JHEP} \textbf{\bibinfo{volume}{07}}, \bibinfo{pages}{001}
  (\bibinfo{year}{2003}), \eprint{hep-ph/0206293}.

\bibitem[{\citenamefont{Alwall et~al.}(2011)\citenamefont{Alwall, Herquet,
  Maltoni, Mattelaer, and Stelzer}}]{Alwall:2011uj}
\bibinfo{author}{\bibfnamefont{J.}~\bibnamefont{Alwall}},
  \bibinfo{author}{\bibfnamefont{M.}~\bibnamefont{Herquet}},
  \bibinfo{author}{\bibfnamefont{F.}~\bibnamefont{Maltoni}},
  \bibinfo{author}{\bibfnamefont{O.}~\bibnamefont{Mattelaer}},
  \bibnamefont{and} \bibinfo{author}{\bibfnamefont{T.}~\bibnamefont{Stelzer}},
  \bibinfo{journal}{JHEP} \textbf{\bibinfo{volume}{1106}}, \bibinfo{pages}{128}
  (\bibinfo{year}{2011}), \eprint{1106.0522}.

\bibitem[{\citenamefont{Krauss et~al.}(2002)\citenamefont{Krauss, Kuhn, and
  Soff}}]{Krauss:2001iv}
\bibinfo{author}{\bibfnamefont{F.}~\bibnamefont{Krauss}},
  \bibinfo{author}{\bibfnamefont{R.}~\bibnamefont{Kuhn}}, \bibnamefont{and}
  \bibinfo{author}{\bibfnamefont{G.}~\bibnamefont{Soff}},
  \bibinfo{journal}{JHEP} \textbf{\bibinfo{volume}{02}}, \bibinfo{pages}{044}
  (\bibinfo{year}{2002}), \eprint{hep-ph/0109036}.

\bibitem[{\citenamefont{Gleisberg and H{\"o}che}(2008)}]{Gleisberg:2008fv}
\bibinfo{author}{\bibfnamefont{T.}~\bibnamefont{Gleisberg}} \bibnamefont{and}
  \bibinfo{author}{\bibfnamefont{S.}~\bibnamefont{H{\"o}che}},
  \bibinfo{journal}{JHEP} \textbf{\bibinfo{volume}{0812}}, \bibinfo{pages}{039}
  (\bibinfo{year}{2008}), \eprint{0808.3674}.

\bibitem[{\citenamefont{Kilian et~al.}(2011)\citenamefont{Kilian, Ohl, and
  Reuter}}]{Kilian:2007gr}
\bibinfo{author}{\bibfnamefont{W.}~\bibnamefont{Kilian}},
  \bibinfo{author}{\bibfnamefont{T.}~\bibnamefont{Ohl}}, \bibnamefont{and}
  \bibinfo{author}{\bibfnamefont{J.}~\bibnamefont{Reuter}},
  \bibinfo{journal}{Eur.Phys.J.} \textbf{\bibinfo{volume}{C71}},
  \bibinfo{pages}{1742} (\bibinfo{year}{2011}), \eprint{0708.4233}.

\bibitem[{\citenamefont{Dai et~al.}(2012)\citenamefont{Dai, Melnikov, and
  Caola}}]{Dai:2012jh}
\bibinfo{author}{\bibfnamefont{L.}~\bibnamefont{Dai}},
  \bibinfo{author}{\bibfnamefont{K.}~\bibnamefont{Melnikov}}, \bibnamefont{and}
  \bibinfo{author}{\bibfnamefont{F.}~\bibnamefont{Caola}},
  \bibinfo{journal}{JHEP} \textbf{\bibinfo{volume}{1204}}, \bibinfo{pages}{095}
  (\bibinfo{year}{2012}), \eprint{1201.1523}.

\bibitem[{\citenamefont{Christiansen and
  Sj{\"o}strand}(2014)}]{Christiansen:2014kba}
\bibinfo{author}{\bibfnamefont{J.~R.} \bibnamefont{Christiansen}}
  \bibnamefont{and}
  \bibinfo{author}{\bibfnamefont{T.}~\bibnamefont{Sj{\"o}strand}}
  (\bibinfo{year}{2014}), \eprint{1401.5238}.

\bibitem[{\citenamefont{Sj{\"o}strand et~al.}(2008)\citenamefont{Sj{\"o}strand,
  Mrenna, and Skands}}]{Sjostrand:2007gs}
\bibinfo{author}{\bibfnamefont{T.}~\bibnamefont{Sj{\"o}strand}},
  \bibinfo{author}{\bibfnamefont{S.}~\bibnamefont{Mrenna}}, \bibnamefont{and}
  \bibinfo{author}{\bibfnamefont{P.~Z.} \bibnamefont{Skands}},
  \bibinfo{journal}{Comput.Phys.Commun.} \textbf{\bibinfo{volume}{178}},
  \bibinfo{pages}{852} (\bibinfo{year}{2008}), \eprint{0710.3820}.

\bibitem[{\citenamefont{Gleisberg et~al.}(2009)\citenamefont{Gleisberg,
  H{\"o}che, Krauss, Sch{\"o}nherr, Schumann et~al.}}]{Gleisberg:2008ta}
\bibinfo{author}{\bibfnamefont{T.}~\bibnamefont{Gleisberg}},
  \bibinfo{author}{\bibfnamefont{S.}~\bibnamefont{H{\"o}che}},
  \bibinfo{author}{\bibfnamefont{F.}~\bibnamefont{Krauss}},
  \bibinfo{author}{\bibfnamefont{M.}~\bibnamefont{Sch{\"o}nherr}},
  \bibinfo{author}{\bibfnamefont{S.}~\bibnamefont{Schumann}},
  \bibnamefont{et~al.}, \bibinfo{journal}{JHEP}
  \textbf{\bibinfo{volume}{0902}}, \bibinfo{pages}{007} (\bibinfo{year}{2009}),
  \eprint{0811.4622}.

\bibitem[{\citenamefont{Nagy and Soper}(2005)}]{Nagy:2005aa}
\bibinfo{author}{\bibfnamefont{Z.}~\bibnamefont{Nagy}} \bibnamefont{and}
  \bibinfo{author}{\bibfnamefont{D.~E.} \bibnamefont{Soper}},
  \bibinfo{journal}{JHEP} \textbf{\bibinfo{volume}{0510}}, \bibinfo{pages}{024}
  (\bibinfo{year}{2005}), \eprint{hep-ph/0503053}.

\bibitem[{\citenamefont{Schumann and Krauss}(2008)}]{Schumann:2007mg}
\bibinfo{author}{\bibfnamefont{S.}~\bibnamefont{Schumann}} \bibnamefont{and}
  \bibinfo{author}{\bibfnamefont{F.}~\bibnamefont{Krauss}},
  \bibinfo{journal}{JHEP} \textbf{\bibinfo{volume}{0803}}, \bibinfo{pages}{038}
  (\bibinfo{year}{2008}), \eprint{0709.1027}.

\bibitem[{\citenamefont{Winter et~al.}(2004)\citenamefont{Winter, Krauss, and
  Soff}}]{Winter:2003tt}
\bibinfo{author}{\bibfnamefont{J.-C.} \bibnamefont{Winter}},
  \bibinfo{author}{\bibfnamefont{F.}~\bibnamefont{Krauss}}, \bibnamefont{and}
  \bibinfo{author}{\bibfnamefont{G.}~\bibnamefont{Soff}},
  \bibinfo{journal}{Eur.Phys.J.} \textbf{\bibinfo{volume}{C36}},
  \bibinfo{pages}{381} (\bibinfo{year}{2004}), \eprint{hep-ph/0311085}.

\bibitem[{\citenamefont{Alekhin et~al.}(2005)\citenamefont{Alekhin, Altarelli,
  Amapane, Andersen, Andreev et~al.}}]{Alekhin:2005dx}
\bibinfo{author}{\bibfnamefont{S.}~\bibnamefont{Alekhin}},
  \bibinfo{author}{\bibfnamefont{G.}~\bibnamefont{Altarelli}},
  \bibinfo{author}{\bibfnamefont{N.}~\bibnamefont{Amapane}},
  \bibinfo{author}{\bibfnamefont{J.}~\bibnamefont{Andersen}},
  \bibinfo{author}{\bibfnamefont{V.}~\bibnamefont{Andreev}},
  \bibnamefont{et~al.} (\bibinfo{year}{2005}), \eprint{hep-ph/0601012}.

\bibitem[{\citenamefont{Sch{\"o}nherr and Krauss}(2008)}]{Schonherr:2008av}
\bibinfo{author}{\bibfnamefont{M.}~\bibnamefont{Sch{\"o}nherr}}
  \bibnamefont{and} \bibinfo{author}{\bibfnamefont{F.}~\bibnamefont{Krauss}},
  \bibinfo{journal}{JHEP} \textbf{\bibinfo{volume}{0812}}, \bibinfo{pages}{018}
  (\bibinfo{year}{2008}), \eprint{0810.5071}.

\bibitem[{\citenamefont{Lai et~al.}(2010)\citenamefont{Lai, Guzzi, Huston, Li,
  Nadolsky et~al.}}]{Lai:2010vv}
\bibinfo{author}{\bibfnamefont{H.-L.} \bibnamefont{Lai}},
  \bibinfo{author}{\bibfnamefont{M.}~\bibnamefont{Guzzi}},
  \bibinfo{author}{\bibfnamefont{J.}~\bibnamefont{Huston}},
  \bibinfo{author}{\bibfnamefont{Z.}~\bibnamefont{Li}},
  \bibinfo{author}{\bibfnamefont{P.~M.} \bibnamefont{Nadolsky}},
  \bibnamefont{et~al.}, \bibinfo{journal}{Phys.Rev.}
  \textbf{\bibinfo{volume}{D82}}, \bibinfo{pages}{074024}
  (\bibinfo{year}{2010}), \eprint{1007.2241}.

\bibitem[{\citenamefont{Seymour}(1992)}]{Seymour:1991xa}
\bibinfo{author}{\bibfnamefont{M.~H.} \bibnamefont{Seymour}},
  \bibinfo{journal}{Z.Phys.} \textbf{\bibinfo{volume}{C56}},
  \bibinfo{pages}{161} (\bibinfo{year}{1992}).

\bibitem[{\citenamefont{Sj{\"o}strand et~al.}(2006)\citenamefont{Sj{\"o}strand,
  Mrenna, and Skands}}]{Sjostrand:2006za}
\bibinfo{author}{\bibfnamefont{T.}~\bibnamefont{Sj{\"o}strand}},
  \bibinfo{author}{\bibfnamefont{S.}~\bibnamefont{Mrenna}}, \bibnamefont{and}
  \bibinfo{author}{\bibfnamefont{P.~Z.} \bibnamefont{Skands}},
  \bibinfo{journal}{JHEP} \textbf{\bibinfo{volume}{0605}}, \bibinfo{pages}{026}
  (\bibinfo{year}{2006}), \eprint{hep-ph/0603175}.

\bibitem[{\citenamefont{Bahr et~al.}(2008)\citenamefont{Bahr, Gieseke, Gigg,
  Grellscheid, Hamilton et~al.}}]{Bahr:2008pv}
\bibinfo{author}{\bibfnamefont{M.}~\bibnamefont{Bahr}},
  \bibinfo{author}{\bibfnamefont{S.}~\bibnamefont{Gieseke}},
  \bibinfo{author}{\bibfnamefont{M.}~\bibnamefont{Gigg}},
  \bibinfo{author}{\bibfnamefont{D.}~\bibnamefont{Grellscheid}},
  \bibinfo{author}{\bibfnamefont{K.}~\bibnamefont{Hamilton}},
  \bibnamefont{et~al.}, \bibinfo{journal}{Eur.Phys.J.}
  \textbf{\bibinfo{volume}{C58}}, \bibinfo{pages}{639} (\bibinfo{year}{2008}),
  \eprint{0803.0883}.

\bibitem[{\citenamefont{H{\"o}che et~al.}(2010)\citenamefont{H{\"o}che,
  Schumann, and Siegert}}]{Hoeche:2009xc}
\bibinfo{author}{\bibfnamefont{S.}~\bibnamefont{H{\"o}che}},
  \bibinfo{author}{\bibfnamefont{S.}~\bibnamefont{Schumann}}, \bibnamefont{and}
  \bibinfo{author}{\bibfnamefont{F.}~\bibnamefont{Siegert}},
  \bibinfo{journal}{Phys.Rev.} \textbf{\bibinfo{volume}{D81}},
  \bibinfo{pages}{034026} (\bibinfo{year}{2010}), \eprint{0912.3501}.

\bibitem[{\citenamefont{Denner and Hebenstreit}(2006)}]{Denner:2006xxx}
\bibinfo{author}{\bibfnamefont{A.}~\bibnamefont{Denner}} \bibnamefont{and}
  \bibinfo{author}{\bibfnamefont{F.}~\bibnamefont{Hebenstreit}},
  \bibinfo{journal}{unpublished}  (\bibinfo{year}{2006}).

\bibitem[{\citenamefont{Catani et~al.}(2001)\citenamefont{Catani, Krauss, Kuhn,
  and Webber}}]{Catani:2001cc}
\bibinfo{author}{\bibfnamefont{S.}~\bibnamefont{Catani}},
  \bibinfo{author}{\bibfnamefont{F.}~\bibnamefont{Krauss}},
  \bibinfo{author}{\bibfnamefont{R.}~\bibnamefont{Kuhn}}, \bibnamefont{and}
  \bibinfo{author}{\bibfnamefont{B.}~\bibnamefont{Webber}},
  \bibinfo{journal}{JHEP} \textbf{\bibinfo{volume}{0111}}, \bibinfo{pages}{063}
  (\bibinfo{year}{2001}), \eprint{hep-ph/0109231}.

\bibitem[{\citenamefont{H{\"o}che et~al.}(2009)\citenamefont{H{\"o}che, Krauss,
  Schumann, and Siegert}}]{Hoeche:2009rj}
\bibinfo{author}{\bibfnamefont{S.}~\bibnamefont{H{\"o}che}},
  \bibinfo{author}{\bibfnamefont{F.}~\bibnamefont{Krauss}},
  \bibinfo{author}{\bibfnamefont{S.}~\bibnamefont{Schumann}}, \bibnamefont{and}
  \bibinfo{author}{\bibfnamefont{F.}~\bibnamefont{Siegert}},
  \bibinfo{journal}{JHEP} \textbf{\bibinfo{volume}{0905}}, \bibinfo{pages}{053}
  (\bibinfo{year}{2009}), \eprint{0903.1219}.

\bibitem[{\citenamefont{L{\"o}nnblad}(2002)}]{Lonnblad:2001iq}
\bibinfo{author}{\bibfnamefont{L.}~\bibnamefont{L{\"o}nnblad}},
  \bibinfo{journal}{JHEP} \textbf{\bibinfo{volume}{0205}}, \bibinfo{pages}{046}
  (\bibinfo{year}{2002}), \eprint{hep-ph/0112284}.

\bibitem[{\citenamefont{Mangano et~al.}(2007)\citenamefont{Mangano, Moretti,
  Piccinini, and Treccani}}]{Mangano:2006rw}
\bibinfo{author}{\bibfnamefont{M.~L.} \bibnamefont{Mangano}},
  \bibinfo{author}{\bibfnamefont{M.}~\bibnamefont{Moretti}},
  \bibinfo{author}{\bibfnamefont{F.}~\bibnamefont{Piccinini}},
  \bibnamefont{and} \bibinfo{author}{\bibfnamefont{M.}~\bibnamefont{Treccani}},
  \bibinfo{journal}{JHEP} \textbf{\bibinfo{volume}{0701}}, \bibinfo{pages}{013}
  (\bibinfo{year}{2007}), \eprint{hep-ph/0611129}.

\bibitem[{\citenamefont{Alwall et~al.}(2008)\citenamefont{Alwall, Hoche,
  Krauss, Lavesson, Lonnblad et~al.}}]{Alwall:2007fs}
\bibinfo{author}{\bibfnamefont{J.}~\bibnamefont{Alwall}},
  \bibinfo{author}{\bibfnamefont{S.}~\bibnamefont{Hoche}},
  \bibinfo{author}{\bibfnamefont{F.}~\bibnamefont{Krauss}},
  \bibinfo{author}{\bibfnamefont{N.}~\bibnamefont{Lavesson}},
  \bibinfo{author}{\bibfnamefont{L.}~\bibnamefont{Lonnblad}},
  \bibnamefont{et~al.}, \bibinfo{journal}{Eur.Phys.J.}
  \textbf{\bibinfo{volume}{C53}}, \bibinfo{pages}{473} (\bibinfo{year}{2008}),
  \eprint{0706.2569}.

\bibitem[{\citenamefont{Hamilton et~al.}(2009)\citenamefont{Hamilton,
  Richardson, and Tully}}]{Hamilton:2009ne}
\bibinfo{author}{\bibfnamefont{K.}~\bibnamefont{Hamilton}},
  \bibinfo{author}{\bibfnamefont{P.}~\bibnamefont{Richardson}},
  \bibnamefont{and} \bibinfo{author}{\bibfnamefont{J.}~\bibnamefont{Tully}},
  \bibinfo{journal}{JHEP} \textbf{\bibinfo{volume}{0911}}, \bibinfo{pages}{038}
  (\bibinfo{year}{2009}), \eprint{0905.3072}.

\bibitem[{\citenamefont{L{\"o}nnblad and Prestel}(2012)}]{Lonnblad:2011xx}
\bibinfo{author}{\bibfnamefont{L.}~\bibnamefont{L{\"o}nnblad}}
  \bibnamefont{and} \bibinfo{author}{\bibfnamefont{S.}~\bibnamefont{Prestel}},
  \bibinfo{journal}{JHEP} \textbf{\bibinfo{volume}{1203}}, \bibinfo{pages}{019}
  (\bibinfo{year}{2012}), \eprint{1109.4829}.

\bibitem[{\citenamefont{L{\"o}nnblad and Prestel}(2013)}]{Lonnblad:2012ng}
\bibinfo{author}{\bibfnamefont{L.}~\bibnamefont{L{\"o}nnblad}}
  \bibnamefont{and} \bibinfo{author}{\bibfnamefont{S.}~\bibnamefont{Prestel}},
  \bibinfo{journal}{JHEP} \textbf{\bibinfo{volume}{1302}}, \bibinfo{pages}{094}
  (\bibinfo{year}{2013}), \eprint{1211.4827}.

\bibitem[{\citenamefont{Pl{\"a}tzer}(2013)}]{Platzer:2012bs}
\bibinfo{author}{\bibfnamefont{S.}~\bibnamefont{Pl{\"a}tzer}},
  \bibinfo{journal}{JHEP} \textbf{\bibinfo{volume}{1308}}, \bibinfo{pages}{114}
  (\bibinfo{year}{2013}), \eprint{1211.5467}.

\bibitem[{\citenamefont{Carli et~al.}(2010)\citenamefont{Carli, Gehrmann, and
  H{\"o}che}}]{Carli:2010cg}
\bibinfo{author}{\bibfnamefont{T.}~\bibnamefont{Carli}},
  \bibinfo{author}{\bibfnamefont{T.}~\bibnamefont{Gehrmann}}, \bibnamefont{and}
  \bibinfo{author}{\bibfnamefont{S.}~\bibnamefont{H{\"o}che}},
  \bibinfo{journal}{Eur.Phys.J.} \textbf{\bibinfo{volume}{C67}},
  \bibinfo{pages}{73} (\bibinfo{year}{2010}), \eprint{0912.3715}.

\bibitem[{\citenamefont{Catani and Seymour}(1997)}]{Catani:1996vz}
\bibinfo{author}{\bibfnamefont{S.}~\bibnamefont{Catani}} \bibnamefont{and}
  \bibinfo{author}{\bibfnamefont{M.}~\bibnamefont{Seymour}},
  \bibinfo{journal}{Nucl.Phys.} \textbf{\bibinfo{volume}{B485}},
  \bibinfo{pages}{291} (\bibinfo{year}{1997}), \eprint{hep-ph/9605323}.

\bibitem[{\citenamefont{Catani et~al.}(2002)\citenamefont{Catani, Dittmaier,
  Seymour, and Trocsanyi}}]{Catani:2002hc}
\bibinfo{author}{\bibfnamefont{S.}~\bibnamefont{Catani}},
  \bibinfo{author}{\bibfnamefont{S.}~\bibnamefont{Dittmaier}},
  \bibinfo{author}{\bibfnamefont{M.~H.} \bibnamefont{Seymour}},
  \bibnamefont{and}
  \bibinfo{author}{\bibfnamefont{Z.}~\bibnamefont{Trocsanyi}},
  \bibinfo{journal}{Nucl.Phys.} \textbf{\bibinfo{volume}{B627}},
  \bibinfo{pages}{189} (\bibinfo{year}{2002}), \eprint{hep-ph/0201036}.

\bibitem[{\citenamefont{Nagy and Soper}(2008)}]{Nagy:2008eq}
\bibinfo{author}{\bibfnamefont{Z.}~\bibnamefont{Nagy}} \bibnamefont{and}
  \bibinfo{author}{\bibfnamefont{D.~E.} \bibnamefont{Soper}},
  \bibinfo{journal}{JHEP} \textbf{\bibinfo{volume}{0807}}, \bibinfo{pages}{025}
  (\bibinfo{year}{2008}), \eprint{0805.0216}.

\bibitem[{\citenamefont{Hoeche et~al.}(2012)\citenamefont{Hoeche, Krauss,
  Schonherr, and Siegert}}]{Hoeche:2011fd}
\bibinfo{author}{\bibfnamefont{S.}~\bibnamefont{Hoeche}},
  \bibinfo{author}{\bibfnamefont{F.}~\bibnamefont{Krauss}},
  \bibinfo{author}{\bibfnamefont{M.}~\bibnamefont{Schonherr}},
  \bibnamefont{and} \bibinfo{author}{\bibfnamefont{F.}~\bibnamefont{Siegert}},
  \bibinfo{journal}{JHEP} \textbf{\bibinfo{volume}{1209}}, \bibinfo{pages}{049}
  (\bibinfo{year}{2012}), \eprint{1111.1220}.

\bibitem[{\citenamefont{Cacciari et~al.}(2008)\citenamefont{Cacciari, Salam,
  and Soyez}}]{Cacciari:2008gp}
\bibinfo{author}{\bibfnamefont{M.}~\bibnamefont{Cacciari}},
  \bibinfo{author}{\bibfnamefont{G.~P.} \bibnamefont{Salam}}, \bibnamefont{and}
  \bibinfo{author}{\bibfnamefont{G.}~\bibnamefont{Soyez}},
  \bibinfo{journal}{JHEP} \textbf{\bibinfo{volume}{0804}}, \bibinfo{pages}{063}
  (\bibinfo{year}{2008}), \eprint{0802.1189}.

\bibitem[{\citenamefont{Cacciari et~al.}(2012)\citenamefont{Cacciari, Salam,
  and Soyez}}]{Cacciari:2011ma}
\bibinfo{author}{\bibfnamefont{M.}~\bibnamefont{Cacciari}},
  \bibinfo{author}{\bibfnamefont{G.~P.} \bibnamefont{Salam}}, \bibnamefont{and}
  \bibinfo{author}{\bibfnamefont{G.}~\bibnamefont{Soyez}},
  \bibinfo{journal}{Eur.Phys.J.} \textbf{\bibinfo{volume}{C72}},
  \bibinfo{pages}{1896} (\bibinfo{year}{2012}), \eprint{1111.6097}.

\bibitem[{\citenamefont{Buckley et~al.}(2013)\citenamefont{Buckley,
  Butterworth, L{\"o}nnblad, Grellscheid, Hoeth et~al.}}]{Buckley:2010ar}
\bibinfo{author}{\bibfnamefont{A.}~\bibnamefont{Buckley}},
  \bibinfo{author}{\bibfnamefont{J.}~\bibnamefont{Butterworth}},
  \bibinfo{author}{\bibfnamefont{L.}~\bibnamefont{L{\"o}nnblad}},
  \bibinfo{author}{\bibfnamefont{D.}~\bibnamefont{Grellscheid}},
  \bibinfo{author}{\bibfnamefont{H.}~\bibnamefont{Hoeth}},
  \bibnamefont{et~al.}, \bibinfo{journal}{Comput.Phys.Commun.}
  \textbf{\bibinfo{volume}{184}}, \bibinfo{pages}{2803} (\bibinfo{year}{2013}),
  \eprint{1003.0694}.

\bibitem[{\citenamefont{Dokshitzer et~al.}(1997)\citenamefont{Dokshitzer,
  Leder, Moretti, and Webber}}]{Dokshitzer:1997in}
\bibinfo{author}{\bibfnamefont{Y.~L.} \bibnamefont{Dokshitzer}},
  \bibinfo{author}{\bibfnamefont{G.}~\bibnamefont{Leder}},
  \bibinfo{author}{\bibfnamefont{S.}~\bibnamefont{Moretti}}, \bibnamefont{and}
  \bibinfo{author}{\bibfnamefont{B.}~\bibnamefont{Webber}},
  \bibinfo{journal}{JHEP} \textbf{\bibinfo{volume}{9708}}, \bibinfo{pages}{001}
  (\bibinfo{year}{1997}), \eprint{hep-ph/9707323}.

\bibitem[{\citenamefont{Rehermann and Tweedie}(2011)}]{Rehermann:2010vq}
\bibinfo{author}{\bibfnamefont{K.}~\bibnamefont{Rehermann}} \bibnamefont{and}
  \bibinfo{author}{\bibfnamefont{B.}~\bibnamefont{Tweedie}},
  \bibinfo{journal}{JHEP} \textbf{\bibinfo{volume}{1103}}, \bibinfo{pages}{059}
  (\bibinfo{year}{2011}), \eprint{1007.2221}.

\bibitem[{\citenamefont{Junk}(1999)}]{Junk:1999kv}
\bibinfo{author}{\bibfnamefont{T.}~\bibnamefont{Junk}},
  \bibinfo{journal}{Nucl.Instrum.Meth.} \textbf{\bibinfo{volume}{A434}},
  \bibinfo{pages}{435} (\bibinfo{year}{1999}), \eprint{hep-ex/9902006}.

\bibitem[{\citenamefont{Brun and Rademakers}(1997)}]{Brun:1997pa}
\bibinfo{author}{\bibfnamefont{R.}~\bibnamefont{Brun}} \bibnamefont{and}
  \bibinfo{author}{\bibfnamefont{F.}~\bibnamefont{Rademakers}},
  \bibinfo{journal}{Nucl.Instrum.Meth.} \textbf{\bibinfo{volume}{A389}},
  \bibinfo{pages}{81} (\bibinfo{year}{1997}).

\end{thebibliography}

\end{document}